\documentclass[dvips,%twocolumn,10pt
]{article}
\usepackage[english]{babel}
\usepackage{times}
\usepackage{pstricks}
\usepackage{fullpage}

\newdimen\proofrulebreadth \proofrulebreadth=.05em
\newdimen\proofdotseparation \proofdotseparation=1.25ex
\newdimen\proofrulebaseline \proofrulebaseline=2ex
\newcount\proofdotnumber \proofdotnumber=3
\let\then\relax
\def\hfi{\hskip0pt plus.0001fil}
\mathchardef\squigto="3A3B
\newif\ifinsideprooftree\insideprooftreefalse
\newif\ifonleftofproofrule\onleftofproofrulefalse
\newif\ifproofdots\proofdotsfalse
\newif\ifdoubleproof\doubleprooffalse
\let\wereinproofbit\relax
\newdimen\shortenproofleft
\newdimen\shortenproofright
\newdimen\proofbelowshift
\newbox\proofabove
\newbox\proofbelow
\newbox\proofrulename
\def\shiftproofbelow{\let\next\relax\afterassignment\setshiftproofbelow\dimen0 }
\def\shiftproofbelowneg{\def\next{\multiply\dimen0 by-1 }%
\afterassignment\setshiftproofbelow\dimen0 }
\def\setshiftproofbelow{\next\proofbelowshift=\dimen0 }
\def\setproofrulebreadth{\proofrulebreadth}

\def\prooftree{% NESTED ZERO (\ifonleftofproofrule)
\ifnum  \lastpenalty=1
\then   \unpenalty
\else   \onleftofproofrulefalse
\fi
\ifonleftofproofrule
\else   \ifinsideprooftree
        \then   \hskip.5em plus1fil
        \fi
\fi
\bgroup% NESTED ONE (\proofbelow, \proofrulename, \proofabove,
\setbox\proofbelow=\hbox{}\setbox\proofrulename=\hbox{}%
\let\justifies\proofover\let\leadsto\proofoverdots\let\Justifies\proofoverdbl
\let\using\proofusing\let\[\prooftree
\ifinsideprooftree\let\]\endprooftree\fi
\proofdotsfalse\doubleprooffalse
\let\thickness\setproofrulebreadth
\let\shiftright\shiftproofbelow \let\shift\shiftproofbelow
\let\shiftleft\shiftproofbelowneg
\let\ifwasinsideprooftree\ifinsideprooftree
\insideprooftreetrue
\setbox\proofabove=\hbox\bgroup$\displaystyle % NESTED TWO
\let\wereinproofbit\prooftree
\shortenproofleft=0pt \shortenproofright=0pt \proofbelowshift=0pt
\onleftofproofruletrue\penalty1
}

\def\eproofbit{% NESTED TWO
\ifx    \wereinproofbit\prooftree
\then   \ifcase \lastpenalty
        \then   \shortenproofright=0pt  % 0: some other object, no indentation
        \or     \unpenalty\hfil         % 1: empty hypotheses, just glue
        \or     \unpenalty\unskip       % 2: just had a tree, remove glue
        \else   \shortenproofright=0pt  % eh?
        \fi
\fi
\global\dimen0=\shortenproofleft
\global\dimen1=\shortenproofright
\global\dimen2=\proofrulebreadth
\global\dimen3=\proofbelowshift
\global\dimen4=\proofdotseparation
\global\count255=\proofdotnumber
$\egroup  % NESTED ONE
\shortenproofleft=\dimen0
\shortenproofright=\dimen1
\proofrulebreadth=\dimen2
\proofbelowshift=\dimen3
\proofdotseparation=\dimen4
\proofdotnumber=\count255
}

\def\proofover{% NESTED TWO
\eproofbit % NESTED ONE
\setbox\proofbelow=\hbox\bgroup % NESTED TWO
\let\wereinproofbit\proofover
$\displaystyle
}%
\def\proofoverdbl{% NESTED TWO
\eproofbit % NESTED ONE
\doubleprooftrue
\setbox\proofbelow=\hbox\bgroup % NESTED TWO
\let\wereinproofbit\proofoverdbl
$\displaystyle
}%
\def\proofoverdots{% NESTED TWO
\eproofbit % NESTED ONE
\proofdotstrue
\setbox\proofbelow=\hbox\bgroup % NESTED TWO
\let\wereinproofbit\proofoverdots
$\displaystyle
}%
\def\proofusing{% NESTED TWO
\eproofbit % NESTED ONE
\setbox\proofrulename=\hbox\bgroup % NESTED TWO
\let\wereinproofbit\proofusing
\kern0.3em$
}

\def\endprooftree{% NESTED TWO
\eproofbit % NESTED ONE
  \dimen5 =0pt% spread of hypotheses
\dimen0=\wd\proofabove \advance\dimen0-\shortenproofleft
\advance\dimen0-\shortenproofright
\dimen1=.5\dimen0 \advance\dimen1-.5\wd\proofbelow
\dimen4=\dimen1
\advance\dimen1\proofbelowshift \advance\dimen4-\proofbelowshift
\ifdim  \dimen1<0pt
\then   \advance\shortenproofleft\dimen1
        \advance\dimen0-\dimen1
        \dimen1=0pt
        \ifdim  \shortenproofleft<0pt
        \then   \setbox\proofabove=\hbox{%
                        \kern-\shortenproofleft\unhbox\proofabove}%
                \shortenproofleft=0pt
        \fi
\fi
\ifdim  \dimen4<0pt
\then   \advance\shortenproofright\dimen4
        \advance\dimen0-\dimen4
        \dimen4=0pt
\fi
\ifdim  \shortenproofright<\wd\proofrulename
\then   \shortenproofright=\wd\proofrulename
\fi
\dimen2=\shortenproofleft \advance\dimen2 by\dimen1
\dimen3=\shortenproofright\advance\dimen3 by\dimen4
\ifproofdots
\then
        \dimen6=\shortenproofleft \advance\dimen6 .5\dimen0
        \setbox1=\vbox to\proofdotseparation{\vss\hbox{$\cdot$}\vss}%
        \setbox0=\hbox{%
                \advance\dimen6-.5\wd1
                \kern\dimen6
                $\vcenter to\proofdotnumber\proofdotseparation
                        {\leaders\box1\vfill}$%
                \unhbox\proofrulename}%
\else   \dimen6=\fontdimen22\the\textfont2 % height of maths axis
        \dimen7=\dimen6
        \advance\dimen6by.5\proofrulebreadth
        \advance\dimen7by-.5\proofrulebreadth
        \setbox0=\hbox{%
                \kern\shortenproofleft
                \ifdoubleproof
                \then   \hbox to\dimen0{%
                        $\mathsurround0pt\mathord=\mkern-6mu%
                        \cleaders\hbox{$\mkern-2mu=\mkern-2mu$}\hfill
                        \mkern-6mu\mathord=$}%
                \else   \vrule height\dimen6 depth-\dimen7 width\dimen0
                \fi
                \unhbox\proofrulename}%
        \ht0=\dimen6 \dp0=-\dimen7
\fi
\let\doll\relax
\ifwasinsideprooftree
\then   \let\VBOX\vbox
\else   \ifmmode\else$\let\doll=$\fi
        \let\VBOX\vcenter
\fi
\VBOX   {\baselineskip\proofrulebaseline \lineskip.2ex
        \expandafter\lineskiplimit\ifproofdots0ex\else-0.6ex\fi
        \hbox   spread\dimen5   {\hfi\unhbox\proofabove\hfi}%
        \hbox{\box0}%
        \hbox   {\kern\dimen2 \box\proofbelow}}\doll%
\global\dimen2=\dimen2
\global\dimen3=\dimen3
\egroup % NESTED ZERO
\ifonleftofproofrule
\then   \shortenproofleft=\dimen2
\fi
\shortenproofright=\dimen3
\onleftofproofrulefalse
\ifinsideprooftree
\then   \hskip.5em plus 1fil \penalty2
\fi
}

\usepackage{savesym}
\usepackage{amsfonts}
\usepackage{stmaryrd}

\usepackage{amsmath}
\savesymbol{iint}
\usepackage{txfonts}
\restoresymbol{TXF}{iint}

\usepackage{bbold}
\usepackage{latexsym}
\usepackage{enumerate}

\newcommand{\Set}{{\sf Set}}
\newcommand{\FSet}{{\sf FSet}}

\newcommand{\Rel}{{\sf Rel}}

\newcommand{\FHilb}{{\sf FHilb}}

\newcommand{\id}{{\rm id}}

\renewcommand{\to}{\longrightarrow}
\newcommand{\ot}{\longleftarrow}

\newcommand{\tto}[1]{\xrightarrow{#1}}

\newcommand{\stto}[1]{\stackrel{#1}\longrightarrow}

\newcommand{\soot}[1]{\stackrel{#1}\longleftarrow}

\newcommand{\BBB}{{\cal B}}
\newcommand{\CCC}{{\cal C}}

\newcommand{\LLL}{{\cal L}}

\newcommand{\SSS}{{\cal S}}

\renewcommand{\Bbb}{\mathbb}

\newcommand{\BBb}{{\Bbb B}}
\newcommand{\CCc}{{\Bbb C}}

\newcommand{\ZZz}{{\Bbb Z}}

\newcounter{countroman}
\newenvironment{rnumerate}%
{\begin{list}{{\rm (\roman{countroman})}}{\usecounter{countroman}}}%
{\end{list}}

\newcounter{countalpha}
\newenvironment{anumerate}%
{\begin{list}{(\alph{countalpha})}{\usecounter{countalpha}}}%
{\end{list}}

\newcounter{countalphabf}
{\protect\begin{list}{{\rm (}{\bf \protect\alph{countalphabf}}{\rm%
)}}{\protect\usecounter{countalphabf}}}% 
{\end{list}}

\newcommand{\bitmz}{\vspace{-.5\baselineskip}\begin{itemize}}
\newcommand{\eitmz}{\end{itemize}\vspace{-.25\baselineskip}}

\newcommand{\bdesc}{\vspace{-.5\baselineskip}\begin{description}}
\newcommand{\edesc}{\end{description}\vspace{-.25\baselineskip}}

\mathcode`\<="4268 %left delimiter
\mathcode`\>="5269 %right delimiter
\mathchardef\gt="313E %relation >
\mathchardef\lt="313C %relation <
\newsavebox{\barr}
\savebox{\barr}{\hspace*{-9.5pt}\raisebox{1.25pt}{$\scriptscriptstyle%
|$}\hspace*{4.5pt}} 
\newsavebox{\barrleft}
\savebox{\barrleft}{\hspace*{-8.5pt}\raisebox{1.25pt}{$\scriptscriptstyle%
|$}\hspace*{10pt}}

 \def\pushright#1{{%              set up
    \parfillskip=0pt            % so \par doesnt push \square to left
    \widowpenalty=10000         % so we dont break the page before \square
    \displaywidowpenalty=10000  % ditto
    \finalhyphendemerits=0      % TeXbook exercise 14.32
    \leavevmode                 % \nobreak means lines not pages
    \unskip                     % remove previous space or glue
    \nobreak                    % don't break lines
    \hfil                       % ragged right if we spill over
    \penalty50                  % discouragement to do so
    \hskip.2em                  % ensure some space
    \null                       % anchor following \hfill
    \hfill                      % push \square to right
    {#1}                        % the end-of-proof mark (or whatever)
    \par}}                      % build paragraph

 \def\qed{\pushright{$\square$}\penalty-700 \smallskip}

\newenvironment{prf}[1]{\begin{trivlist} \item[{\bf ~Proof}#1.]}%
{\qed\end{trivlist}}

\newcommand{\be}[1]{\begin{#1}}

\newcommand{\ee}[1]{\end{#1}}
\newcommand{\beq}{\begin{equation}}
\newcommand{\eeq}{\end{equation}}
\newcommand{\ba}[1]{\begin{array}{#1}}
\newcommand{\ea}{\end{array}}
\newcommand{\bea}{\begin{eqnarray}}
\newcommand{\eea}{\end{eqnarray}}
\newcommand{\bear}{\begin{eqnarray*}}
\newcommand{\eear}{\end{eqnarray*}}
\newcommand{\bpr}{\begin{prf}{}}
\newcommand{\epr}{\end{prf}}
\newcommand{\bprf}[1]{\begin{prf}{#1}}
\newcommand{\eprf}{\end{prf}}

\newtheorem{thm}{Theorem}[section]

\newtheorem{prop}[thm]{Proposition}
\newtheorem{lemma}[thm]{Lemma}
\newtheorem{corollary}[thm]{Corollary}
\newtheorem{cond}{}[thm]

\newtheorem{prenumb}[thm]{\hspace{-1ex}}

\newcommand{\sto}{\rightarrow}

\newcommand{\tp}[2]{{#1}\hspace{-.45ex} :\hspace{-.55ex}{#2}}

\newcommand{\rect}{\Xi}
\newcommand{\mnd}{{\scriptstyle \nabla}}
\newcommand{\unt}{{\scriptstyle \bot}}
\newcommand{\cmn}{{\scriptstyle \Delta}}
\newcommand{\cun}{{\scriptstyle \top}}
\newcommand{\starr}{\bullet}%{\star}
\newcommand{\yon}[1]{\widehat{#1}}

\newcommand{\mplus}{+}
\newcommand{\munit}{o}
\newcommand{\opr}[1]{\widetilde{#1}}

\newcommand{\pw}[1]{{#1}_\wp}

\newcommand{\ab}{{\sf ab}}
\newcommand{\adx}{{\sf ad}}
\newcommand{\abx}{{\sf ab}}
\newcommand{\adxE}{{\sf ad}}
\newcommand{\abxE}{{\sf ab}}

\newcommand{\Mes}[2]{{#1}{\left\{#2\right\}}}
\newcommand{\Tr}{{\rm Tr}}

\newcommand{\Ext}[1]{{\sf Ext}_{#1}}
\newcommand{\Abs}[1]{{\sf Abs}_{#1}}
\newcommand{\dAbs}[1]{\ddag\mbox{-}{\sf Abs}_{#1}}
\newcommand{\bs}{{\BBB}}
\newcommand{\bsAbs}[1]{\mbox{\small $\bs$-}{\sf Abs}_{#1}}
\newcommand{\dExt}[1]{\ddag\mbox{-}{\sf Ext}_{#1}}
\newcommand{\Kl}[2]{{#1}_{[#2]}}
\newcommand{\EM}[2]{{#1}^{[#2]}}
\newcommand{\Ac}[2]{{#1}^{\{#2\}}}
\newcommand{\Nat}{{\rm Nat}}

\title{Geometry of abstraction\\
in quantum computation}
\author{Dusko~Pavlovic\\
Oxford University and Kestrel Institute}

\date{}  
  
\begin{document}

\maketitle

\begin{abstract}
Quantum algorithms are sequences of abstract operations, performed on non-existent computers. They are in obvious need of categorical semantics. We present some steps in this direction, following earlier contributions of Abramsky, Coecke and Selinger. In particular, we analyze function abstraction in quantum computation, which turns out to characterize its classical interfaces. 

Some quantum algorithms provide feasible solutions of important hard problems, such as factoring and discrete log (which are the building blocks of modern cryptography). It is of a great practical interest to precisely characterize the computational resources needed to execute such quantum algorithms. There are many ideas how to build a quantum computer. Can we prove some necessary conditions? Categorical semantics help with such questions. We show how to implement an important family of quantum algorithms using just abelian groups and relations.

%Semantics of computation is a family of mathematical theories used to design and analyze programming structures and languages. Calculi of variable abstraction and application, and the corresponding categorical structures, are its conceptual centerpiece, and the conceptual core of functional programming.

%In their LICS 2004 paper, Abramsky and Coecke extended the semantical toolkit to capture quantum protocols. Continuing in this direction, we examine the notion of variable abstraction that arises in quantum computation, and encapsulates classical data and controls. The laws of abstraction turn out to correspond to interesting transformations in the underlying graphical calculus, formalized by Selinger. The resulting geometric structures provide a convenient, often surprisingly intuitive tool for reasoning about quantum algorithms. 

%We analyze the design pattern of an important family of quantum algorithms (including Shor's algorithms for factoring and discrete log). The obtained description of the application of quantum resources shows that these algorithms are not limited to the standard physical model of a quantum computer: we sketch an implementation using finite groups and relations.
\end{abstract}

% \tableofcontents

\section{Introduction}\label{what}
%Computer science and technology have been developed over a very robust idea of computation: all models of computation provided the same notion of computability (Church's thesis). Different computational resources, however, turn out to provide different notions of feasibility.

%Very few resources provide an exponential speedup. Quantum entanglement is one such resource. The examples are Simon's and Shor's algorithms.

%Beyond Church's thesis: cell, society, the Web, networks. Computational power depends on the resources. Ants.

\paragraph{What do quantum programmers do?}
They do a variety of things, but there is a "design pattern" that they often follow, based on the {\em Hidden Subgroup Problem (HSP)} \cite[sec.~5.4]{Lomonaco,Nielsen-Chuang}. Shor's factoring and discrete log algorithms \cite{Shor97} are examples of this pattern, as well as Hallgren's algorithm for the Pell equation \cite{HallgrenS:Pell}. They all provide an exponential speedup with respect to the best known classical algorithms. The simplest member of the family is Simon's algorithm for period finding \cite{Simon}, which we use as the running example. The other HSP algorithms only differ in "domain specific" details, but yield to the same semantics.  

The input for Simon's algorithm is an arbitrary function $f:\ZZz_2^m \sto \ZZz_2^n$, where $(\ZZz_2,\oplus,0)$ is the group with two elements, and $\oplus$ is the "exclusive or" operation. The task is to find the period of $f$, if it exists, i.e. a bitstring $c\in \ZZz_2^m$ such that $f(x\oplus c) = f(x)$ for all $x\in \ZZz_2^m$. For simplicity, let us assume that there is exactly one such $c$, as the discussion of the other cases does not bring in anything essential.

Since $f$ is arbitrary, one cannot ascertain that a bitstring $c$ is a solution without computing the value of $f(x)$ for every $x\in \ZZz_2^m$. But a quantum computer can compute all such values at once! This is called {\em quantum parallelism}, and is one of the first things explained to quantum programmers' apprentices \cite[sec.~1.4.2]{Nielsen-Chuang}. 

Mathematically speaking, the main capability of a quantum computer is that it can evaluate unitary operators. If the inputs of a function are represented as the basis vectors of a Hilbert space, and the function itself is captured as a unitary operator over it, then the quantum computer can compute all values of the function at once, by evaluating this unitary over a suitably generated combination of the basis vectors. Simon's algorithm shows how to extract the information about the period of the function from the projections of the resulting mixture.

%
%It can thus compute all values of a function if is fed a superposition of all inputs for the function. It outputs a projection of the superposition of all of the outputs. The task is then to extract from this superposition the desired information about the function. Simon shows how to extract the information about its period.

But how do we represent a function $f:\ZZz_2^m \sto \ZZz_2^n$ by a unitary operator? For an involutive function $g:B\sto B$, the answer is easy: define $U_g:\CCc^B\to \CCc^B$  by setting $U_g|b> = \left|g(b)\right>$, where $|b>\in \CCc^B$ are the basis vectors indexed by $b\in B$. The fact that $U_g$ is unitary follows from $g\circ g=\id_B$. For a general $f:\ZZz_2^m \sto \ZZz_2^n$, first define a corresponding involution $f'$, and then extract the unitary $U_f$:
\[
\prooftree
\prooftree
f:\ZZz_2^m \to \ZZz_2^n: x\mapsto f(x)
\justifies
f':  \ZZz_2^{m+n} \to \ZZz_2^{m+n}: x,y\mapsto x,y\oplus f(x) 
\endprooftree
\justifies
U_f : \CCc^{\ZZz_2^{m+n}}\to \CCc^{\ZZz_2^{m+n}}: |x,y\rangle\mapsto |x,y\oplus f(x)\rangle 
\endprooftree
\]
where the basis vectors $|x,y>$ of $\CCc^{\ZZz_2^{m+n}}$ are indexed by the bitstrings $x$ of length $m$ concatenated with the bitstrings $y$ of length $n$. The values of the function $f$ are recovered from $U_f|x,0> = \left|x,f(x)\right>$.

The other conceptual component of Simon's algorithm, and of all HSP-algorithms, is a standard application of transform theory \cite{Widder:transform}: transform the inputs into another domain, where the computation is easier, compute the outputs there, and then transform them back\footnote{E.g., Laplace's transform maps a differential equation into a polynomial equation over the field, generated by the convolution ring in which the original equation was stated  \cite{PavlovicD:LAPL}. The solutions of the polynomial equation are then mapped back by the inverse Laplace transform.}. In our special case, $U_f$ is thus precomposed and postcomposed with a suitable version of the Fourrier transform, which for $\ZZz_2$ boils down to the Hadamard-Walsh tranform $H^{\otimes m} |z> = \sum_{x\in \ZZz_2^m} (-1)^{x\cdot z} |x>$. Here $x\cdot z$ denotes the inner product in $\ZZz_2^m$, and we ignore the renormalizing factor $2^{-\frac{m}{2}}$. This transform is applied to the first $m$ arguments of $U_f$, to generate the desired superposition of all inputs of $f$. The quantum computer thus computes the following vector\footnote{We ignore the renormalizing factors throughout.}:
\bear
Simon & = &  (H^{\otimes m}\otimes \id) U_f (H^{\otimes m}\otimes \id)\  |0,0> \\
& = & \sum_{z,x\in \ZZz_2^m} (-1)^{x\cdot z} |z,f(x)>
\eear
When we measure the first component of this vector, it collapses to a single $|z>$, i.e. we get $\gamma_{z} = |z>\otimes  \sum_{x\in \ZZz_2^m} (-1)^{x\cdot z} \left |f(x)\right>$. By assumption, there is exactly one $c\in \ZZz_2^m$ such that $f(x\oplus c) = f(x)$ holds for all $x$. The coefficient of each of the basis vectors $|z,f(x)>=|z,f(x\oplus c>$ is thus $\gamma_{z}^x = (-1)^{x\cdot z} + (-1)^{(x\oplus c)\cdot z} =(-1)^{x\cdot z}\left(1+(-1)^{c\cdot z}\right)$. It follows that $\left(\forall x\in \ZZz_2^m.\ \gamma_{z}^x\neq 0\right) \ \iff\ c\cdot z = 0$. Each time that we run the algorithm, we can thus extract a linear equation in $c$. After $m$ runs, we can thus compute $c$. (The probability that at some step $k\leq m$ we may get  an equation dependent on the previous ones is 0, because $z$ are chosen randomly, and the measure of every proper linear subspace of $\ZZz_2^m$ is 0.) On the other hand, in order to convince ourselves classically that $c$ is the period of $f$, we should to compute all values $f$, which requires $2^m$ steps, since $f$ is an arbitrary function.

The core of Shor's factoring algorithm follows the same pattern, adapted for $f:\ZZz_k\sto \ZZz_k$, where $f(x) = a^x \mod k$. The factored integer is $k$, and $a$ is randomly selected to be tested for common factors with it, which can be derived by finding a period of $f$.

%\subsubsection*
\paragraph{Summary of the paper.}
A program generally describes a family of computations over a family of input data. The various input data to be computed with are denoted by variables. E.g., the polynomial $x^2+x + 2$ can be construed as a program, describing the family of computations that can be performed for the various values of $x$. It is tacitly assumed that the possible values of $x$ can be copied, so that one copy can be substituted for each occurrence of $x$ in the polynomial $x^2+x$; and that these data can also be deleted, if the polynomial is just $2$, and $x$ does not occur in it. 

The first problem with quantum programming is that quantum data cannot be manipulated in this way: it is a fundamental property of quantum states that they generally cannot be copied \cite{Wooters:NoCloning,Dieks:NoCloning}, or even deleted \cite{BraunsteinS:NoDeleting,AbramskyS:NoCloning}. So how do we write quantum programs? In particular, given a program $f(x)$ for a function $f$, what kind of a program transformation leads to the quantum program $U_f|x,y>$, that we used to specify the unitary $U_f$ above? This question is analyzed and answered in sections \ref{MonAbsSec} and \ref{DagAbsSec}. It turns out that the needed copying and deleting operations are closely related with the abstraction.

On the other hand, copying, deleting and abstraction capabilities can be viewed as the characteristics of classical computation. In a quantum computer, a structure that supports copying, deleting and abstraction can be construed as its classical interface. This is what we call a {\em classical structure}. An early analysis of this structure was in \cite{PavlovicD:QMWS}. In the meantime, there are several  versions, and many applications \cite{Coecke-Duncan,PavlovicD:QI09,PavlovicD:CQStruct}. In recent work, Coecke \cite{Coecke-Edwards} uses the term {\em basis structures}\/ for the same concept, because a classical structure over a finitely dimensional Hilbert space precisely correspond to a choice of a basis \cite{PavlovicD:MSCS08}, and can be viewed as a purely categorical, element-free version of this notion. While the simple basis intuitions are attractive, I stick here with the original terminology. One reason is that the correspondence of classical structures and the induced bases is not always as simple as it is in the category of finitely dimensional Hilbert spaces \cite{PavlovicD:QI09}, and it is useful to keep the distinction. A more important reason is that classical structures express the fact that {\em classicality is relative\/} as an algebraic structure. The fact that classical data with respect to one classical structure may be entangled with respect to another one is the fundamental feature of quantum computation. This is usually captured through change of basis. Classical structures provide an algebraic framework for such transforms. This is summarized in section \ref{Unbiased}.

The final step of the described algorithm pattern, measurement, is modeled in section \ref{Measurement}. The resulting categorical semantics is supported not only by the standard Hilbert space model, but also by non-standard models. We spell out a relational interpretation, based on \cite{Coecke-Edwards,PavlovicD:QI09}. In particular, Simon's algorithm turns out to have an effective relational implementation, using an abelian group as the computational resource supplying the power of a quantum computer.

%\paragraph{Conventions and prerequisites.}
Section \ref{Prerequisites} provides a brief summary of the basic semantical prerequisites, notations and terminology. 

\section{Preliminaries}\label{Prerequisites}
\subsection{Monoidal categories}
We assume that the reader has some understanding of the basic categorical concepts and terminology \cite{MacLaneS:CWM}, and work with symmetric monoidal categories $(\CCC,\otimes,I)$ \cite{KellyM:Enriched,Joyal-Street:geometry}. 

\paragraph{Strictness.} For simplicity, and without loss of generality, we tacitly assume that each of our monoidal categories is {\em strictly associative and unitary}, i.e. that the objects form a monoid in the usual sense. This causes no loss of generality because every monoidal category is equivalent to a strictly associative and unitary one, along a monoidal equivalence. But note that the tensor symmetry cannot be "strictified" without essentially changing the category; the canonical isomorphisms  $A\otimes B\stackrel{c}{\to}B\otimes A$ are thus generally {\em not\/} identities. 

On the other hand, just like the tensors, we strictify functors: a {\em monoidal\/} functor $F$ is always assumed to be strict, i.e. it preserves the monoidal structure on the nose: $F(A\otimes B) = FA\otimes FB$ and $FI = I$.

The arrows from $I$ are sometimes called {\em vectors}, or {\em elements}. The abstract "vector spaces" are thus written $\CCC(X) = \CCC(I,X)$. When confusion is unlikely, we elide the tensor symbol and write $XAf$ instead of $X\otimes A\otimes f$.

\subsubsection{String diagrams}
Calculations in monoidal categories are supported by a simple and intuitive graphical language: the string diagrams. This language has its roots in Penrose's diagrammatic notation \cite{Penrose}, and it has been formally developed in categorical {\em coherence theory}, and in particular in Joyal and Street's {\em geometry of tensor calculus} \cite{Joyal-Street:geometry}. The objects  are drawn as strings, and the morphisms as boxes attached on these strings. One can think that the information flows through the strings, and is processed in the boxes. A direction of this flow is chosen by convenience. We shall assume that the information flows up, so that the strings at the bottom of a box denote the domain of the corresponding morphism; the threads at the top the codomain. Drawing the strings $A$ and $B$ next to each other represents $A\otimes B$; similarly with the boxes. Drawing a thread from one box to another  denotes the composition of the corresponding morphisms. 
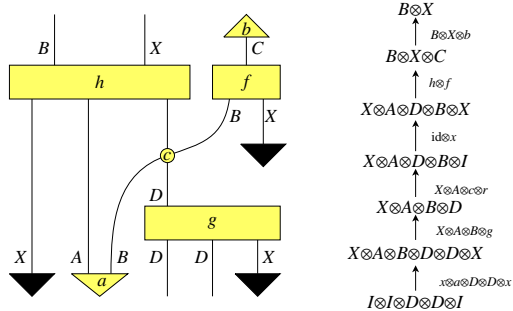
\begin{center}
\def\JPicScale{0.75}
\ifx\JPicScale\undefined\def\JPicScale{1}\fi
\psset{unit=\JPicScale mm}
\psset{linewidth=0.3,dotsep=1,hatchwidth=0.3,hatchsep=1.5,shadowsize=1,dimen=middle}
\psset{dotsize=0.7 2.5,dotscale=1 1,fillcolor=black}
\psset{arrowsize=1 2,arrowlength=1,arrowinset=0.25,tbarsize=0.7 5,bracketlength=0.15,rbracketlength=0.15}
\begin{pspicture}(0,0)(90.62,59)
\newrgbcolor{userFillColour}{1 1 0.4}
\pspolygon[linewidth=0.12,fillcolor=userFillColour,fillstyle=solid](8,43)(40,43)(40,49)(8,49)
\pspolygon[linewidth=0.12](48,16)(48,16)(48,16)(48,16)
\rput(24,46){$\scriptstyle h$}
\psline[linewidth=0.12](22,43)(22,12)
\newrgbcolor{userFillColour}{1 1 0.4}
\pspolygon[linewidth=0.12,fillcolor=userFillColour,fillstyle=solid](32,18)(56,18)(56,24)(32,24)
\pspolygon[linewidth=0.12](56,21)(56,21)(56,21)(56,21)
\psline[linewidth=0.12](52,12)(52,18)
\psline[linewidth=0.12](36,8)(36,18)
\psline[linewidth=0.12](36,24)(36,43)
\psline[linewidth=0.12](12,12)(12,43)
\psbezier[linewidth=0.12](26,12)(26,21.34)(27,30)(36,33)
\psbezier[linewidth=0.12](36,33)(42,35)(46,36)(47,43)
\rput(10,15){$\scriptstyle X$}
\rput(54,15){$\scriptstyle X$}
\rput(20,15){$\scriptstyle A$}
\rput(44,21){$\scriptstyle g$}
\newrgbcolor{userFillColour}{1 1 0.4}
\pspolygon[linewidth=0.12,fillcolor=userFillColour,fillstyle=solid](44,43)(56,43)(56,49)(44,49)
\rput(50,46){$\scriptstyle f$}
\psline[linewidth=0.12](44,8)(44,18)
\psline[linewidth=0.12](53,35)(53,43)
\rput(54.38,40){$\scriptstyle X$}
\psline[linewidth=0.12](16,49)(16,58)
\psline[linewidth=0.12](50,49)(50,54)
\psline[linewidth=0.12](32,49)(32,58)
\rput(14,52){$\scriptstyle B$}
\rput(34,52){$\scriptstyle X$}
\rput(52,52){$\scriptstyle C$}
\newrgbcolor{userFillColour}{1 1 0.4}
\rput{0}(36,33){\psellipse[linewidth=0.12,fillcolor=userFillColour,fillstyle=solid](0,0)(1.25,1.25)}
\rput(36,33){$\scriptstyle c$}
\rput(34,15){$\scriptstyle D$}
\rput(42,15){$\scriptstyle D$}
\rput(48,40){$\scriptstyle B$}
\rput(28,15){$\scriptstyle B$}
\rput(34,26){$\scriptstyle D$}
\rput(80,7){$\scriptstyle I\otimes I\otimes D\otimes D\otimes I$}
\rput(65.5,9.5){}
\rput(80,15.63){$\scriptstyle X\otimes A\otimes B\otimes D\otimes D \otimes X$}
\rput(80.63,23.75){$\scriptstyle X\otimes A\otimes B\otimes D$}
\rput(80,31.88){$\scriptstyle X\otimes A\otimes D\otimes B\otimes I$}
\rput(80,50.63){$\scriptstyle B\otimes X\otimes C$}
\psline[linewidth=0.12]{->}(80,9)(80,13.5)
\psline[linewidth=0.12]{->}(80,18)(80,22.5)
\psline[linewidth=0.12]{->}(80,25.62)(80,30.63)
\psline[linewidth=0.12]{->}(80,33.75)(80,39.25)
\psline[linewidth=0.12]{->}(80,43.13)(80,48.13)
\rput(88.13,26.88){$\scriptscriptstyle X\otimes A\otimes c\otimes r$}
\rput(90.62,10.62){$\scriptscriptstyle x\otimes a\otimes D\otimes D\otimes x$}
\newrgbcolor{userFillColour}{1 1 0.4}
\psline[linewidth=0.12,fillcolor=userFillColour,fillstyle=solid](46,54)
(50,58)
(54,54)(46,54)
\rput(50,55.5){$\scriptstyle b$}
\psline[linewidth=0.12]{->}(80,52.5)(80,57.5)
\rput(80,59){$\scriptstyle B\otimes X$}
\rput(86.25,54.38){$\scriptscriptstyle B\otimes X\otimes b$}
\rput(88.75,19.38){$\scriptscriptstyle X\otimes A\otimes B\otimes g$}
\rput(80,41.25){$\scriptstyle X\otimes A\otimes D\otimes B\otimes X$}
\rput(85,36.25){$\scriptscriptstyle {\rm id}\otimes x$}
\rput(84.38,45.63){$\scriptscriptstyle h\otimes f$}
\newrgbcolor{userFillColour}{1 1 0.4}
\psline[linewidth=0.12,fillcolor=userFillColour,fillstyle=solid](19,12)
(24,8)
(29,12)(19,12)
\psline[linewidth=0.12,fillstyle=solid](8,12)
(12,8)
(16,12)(8,12)
\psline[linewidth=0.12,fillstyle=solid](49,35)
(53,31)
(57,35)(49,35)
\psline[linewidth=0.12,fillstyle=solid](48,12)
(52,8)
(56,12)(48,12)
\rput(24.38,10.62){$\scriptstyle a$}
\end{pspicture}

%\caption{default}
%\label{default}
\end{center}
One of the salient features if this notation is that the associativity is implicit, and automatic, both of the tensor and of the composition. The tensor symmetry $c: B\otimes D\to D\otimes B$ is denoted above by a circle. The circle is usually omitted, so that symmetry boils down to crossing the strings. The identity morphisms are the "invisible boxes", that can be placed on any thread.  The tensor unit $I$ is the "invisible thread", that can be added to any diagram. This means that a box representing a vector $a\in \CCC(I,AB)$ does not have any visible threads coming in from below. This is often emphasized by reducing the bottom of such a box to a point: e.g., the vector $I\stackrel a \to AB$ is denoted by a triangle. The box representing a covector $b\in \CCC(C,I)$ does not have any visible threads coming out, and boils down to a triangle pointing up.  The black triangles denote the vector indeterminates $I\stackrel{x}\to X$, freely adjointed to monoidal categories to form polynomials. Such polynomial constructions will be discussed in Sec.~\ref{MonAbsSec}.

\subsubsection{Monoids and comonoids}\label{MonComon}
A monoid in a monoidal category is a pair of arrows $X\otimes X\stto{\mnd} X \soot{\unt} I$ such that
\begin{gather*}
\mnd\circ (\mnd \otimes X)  =  \mnd \circ (X\otimes \mnd)\\
 \mnd \circ (\unt \otimes X)  = \mnd \circ (X\otimes \unt)  = \id_X
\end{gather*}
When the tensor is the cartesian product, this captures the usual notion of monoid. 

A comonoid in a monoidal category is dual to a monoid: it is a pair of arrows  $X\otimes X\stackrel{\cmn}\ot X \stackrel{\cun}{\to}I$ such that
\begin{gather*}
(\cmn \otimes X) \circ \cmn  =  (X\otimes \cmn)\circ \cmn\\
(\cun \otimes X) \circ \cmn  = (X\otimes \cun) \circ \cmn = \id_X
\end{gather*}
In {\bf string diagrams}, we draw the monoid evaluations as trapezoids pointing up, whereas their units are little triangles pointing down. The comonoids are represented by the trapezoids and the little triangles in the opposite directions. E.g., the comonoid laws correspond to the following graph transformations
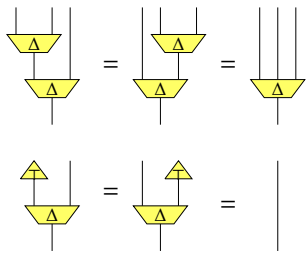
\newcommand{\comonoidd}{\cmn}
\newcommand{\comonunn}{\cun}
\newcommand{\comontwo}{\cmn}
\begin{center}
\def\JPicScale{0.6}
\ifx\JPicScale\undefined\def\JPicScale{1}\fi
\psset{unit=\JPicScale mm}
\psset{linewidth=0.3,dotsep=1,hatchwidth=0.3,hatchsep=1.5,shadowsize=1,dimen=middle}
\psset{dotsize=0.7 2.5,dotscale=1 1,fillcolor=black}
\psset{arrowsize=1 2,arrowlength=1,arrowinset=0.25,tbarsize=0.7 5,bracketlength=0.15,rbracketlength=0.15}
\begin{pspicture}(0,0)(66,40)
\rput(40.75,-6){}
\psline[linewidth=0.22](30,24)(30,40)
\newrgbcolor{userFillColour}{1 1 0.4}
\pspolygon[linewidth=0.22,fillcolor=userFillColour,fillstyle=solid](28,24)
(40,24)
(37,20)
(31,20)(28,24)
\psline[linewidth=0.22](34,20)(34,14)
\psline[linewidth=0.22](6,-4)(6,2)
\newrgbcolor{userFillColour}{1 1 0.4}
\psline[linewidth=0.22,fillcolor=userFillColour,fillstyle=solid](3,2)
(6,6)
(9,2)(3,2)
\psline[linewidth=0.22](38,24)(38,30)
\psline[linewidth=0.22](34,34)(34,40)
\newrgbcolor{userFillColour}{1 1 0.4}
\pspolygon[linewidth=0.22,fillcolor=userFillColour,fillstyle=solid](32,34)
(44,34)
(41,30)
(35,30)(32,34)
\psline[linewidth=0.22](42,34)(42,40)
\rput(23,27){$=$}
\psline[linewidth=0.22](14,24)(14,40)
\newrgbcolor{userFillColour}{1 1 0.4}
\pspolygon[linewidth=0.22,fillcolor=userFillColour,fillstyle=solid](4,24)
(16,24)
(13,20)
(7,20)(4,24)
\psline[linewidth=0.22](10,20)(10,14)
\psline[linewidth=0.22](6,24)(6,30)
\psline[linewidth=0.22](2,34)(2,40)
\newrgbcolor{userFillColour}{1 1 0.4}
\pspolygon[linewidth=0.22,fillcolor=userFillColour,fillstyle=solid](0,34)
(12,34)
(9,30)
(3,30)(0,34)
\psline[linewidth=0.22](10,34)(10,40)
\psline[linewidth=0.22](56,24)(56,40)
\newrgbcolor{userFillColour}{1 1 0.4}
\pspolygon[linewidth=0.22,fillcolor=userFillColour,fillstyle=solid](54,24)
(66,24)
(63,20)
(57,20)(54,24)
\psline[linewidth=0.22](60,20)(60,14)
\psline[linewidth=0.22](64,24)(64,40)
\psline[linewidth=0.22](60,24)(60,40)
\rput(49,27){$=$}
\psline[linewidth=0.22](30,-4)(30,6)
\newrgbcolor{userFillColour}{1 1 0.4}
\pspolygon[linewidth=0.22,fillcolor=userFillColour,fillstyle=solid](28,-4)
(40,-4)
(37,-8)
(31,-8)(28,-4)
\psline[linewidth=0.22](34,-8)(34,-14)
\rput(23,-1){$=$}
\psline[linewidth=0.22](14,-4)(14,6)
\newrgbcolor{userFillColour}{1 1 0.4}
\pspolygon[linewidth=0.22,fillcolor=userFillColour,fillstyle=solid](4,-4)
(16,-4)
(13,-8)
(7,-8)(4,-4)
\psline[linewidth=0.22](10,-8)(10,-14)
\psline[linewidth=0.22](60,-14)(60,6)
\rput(49,-4){$=$}
\psline[linewidth=0.22](38,-4)(38,2)
\newrgbcolor{userFillColour}{1 1 0.4}
\psline[linewidth=0.22,fillcolor=userFillColour,fillstyle=solid](35,2)
(38,6)
(41,2)(35,2)
\rput(6,32){$\comonoidd$}
\rput(10,22){$\comonoidd$}
\rput(38,32){$\comonoidd$}
\rput(34,22){$\comonoidd$}
\rput(34,-6){$\comonoidd$}
\rput(10,-6){$\comonoidd$}
\rput(6,3){$\comonunn$}
\rput(38,3){$\comonunn$}
\rput(60,22){$\comontwo$}
\end{pspicture}

\end{center}
\vspace{1.5\baselineskip}
A monoid is {\em commutative\/} if $\mnd\circ c_{XX} = \mnd$. A comonoid is commutative if $c_{XX} \circ \cmn = \cmn$. In string diagrams, this means that the value of the output of $\mnd$ does not change if the strings that come into it cross; and that the output of $\cmn$ does not change if the strings coming out of it cross.

\subsubsection{Cartesian categories} 
A monoidal category $(\CCC,\otimes,I)$ is {\em cartesian} when it comes with natural transformations
\[
X\otimes X\stackrel{\delta_X}\ot X \stackrel {!_X}\to I
\]
which make every object $X$ into a comonoid. The naturality of this structure means that every morphism $X\stackrel{f}\to Y$ in $\CCC$ is a comonoid homomorphism. It is easy to see that this makes the tensor $X\otimes Y$ into a product $X\otimes Y$, such that any pair of arrows $A\stackrel g \to X$ and $A\stackrel h \to Y$ corresponds to a unique arrow $A\stackrel{<g,h>} \to A\times B$, and the tensor unit $I$ into the final object $1$, with a unique arrow from each object. Cartesian structure is thus written in the form $(\CCC,\times, 1)$. 

\subsubsection{Monads and comonads}
A {\em monad}  on a category $\CCC$ can be defined as a functor $T:\CCC\to \CCC$ together with a monoid structure $TT\stackrel m \to T \stackrel h \ot {\rm Id}$ in the category of endofunctors on $\CCC$. With the corresponding monoid homomorphisms, monads form a category on their own \cite{BarrM:TTT}. Dually, {\em comonads\/} on $\CCC$ can be defined as comonoids in the category of endofunctors over $\CCC$, and accomodate similar developments. 

The categories of algebras for a monad and coalgebras for a comonad, and in particular the Kleisli and the Eilenberg-Moore constructions that will be used below, are presented in detail in \cite{MacLaneS:CWM,BarrM:TTT}, and in many other books.

The following observation is the starting point for most of the constructions in this paper. The proof is left as an easy exercise.

\be{prop}\label{monoid-monad}
Every (co)monoid $X$ in a monoidal category $\CCC$ induces a (co)monad $X\otimes (-):\CCC\to \CCC$. The corresponding Kleisli category $\Kl{\CCC}{X}$ is monoidal if and only if the (co)monoid $X$ is commutative.

More precisely, the category of monoids in a monoidal category $\CCC$ is equivalent with the category of monads $T$ on $\CCC$ such that $T(A\otimes B) = T(A)\otimes B$ and moreover $h_B = h_I\otimes B$ and $m_B = m_I \otimes B$ hold for all $A,B\in \CCC$. The dual statement holds for comonoids and comonads.
\end{prop}

\subsubsection{Convolution and representation} \label{Convolution-Representation}
Any monoid $(X,\mnd,\unt)$ in a monoidal category $(\CCC,\otimes,I)$ induces the ordinary monoid $\left(\CCC(X),\starr,\unt\right)$, whose operation
\bea\label{def-convolution}
a\starr b & = & \mnd \circ (a \otimes b)
\eea
is often called {\em convolution}. A Cayley representation (or Yoneda embedding) of the monoid $(X,\mnd,\unt)$ is a map
\bea\label{Cayley}
\yon{(-)} :\  \CCC(X) & \to &  \CCC(X,X)\\
\left(I\stackrel{a}{\rightarrow} X\right) &\longmapsto & \left(X\stackrel{a\otimes X}\longrightarrow X\otimes X \stackrel{\nabla}\rightarrow X\right)\notag
\eea
furthermore represents the vectors $a\in \CCC(X)$ as endomorphisms $\yon{a} \in \CCC(X,X)$. 

\begin{lemma}(Cayley, Yoneda) The Cayley representation is a monoid isomorphism between the convolution monoid $\left(\CCC(X),\starr,\unt\right)$ and the monoid  $(\Nat(X,X), \circ, \id_X)$ of {\em natural\/} endomorphisms
\bear
\Nat(X,X) & = & \left\{f\in \CCC(X,X)\ |\ \forall ab\in \CCC(X).\ f \circ(a\starr b)  =  (f\circ a)\starr b\right\}
\eear
\end{lemma}

A comonoid structure on $X$ induces a convolution monoid on $\CCC(X,I)$, with $c\starr d  = (c \otimes d)\circ \cmn$, and with a similar Cayley representation. In general, a convolution monoid can be defined over any hom-set $\CCC(X,Y)$, where $X$ is a comonoid and $Y$ a monoid, by setting $f\starr g  = \mnd_Y\circ (f \otimes g)\circ \cmn_X$.

\paragraph{Scalars.} The canonical isomorphism $I\otimes I\cong I$ makes the tensor unit $I$ of $\CCC$ into a commutative monoid and comonoid; the tensor associativity is the associativity law of this (co)monoid; the tensor commutativity makes the (co)monoid commutative; the coherence conditions tell that this is the only (co)monoid structure on $I$. The convolution monoid $(\CCC(I,I),\starr,\id_I)$ is the abstract {\em scalar algebra\/} of the monoidal category $\CCC$. The coherence conditions imply that there is only one monoid structure on $I$, hence $s\starr t = s\circ t = s\otimes t$ holds for all scalars $s,t\in \CCC(I,I)$. 

Abusing notation, the scalar action $\starr: \CCC(I,I)\times \CCC(A,B)\to \CCC(A,B)$ is defined by $s\starr f = s\otimes f$. If the tensor unit $I$ is not strict, then $s\otimes f$ needs to be precomposed by $A\cong I\otimes A$ and postcomposed by $I\otimes B\cong B$. 

\subsection{Duals with daggers}\label{Duality}
\subsubsection{Dualities}
A {\em duality\/} structure in a monoidal category $\CCC$ consists of two objects $X$ and $X^\ast$ and two arrows, the pairing $X\otimes X^\ast \tto{\varepsilon}I$ and the copairing $I\tto{\eta}X^\ast \otimes X$, such that
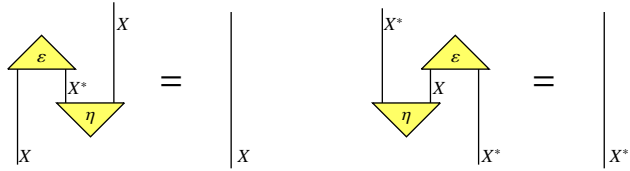
\[ (\varepsilon\otimes X)(X\otimes \eta) = X \qquad \qquad(X^\ast\otimes \varepsilon)(\eta\otimes X^\ast) = X^\ast\]
\newcommand{\obj}{\scriptstyle X}
\newcommand{\xobj}{\scriptstyle X^\ast}
\newcommand{\eqls}{$\mbox{\Large =}$}
\newcommand{\comonunit}{\scriptstyle\varepsilon}
\newcommand{\monunit}{\scriptstyle\eta}
\begin{center}
\def\JPicScale{0.8}
\ifx\JPicScale\undefined\def\JPicScale{1}\fi
\psset{unit=\JPicScale mm}
\psset{linewidth=0.3,dotsep=1,hatchwidth=0.3,hatchsep=1.5,shadowsize=1,dimen=middle}
\psset{dotsize=0.7 2.5,dotscale=1 1,fillcolor=black}
\psset{arrowsize=1 2,arrowlength=1,arrowinset=0.25,tbarsize=0.7 5,bracketlength=0.15,rbracketlength=0.15}
\begin{pspicture}(0,0)(106.88,28.62)
\psline[linewidth=0.22](69.12,11.62)(69.12,27.62)
\psline[linewidth=0.22](77.12,11.62)(77.12,17.62)
\psline[linewidth=0.22](85.12,1.62)(85.12,17.62)
\psline[linewidth=0.22](44,27)(44,1)
\rput(34,15){$\eqls$}
\psline[linewidth=0.22](24.51,11.55)(24.5,28.62)
\psline[linewidth=0.22](8.5,1.62)(8.5,17.62)
\psline[linewidth=0.22](16.5,11.62)(16.5,17.62)
\newrgbcolor{userFillColour}{1 1 0.4}
\psline[linewidth=0.22,fillcolor=userFillColour,fillstyle=solid](75.62,17.5)
(81.25,23.12)
(86.88,17.5)(75.62,17.5)
\rput[l](45,3.12){$\obj$}
\rput(81.25,19.38){$\comonunit$}
\newrgbcolor{userFillColour}{1 1 0.4}
\psline[linewidth=0.22,fillcolor=userFillColour,fillstyle=solid](6.88,17.5)
(12.5,23.12)
(18.12,17.5)(6.88,17.5)
\newrgbcolor{userFillColour}{1 1 0.4}
\psline[linewidth=0.22,fillcolor=userFillColour,fillstyle=solid](78.75,11.88)
(73.13,6.25)
(67.5,11.87)(78.75,11.88)
\rput(12.5,19.38){$\comonunit$}
\newrgbcolor{userFillColour}{1 1 0.4}
\psline[linewidth=0.22,fillcolor=userFillColour,fillstyle=solid](26.25,11.88)
(20.63,6.25)
(15,11.87)(26.25,11.88)
\rput(20.62,9.38){$\monunit$}
\rput[l](25,25){$\obj$}
\psline[linewidth=0.22](106,27)(106,1)
\rput(96,15){$\eqls$}
\rput[l](106.88,3.12){$\xobj$}
\rput[l](69.38,25){$\xobj$}
\rput[l](85.62,3.12){$\xobj$}
\rput[l](8.75,3.12){$\obj$}
\rput(73.12,9.38){$\monunit$}
\rput[l](16.88,14.38){$\xobj$}
\rput[l](77.5,14.38){$\obj$}
\end{pspicture}

\end{center}
A duality structure is written $(\eta,\varepsilon): X\dashv X^\ast$. Note that $X^{\ast\ast} = X$, because $(c\eta, \varepsilon c):X^\ast \dashv X$ is also a duality structure. If every object $X\in \CCC$ has a chosen duality structure, then such choices induce a {\em duality functor} $\ast:\CCC^{op}\to \CCC$, which maps $A\tto{f}B$ to 
\[f^\ast: B^\ast\tto{\eta B^\ast} A^\ast AB^\ast \tto{AfB^\ast} A^\ast BB^\ast \tto{A^\ast \varepsilon} A^\ast\]
\vspace{-1\baselineskip}
\renewcommand{\obj}{\scriptstyle B^\ast}
\newcommand{\bobj}{\scriptstyle B}
\renewcommand{\xobj}{\scriptstyle A^\ast}
\newcommand{\aobj}{\scriptstyle A}
\newcommand{\eff}{\scriptstyle f}
\begin{center}
\def\JPicScale{0.8}
\ifx\JPicScale\undefined\def\JPicScale{1}\fi
\psset{unit=\JPicScale mm}
\psset{linewidth=0.3,dotsep=1,hatchwidth=0.3,hatchsep=1.5,shadowsize=1,dimen=middle}
\psset{dotsize=0.7 2.5,dotscale=1 1,fillcolor=black}
\psset{arrowsize=1 2,arrowlength=1,arrowinset=0.25,tbarsize=0.7 5,bracketlength=0.15,rbracketlength=0.15}
\begin{pspicture}(0,0)(19.88,30)
\psline[linewidth=0.22](2,5)(2,30)
\psline[linewidth=0.22](10,6)(10,12)
\psline[linewidth=0.22](18,0)(18,24)
\newrgbcolor{userFillColour}{1 1 0.4}
\psline[linewidth=0.22,fillcolor=userFillColour,fillstyle=solid](8.62,23.75)
(14.25,29.37)
(19.88,23.75)(8.62,23.75)
\rput(14.25,25.63){$\comonunit$}
\newrgbcolor{userFillColour}{1 1 0.4}
\psline[linewidth=0.22,fillcolor=userFillColour,fillstyle=solid](11.75,5.63)
(6.13,0)
(0.5,5.62)(11.75,5.63)
\rput[r](1.25,25.12){$\xobj$}
\rput(6.12,3.13){$\monunit$}
\rput[l](18.75,4.5){$\obj$}
\psline[linewidth=0.22](10,18)(10,23.88)
\newrgbcolor{userFillColour}{1 1 0.4}
\pspolygon[fillcolor=userFillColour,fillstyle=solid](6,18)(14,18)(14,12)(6,12)
\rput(10,15.12){$\eff$}
\rput[r](9.38,8.75){$\aobj$}
\rput[l](10.62,20.62){$\bobj$}
\end{pspicture}

\end{center}
Using a duality $(\eta,\varepsilon): X\dashv X^\ast$, the abstract trace operators %\cite{Joyal-Street-Verity}  
$\Tr^{AB}_X: \CCC(XA,XB) \to\CCC(A,B)$ can be defined as follows:
\[
\Tr^{AB}_X g : A\xrightarrow{\eta_X A} X^\ast XA \xrightarrow{X^\ast g} X^\ast XB \xrightarrow{\varepsilon_{X^\ast} B} B
\]
\vspace{-1.3\baselineskip}
\renewcommand{\obj}{\scriptstyle X}
\renewcommand{\xobj}{\scriptstyle X^\ast}
\renewcommand{\eff}{\scriptstyle g}
\begin{center}
\def\JPicScale{0.8}
\ifx\JPicScale\undefined\def\JPicScale{1}\fi
\psset{unit=\JPicScale mm}
\psset{linewidth=0.3,dotsep=1,hatchwidth=0.3,hatchsep=1.5,shadowsize=1,dimen=middle}
\psset{dotsize=0.7 2.5,dotscale=1 1,fillcolor=black}
\psset{arrowsize=1 2,arrowlength=1,arrowinset=0.25,tbarsize=0.7 5,bracketlength=0.15,rbracketlength=0.15}
\begin{pspicture}(0,0)(21.25,30)
\psline[linewidth=0.22](1.88,5)(1.88,25)
\psline[linewidth=0.22](10,5)(10,12)
\psline[linewidth=0.22](18,0)(18,30)
\newrgbcolor{userFillColour}{1 1 0.4}
\psline[linewidth=0.22,fillcolor=userFillColour,fillstyle=solid](0.5,23.75)
(6.12,29.37)
(11.75,23.75)(0.5,23.75)
\rput(6.12,25.63){$\comonunit$}
\newrgbcolor{userFillColour}{1 1 0.4}
\psline[linewidth=0.22,fillcolor=userFillColour,fillstyle=solid](11.75,5.63)
(6.13,0)
(0.5,5.62)(11.75,5.63)
\rput[r](0.62,14.38){$\xobj$}
\rput(6.12,3.13){$\monunit$}
\psline[linewidth=0.22](10,18)(10,23.88)
\newrgbcolor{userFillColour}{1 1 0.4}
\pspolygon[fillcolor=userFillColour,fillstyle=solid](6.25,18.12)(21.25,18.12)(21.25,12)(6.25,12)
\rput(13.75,15){$\eff$}
\rput[l](20,3){$\aobj$}
\rput[l](19,26){$\bobj$}
\rput[l](10.62,20.62){$\obj$}
\rput[l](10.62,8.75){$\obj$}
\end{pspicture}

\end{center}

\subsubsection{Dagger-monoidal categories}\label{dagmondefs}
A {\em dagger\/} over a category $\CCC$ is an involutive ioof $\ddag : \CCC^{op} \to \CCC$. In other words, it satisfies $A^\ddag = A$ on the objects and $f^{\ddag\ddag} = f$ on the arrows. This very basic structure turns out to suffice for some crucial concepts.

\be{defn}
 A morphism $u\in \CCC(A,B)$ {\em unitary\/} if $u^\ddag\circ u = \id_A$ and $u\circ u^\ddag = \id_B$.  An endomorphism $p\in \CCC(A,A)$ is a {\em projector\/} if $p = p^\ddag = p\circ p$. A projector is {\em pure\/} if moreover $\Tr^{II}_A(p) = \id_I$.
\ee{defn}

\paragraph{Remarks.} Note that the abstract trace operators, given above, require a monoidal structure in $\CCC$. The interactions between the dagger with the monoidal structure, and in particular with the duals, has been recognized and analyzed in \cite{Abramsky-Coecke:LICS,SelingerP:CP,SelingerP:idempot}.  A {\em dagger-monoidal\/} category $(\CCC,\otimes, I, \ddag)$ is a dagger-category with a monoidal structure where all canonical isomorphisms, that form the monoidal structure, are unitary. When the monoidal structure is strict, this boils down to the requirement that the symmetry $c:A\otimes B\sto B\otimes A$ is unitary.

In the {\bf string diagrams}, the morphism $f^\ddag$ is represented by flipping the box $f$ around its horizontal axis. The morphism boxes thus need to be made asymmetric to record this flipping: in \cite{SelingerP:CP}, a corner of the box is filled; in \cite{PavlovicD:QMWS}, a corner is cut off. 

\subsubsection{Abstract conjugates and reals}
Since the dagger and the duality functors $(-)^\ddag, (-)^\ast:\CCC^{op}\to \CCC$ commute, their composite defines the {\em conjugation\/} ioof $(-)_\ast:\CCC^{op}\to \CCC$, which maps $f$ to
$
f_\ast  = f^{\ast \ddag}= f^{\ddag\ast}
$.
In the category of complex Hilbert spaces, the conjugation ioof corresponds is induced by the conjugation of the complex numbers. In the category of real Hilbert spaces, it degenerates into the identity functor. 

\be{defn} A morphism $f$ is said to be {\em real\/} if $f = f_\ast$ (or equivalently $f^\ddag = f^\ast$).
\ee{defn}

\paragraph{Remarks.} Pursuing the Hilbert space intuitions, the arrows $f$ and $f^\ddag$ are sometimes thought of as each other's adjoints. On the other hand, in a completely different sense, the dual objects $A$ and $A^\ast$ are each other's adjoints, if the monoidal category is viewed as a bicategory with one object.

\subsubsection{Inner products and entanglement}\label{entangled}
The dagger-monoidal structure has been proposed as a framework for categorical semantics of quantum computation \cite{Abramsky-Coecke:LICS,SelingerP:CP}. It turns out that this modest structure suffices for deriving many important notions:
\begin{itemize}
\item {\em inner product}
%{\small
\bea\label{inner-product}
<-|->_A\ :\ \CCC(A)\times \CCC(A) & \to & \CCC(I)\\
(I\tto{a,b} A) & \longmapsto &  \left(I \stackrel{a}{\rightarrow} A\stackrel{b^\ddag}{\rightarrow} I\right)\notag
%(\psi,\varphi:I\to A) & \longmapsto &  \left(I \stackrel{\varphi}{\rightarrow} A\stackrel{\psi^\ddag}{\rightarrow} I\right)
\eea

\item {\em partial inner product}
%{\footnotesize 
\bea\label{partial-inner}
<-|->^B_{A}\ :\ \CCC(A)\times \CCC(AB) & \to & \CCC(B)\\
\left(I\stto a A, I\stto b AB\right) & \longmapsto & \left(I\stackrel{a}{\rightarrow}AB\tto{b^\ddag\otimes B} B\right)\notag
%\left(\psi:I\rightarrow A, \varphi:I\rightarrow A\otimes B\right) & \longmapsto & \left(I\stackrel{\varphi}{\rightarrow}A\otimes B\stackrel{\psi^\ddag\otimes B}{\to} B\right)
\eea

\item {\em weakly entangled vectors} $\eta\in \CCC(A\otimes A)$, such that for all $a \in \CCC(A)$ holds
\bea\label{weakly-entangled}
<a_\ast\ |\ \eta>^A_{A} & = & a
\eea
\end{itemize}
Furthermore, an abstract version of strong entanglement can be defined as self-duality.

\be{defn}\label{strong-entanglement}
A vector $\eta\in \CCC(X\otimes X)$ is said to be (strongly) entangled if $(\eta,\eta^\ddag):X\dashv X$ is a duality, i.e. satisfies $(\eta^\ddag \otimes X)(X\otimes \eta) = X = (X \otimes \eta^\ddag)(\eta\otimes X)$, and thus $X^\ast = X$. 
\ee{defn}

\begin{prop}\label{entangprop}
For every object $X$ in a dagger-monoidal category $\CCC$ holds (a) $\iff$ (b) $\Longleftarrow$ (c), 
where
\begin{anumerate}
\item $\eta \in \CCC(X\otimes X)$ is weakly entangled
\item $\eta^\ddag \in \CCC(X\otimes X,I)$ internalizes the inner product, as  $<a |b > =
\eta^\ddag\circ(a_\ast \otimes b) $
\item $\eta \in \CCC(X\otimes X)$ is strongly entangled.
\end{anumerate}
The three conditions are equivalent if $I$ generates $\CCC$, in the sense that whenever $fa = ga$ for all $a\in \CCC(X)$, then $f=g$.
\end{prop}

\noindent A {\bf proof} can be conveniently built from transformations among the string diagrams of the conditions:
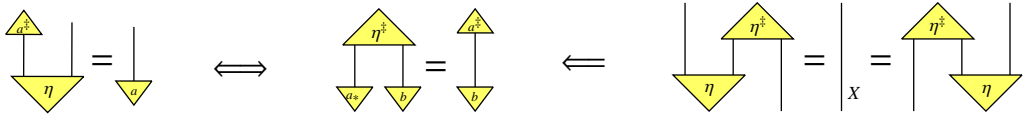
\renewcommand{\comonunit}{\scriptstyle \eta^\ddag}
\renewcommand{\monunit}{\scriptstyle \eta}
\newcommand{\first}{\scriptscriptstyle a}
\newcommand{\firstconj}{\scriptscriptstyle a_\ast}
\newcommand{\firstdag}{\scriptscriptstyle a^\ddag}
\newcommand{\second}{\scriptscriptstyle b}
\newcommand{\egvnt}{\mbox{\large $\iff$}}
\newcommand{\onlyif}{\mbox{\large $\Longleftarrow$}}
\begin{center}
\def\JPicScale{.8}
\ifx\JPicScale\undefined\def\JPicScale{1}\fi
\psset{unit=\JPicScale mm}
\psset{linewidth=0.3,dotsep=1,hatchwidth=0.3,hatchsep=1.5,shadowsize=1,dimen=middle}
\psset{dotsize=0.7 2.5,dotscale=1 1,fillcolor=black}
\psset{arrowsize=1 2,arrowlength=1,arrowinset=0.25,tbarsize=0.7 5,bracketlength=0.15,rbracketlength=0.15}
\begin{pspicture}(0,0)(168.88,18)
\psline[linewidth=0.22](112.88,6)(112.88,18)
\psline[linewidth=0.22](120.88,6)(120.88,12)
\psline[linewidth=0.22](128.88,0)(128.88,12)
\psline[linewidth=0.22](138.88,18)(138.88,0)
\rput(134.38,8){$\eqls$}
\psline[linewidth=0.22](166.88,5.92)(166.88,18)
\psline[linewidth=0.22](150.88,0)(150.88,12)
\psline[linewidth=0.22](158.88,6)(158.88,12)
\rput(145.38,8){$\eqls$}
\newrgbcolor{userFillColour}{1 1 0.4}
\psline[linewidth=0.22,fillcolor=userFillColour,fillstyle=solid](148.88,12)
(154.88,18)
(160.88,12)(148.88,12)
\rput(154.94,14.29){$\comonunit$}
\rput[l](139.88,3){$\obj$}
\newrgbcolor{userFillColour}{1 1 0.4}
\psline[linewidth=0.22,fillcolor=userFillColour,fillstyle=solid](118.88,12)
(124.88,18)
(130.88,12)(118.88,12)
\rput(125.12,14.11){$\comonunit$}
\newrgbcolor{userFillColour}{1 1 0.4}
\psline[linewidth=0.22,fillcolor=userFillColour,fillstyle=solid](168.88,6)
(162.88,0)
(156.88,6)(168.88,6)
\rput(162.98,3.39){$\monunit$}
\newrgbcolor{userFillColour}{1 1 0.4}
\psline[linewidth=0.22,fillcolor=userFillColour,fillstyle=solid](122.88,6)
(116.88,0)
(110.88,6)(122.88,6)
\rput(116.91,3.57){$\monunit$}
\psline[linewidth=0.22](58,2)(58,11)
\psline[linewidth=0.22](66,2)(66,11)
\rput(71.5,7){$\eqls$}
\newrgbcolor{userFillColour}{1 1 0.4}
\psline[linewidth=0.22,fillcolor=userFillColour,fillstyle=solid](56,11)
(62,17)
(68,11)(56,11)
\rput(62.14,13.21){$\comonunit$}
\newrgbcolor{userFillColour}{1 1 0.4}
\psline[linewidth=0.22,fillcolor=userFillColour,fillstyle=solid](75,13)
(78,17)
(81,13)(75,13)
\newrgbcolor{userFillColour}{1 1 0.4}
\psline[linewidth=0.22,fillcolor=userFillColour,fillstyle=solid](69,4)
(66,0)
(63,4)(69,4)
\newrgbcolor{userFillColour}{1 1 0.4}
\psline[linewidth=0.22,fillcolor=userFillColour,fillstyle=solid](61,4)
(58,0)
(55,4)(61,4)
\psline[linewidth=0.22](78,13)(78,4)
\rput(66.07,2.5){$\second$}
\newrgbcolor{userFillColour}{1 1 0.4}
\psline[linewidth=0.22,fillcolor=userFillColour,fillstyle=solid](81,4)
(78,0)
(75,4)(81,4)
\rput(77.96,2.5){$\second$}
\rput(78.05,14.46){$\firstdag$}
\psline[linewidth=0.22](11,14.75)(10.97,5.75)
\psline[linewidth=0.22](3,16)(2.97,5.77)
\rput(16,8){$\eqls$}
\newrgbcolor{userFillColour}{1 1 0.4}
\psline[linewidth=0.22,fillcolor=userFillColour,fillstyle=solid](12.97,5.75)
(6.96,-0.24)
(0.97,5.77)(12.97,5.75)
\newrgbcolor{userFillColour}{1 1 0.4}
\psline[linewidth=0.22,fillcolor=userFillColour,fillstyle=solid](-0.01,12.76)
(3,16.75)
(5.99,12.74)(-0.01,12.76)
\psline[linewidth=0.22](21.29,14)(21.29,5)
\newrgbcolor{userFillColour}{1 1 0.4}
\psline[linewidth=0.22,fillcolor=userFillColour,fillstyle=solid](24.29,5)
(21.29,1)
(18.29,5)(24.29,5)
\rput(3.07,14.14){$\firstdag$}
\rput(7.14,3.25){$\monunit$}
\rput(21.32,3.25){$\first$}
\rput(39,7){$\egvnt$}
\rput(96,8){$\onlyif$}
\rput(57.75,2.38){$\firstconj$}
\end{pspicture}

\end{center}

\subsection{Notation and terminology}
To describe relations on finite sets, we often find it convenient to use von Neumann's representation of ordinals, where $0=\emptyset$ is the empty set, and $n = \{0,1,\ldots, n-1\}$. Moreover, the pairs $<i,j>\in n\times n$ are often abbreviated to $ij\in n\times n$. 

When space is constrained and confusion unlikely, we often elide the tensors and write $AfXX$ instead of $A\otimes f\otimes X\otimes X$. 

\paragraph{Ioofs and embeddings.} Many categorical constructions lead to functors where the object part is the identity. They are often called Identity-Onthe-Objects-Functors. I call them {\em ioofs}. If the reader finds this abbreviation objectionable, she is welcome to unfold each of its occurrences, and read out the full phrase
% (like she presumably does with POVMs, CW-complexes, and ssh-protocol)
.  

In a similar development, the functors that are full and faithful are often called Full-and-Faithful-Functors. I call them {\em embeddings}. The reader may notice that every functor can be factored into an ioof followed by an embedding.

\section{Polynomials and abstraction}
\label{MonAbsSec}
In this section we formalize the program transformations needed to implement a classical function in a quantum computer. If a program is an arrow in a category, a program transformation is simply a functor out of it. But the problem with transforming a classical program into a quantum program is that classical data can be copied and deleted, whereas quantum data cannot. So the program transformation must map classical data to classical data, distinguished within a quantum universe. What does this mean? When the classical program $f'(x,y)$ was transformed into the corresponding quantum program $U_f|x,y>$ in the Introduction, the classical inputs were denoted by the variables $x,y$, and mapped to the  basis vector variables $|x,y>$.  The fact that the classical inputs can be copied and deleted was captured as a syntactical property of the variables. 

If the data over which a program will compute are denoted by variables, then the program itself is a polynomial in some suitable algebraic theory. More precisely, a program is an {\em abstraction\/} over the as-yet-undetermined input data, and a computation is an {\em application\/} of the program. More generally, a program transformation can be viewed as a {\em substitution\/} into a polynomial. So we need functorial semantics of polynomial constructions, and of the abstraction and substitution operations. In the framework of cartesian (closed) categories, such a treatment goes back to Lambek and Scott's seminal work \cite{LambekJ:Advances,Lambek-Scott:book}. It was extended to monoidal categories in \cite{PavlovicD:MSCS97}. Here we extend it to dagger-monoidal categories.

\subsection{Polynomial constructions} 
Adjoining an indeterminate $x$ to a ring $R$ leads to the ring of polynomials $R[x]$. Its universal property is that every ring homomorphism $f:R\to S$ extends to a unique ring homomorphism $f_a : R[x]\to S$ for each choice of $a\in S$ to which $x$ is mapped.
\vspace{\baselineskip}
\renewcommand{\ZZz}{R}
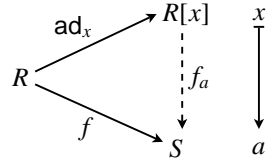
\begin{center}
\def\JPicScale{0.85}
\ifx\JPicScale\undefined\def\JPicScale{1}\fi
\psset{unit=\JPicScale mm}
\psset{linewidth=0.3,dotsep=1,hatchwidth=0.3,hatchsep=1.5,shadowsize=1,dimen=middle}
\psset{dotsize=0.7 2.5,dotscale=1 1,fillcolor=black}
\psset{arrowsize=1 2,arrowlength=1,arrowinset=0.25,tbarsize=0.7 5,bracketlength=0.15,rbracketlength=0.15}
\begin{pspicture}(0,0)(60,17)
\psline[tbarsize=0.15 5]{->}(25,8)(44,17)
\psline[tbarsize=0.15 5]{->}(25,6)(45,-3)
\rput[l](46,-4){$S$}
\psline[linestyle=dashed,dash=1 1,tbarsize=0.15 5]{->}(48,14)(48,-1)
\rput(33,-1){$f$}
\rput(60,17){$x$}
\rput(60,-4){$a$}
\rput(56,-5){}
\psline[tbarsize=0.15 5]{|*->}(60,15)(60,-1)
\rput[l](49,7){$f_a$}
\rput(31,15){${\sf ad}_x$}
\rput[l](45,17){$\ZZz[x]$}
\rput[r](24,7){$\ZZz$}
\end{pspicture}

\end{center}
\renewcommand{\ZZz}{\Bbb Z}
\vspace{1.5\baselineskip}
The same construction applies to other algebraic theories: e.g., one could form polynomial groups, or polynomial lattices. Categorically, for an arbitrary algebraic theory $T$, a polynomial $T$-algebra $A[x]$ can be viewed as the coproduct in the category of $T$-algebras of the $T$-algebra $A$ and the free $T$-algebra over one generator, denoted $x$.

The polynomial construction also applies to algebraic structures over categories, such as cartesian, monoidal, or $\ast$-autonomous; polynomial categories can be built for any algebraic theory $T$ over the category of categories. The polynomial category $\SSS[x]$ is then the free $T$-category obtained by freely adjoining a single generator $x$ to the $T$-category $\SSS$; i.e. as the coproduct of $\SSS$ and the free $T$-category generated by $x$. However, categories are generated over graphs, rather than sets, so the question is what kind of a graph should $x$ be. There seem to be two minimal choices:
\begin{anumerate}
\item $x$ is an object: a graph with one node and no edges; or
\item $x$ is an arrow: a graph with two nodes and an edge between them.
\end{anumerate}
While case (a) leads to the constructions which do not involve the arrows, and thus largely boil down to the polynomial constructions of universal algebra, case (b) involves genuinely categorical aspects. These new aspects are isolated by assuming that {\em only\/} new arrows are adjoined to $\SSS$, and {\em no\/} new objects. More precisely, an indeterminate arrow $A\stackrel{x}\rightarrow B$ is freely adjoined between the extant objects $A, B$ of $\SSS$. In other words, $\SSS[A\stackrel{x}\rightarrow B]$ can be viewed as the following pushout
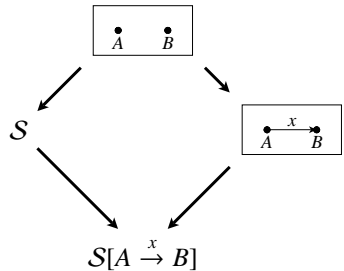
\newcommand{\source}{\scriptstyle A}
\newcommand{\target}{\scriptstyle B}
\newcommand{\indet}{\scriptstyle x}
\newcommand{\category}{\SSS}
\newcommand{\polycat}{\SSS[A\stackrel x \rightarrow B]}
\begin{center}
\def\JPicScale{0.66}
\ifx\JPicScale\undefined\def\JPicScale{1}\fi
\psset{unit=\JPicScale mm}
\psset{linewidth=0.3,dotsep=1,hatchwidth=0.3,hatchsep=1.5,shadowsize=1,dimen=middle}
\psset{dotsize=0.7 2.5,dotscale=1 1,fillcolor=black}
\psset{arrowsize=1 2,arrowlength=1,arrowinset=0.25,tbarsize=0.7 5,bracketlength=0.15,rbracketlength=0.15}
\begin{pspicture}(0,0)(80,50)
\pspolygon[linewidth=0.2](30,50)(50,50)(50,40)(30,40)
\psdots[linewidth=0.2,arrowscale=1 1.1,dotsize=0.9 2.5,dotsize=0.9 2.5,dotstyle=o](35,45)
(45,45)
\rput(35,42.5){$\source$}
\rput(45,42.5){$\target$}
\pspolygon[linewidth=0.2](60,30)(80,30)(80,20)(60,20)
\psline[linewidth=0.2,arrowscale=1 1.1,dotsize=0.9 2.5]{->}(65,25)(75,25)
\psdots[linewidth=0.2,arrowscale=1 1.1,dotsize=0.9 2.5,dotsize=0.9 2.5,dotstyle=o](65,25)
(75,25)
\rput(65,22.5){$\source$}
\rput(75,22.5){$\target$}
\rput(70,26.5){$\indet$}
\rput(15,25){$\category$}
\rput(40,0){$\polycat$}
\psline[linewidth=0.6]{->}(27.5,37.5)(18.75,28.75)
\psline[linewidth=0.6]{->}(57.5,17.5)(45,5)
\psline[linewidth=0.6]{->}(52.5,37.5)(57.5,32.5)
\psline[linewidth=0.6]{->}(18.75,21.25)(35,5)
\end{pspicture}

\end{center}
in the category of $T$-categories. 

Lambek was the first to use polynomial categories in his interpretation of typed $\lambda$-calculus in cartesian closed categories \cite{LambekJ:Advances}. The approach was elaborated in the book \cite{Lambek-Scott:book}, from which categorical semantics branched in many directions.  The terms containing a variable $x$ of type $X$ were represented as the arrows of the polynomial category $\SSS[\tp x X]$, built by adjoining to a cartesian closed category $\SSS$  an indeterminate arrow $1\stto x X$, where $X$ is an object of $\SSS$. The universal property of $\SSS[\tp{x}{X}]$ is the same as before: every structure preserving functor $F:\SSS\to \LLL$ extends to a unique structure-preserving functor $F_a: \SSS[x] \to \LLL$ by mapping $1\stto x X$ to $1\stto a FX$ in $\LLL$. 
\medskip
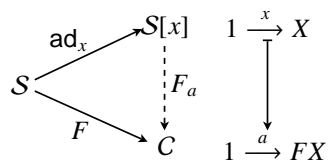
\begin{center}
\def\JPicScale{0.75}
\ifx\JPicScale\undefined\def\JPicScale{1}\fi
\psset{unit=\JPicScale mm}
\psset{linewidth=0.3,dotsep=1,hatchwidth=0.3,hatchsep=1.5,shadowsize=1,dimen=middle}
\psset{dotsize=0.7 2.5,dotscale=1 1,fillcolor=black}
\psset{arrowsize=1 2,arrowlength=1,arrowinset=0.25,tbarsize=0.7 5,bracketlength=0.15,rbracketlength=0.15}
\begin{pspicture}(0,0)(51.25,18.12)
\psline[tbarsize=0.15 5]{->}(8.75,8)(27.75,17)
\psline[tbarsize=0.15 5]{->}(8.75,6)(28.75,-3)
\psline[linestyle=dashed,dash=1 1,tbarsize=0.15 5]{->}(31.75,14)(31.75,-1)
\rput(39.75,-5){}
\psline[tbarsize=0.15 5]{|*->}(50,15)(50,-1)
\rput(14.75,15){${\sf ad}_x$}
\rput(50,18.12){$1\stackrel{x}\to X$}
\rput(51.25,-3.75){$1\stackrel{a}{\to} FX$}
\rput[r](7.75,7){$\SSS$}
\rput(31.88,17.5){$\SSS[x]$}
\rput(16.75,-1){$F$}
\rput[l](32.75,7){$F_a$}
\rput[l](30.62,-3.75){$\CCC$}
\end{pspicture}

\end{center}
\bigskip
Just like a polynomial ring, the category $\SSS[\tp{x}{X}]$ can be constructed syntactically. 
%This is described in \cite{LambekJ:Advances,Lambek-Scott:book}. 
However, the cartesian closed structure allows a more effective and more familiar presentation of $\SSS[\tp{x}{X} ]$.

\begin{thm}{\rm \cite{LambekJ:Advances,Lambek-Scott:book}}
\label{Lambek}
Let $\SSS$ be a cartesian category, $X\in \SSS$ an object  and $\SSS[\tp{x}{X}]$ the free cartesian category generated by $\SSS$ and $1\stto x X$. Then the inclusion functor $\adx : \SSS\to \SSS[\tp{x}{X}]$ has a left adjoint, the\/ {\em abstraction} functor $\abx :\SSS[\tp{x}{X}] \to  \SSS\ :\ A \mapsto X\times A$ 
\newcommand{\blah}{\SSS\big(\abx (A),B\big)}
\newcommand{\picSetxBlahh}{\SSS[\tp{x}{X}]\big(A,\adx(B)\big)}
\newcommand{\one}{\scriptstyle A\stackrel{\varphi(x)}{\to} B}
\newcommand{\two}{\scriptstyle X\times A\xrightarrow{\kappa x.\varphi(x)}B}
\newcommand{\four}{\scriptstyle X\times A\stackrel{f}{\to} B}
\newcommand{\three}{\scriptstyle A\stackrel{<x,\id>}\to X\times A\stackrel{f}{\rightarrow} B}
\medskip
\begin{center}
\def\JPicScale{1.3}
\ifx\JPicScale\undefined\def\JPicScale{1}\fi
\psset{unit=\JPicScale mm}
\psset{linewidth=0.3,dotsep=1,hatchwidth=0.3,hatchsep=1.5,shadowsize=1,dimen=middle}
\psset{dotsize=0.7 2.5,dotscale=1 1,fillcolor=black}
\psset{arrowsize=1 2,arrowlength=1,arrowinset=0.25,tbarsize=0.7 5,bracketlength=0.15,rbracketlength=0.15}
\begin{pspicture}(0,0)(53,14)
\rput(32,1){$\blah$}
\rput(32,13){$\picSetxBlahh$}
\rput(36,7){}
\psline[tbarsize=0.15 5]{|*->}(53,11)(53,4)
\psline[tbarsize=0.15 5]{|*->}(7,4)(7,11)
\rput(53,14){$\one$}
\rput(53,2){$\two$}
\rput(7,14){$\three$}
\rput(7,2){$\four$}
\psbezier{->}(34,10.88)(36.1,8.08)(36.1,5.71)(34,3)
\psbezier{<-}(31,10.88)(28.9,8.16)(28.9,5.8)(31,3)
\end{pspicture}

\end{center}
and $\SSS[\tp{x}{X}]$ is equivalent with the Kleisli category for the comonad $X\times (-):\SSS\to\SSS$.

When $\SSS$ is cartesian closed, then $\SSS[\tp{x}{X}]$ is cartesian closed too. The Kleisli category for the comonad $X\times (-):\SSS\to \SSS$ is isomorphic with the Kleisli category for the monad $(-)^X:\SSS\to \SSS$. The abstraction functor can now be viewed as a \/{\em right} adjoint of the inclusion $\adx : \SSS\to \SSS[\tp{x}{X}]$
\renewcommand{\blah}{\SSS\big( A,\abx(B)\big)}
\renewcommand{\picSetxBlahh}{\SSS[\tp{x}{X}]\big(\adx(A),B\big)}
\renewcommand{\one}{\scriptstyle A\stackrel{\varphi(x)}{\to} B}
\renewcommand{\two}{\scriptstyle A\xrightarrow{\lambda x.\varphi(x)} B^X}
\renewcommand{\four}{\scriptstyle A\stackrel{f}{\to} B^X}
\renewcommand{\three}{\scriptstyle A\stackrel{<f, x>}{\to}B^X\times X\stackrel{\varepsilon}{\rightarrow} B}
\medskip
\begin{center}
\def\JPicScale{1.3}

\end{center}
This latter adjunction provides a categorical model of simply typed lambda-calculus.
\end{thm}

%\paragraph{Polynomials, copying, and abstraction.} The essential feature of all polynomial constructions is that the adjoined indeterminate $x$ can be copied and deleted, so that the polynomials 
%\begin{itemize}
%\item $x\longmapsto x^2 + 2x$ and
%\item $x\longmapsto 3$
%\end{itemize}
%can be formed.

%The copying capability leads to abstraction: see proof of Thm. \ref{abs1}. {\em In a typed universe, a polynomial extension always comes with the abstraction capability, i.e. with an adjunction. The universal property (initiality)  of the polynomial extension then means that it must be equivalent with the Kleisli category.}

%Polynomial categories with abstraction should be contrasted with the Oles-type extensions of categories. Interpreted as polynomials, the arrows of Oles' category only allow a single occurrence of each variable. When two Oles' polynomial arrows are composed, their indeterminates are renamed, to avoid the name clash.

\paragraph{Notion of abstraction.} Function abstraction is what makes programming possible. The first example of program abstraction were probably G\"odel's numberings of primitive recursive functions \cite{GodelK:incompleteness}. G\"odel's construction demonstrated that recursive programs, specifying entire families of computations (of the values of a function for all its inputs), can be stored as data. Von Neumann later explicated this as the fundamental principle of computer architecture. Kleene, on the other side, refined the idea of program abstraction into the fundamental lemma of recursion theory: the s-m-n theorem \cite{KleeneS:smn}. Church, finally\footnote{Although Church's paper appeared three years earlier than Kleene's, Church's proposal is the final step in the conceptual development of function abstraction as the foundation of computation.} proposed the formal operations of function abstraction and data application as the driving force of all computation \cite{ChurchA:1940}. This proposal became the foundation of functional programming. Lawvere's observation that Church's $\lambda$-abstraction could be interpreted as an adjunction transposition \cite{LawvereFW:Dialectica} was a critical step towards categorical semantics of computation. Theorem \ref{Lambek} spells out this observation in terms of polynomial categories. Besides the familiar $\lambda$-abstraction, which uses the right adjoint of the inclusion $\adx : \SSS\to \SSS[\tp{x}{X}]$ to transpose a polynomial into a function which outputs functions
\[
\prooftree
\varphi(x):A\rightarrow B
\justifies
\lambda x.\varphi(x):A\rightarrow B^X
\endprooftree
\]
the theorem points to an analogous abstraction operation which uses the {\em left\/} adjoint to the inclusion $\adx : \SSS\to \SSS[\tp{x}{X}]$, and transposes polynomials into {\em indexed\/} families of functions
\[
\prooftree
\varphi(x):A\rightarrow B
\justifies
\kappa x.\varphi(x):X\times A\rightarrow B
\endprooftree
\]
This form of abstraction does not require higher-order types, and lifts from cartesian to monoidal categories \cite{PavlovicD:MSCS97}. In the present paper, we extend such abstraction operations to monoidal categories with enough structure to support the basic forms of quantum programming. --- In this way, the usual quantum programming constructions can be viewed as a form of functional programming in Hilbert spaces. 

But what kind of functional programming is it? 

The fundamental assumption of functional programming is that all data can be copied and deleted. Theorem \ref{Lambek} shows that this implies a canonical abstraction operation.

The fundamental assumption of quantum programming is that some data ---the quantum data--- cannot be copied or deleted; but they can be entangled. Entanglement is then developed into a powerful computational resource. In-between the data that can be copied and deleted, and the data that can be entangled, there is a rich structure of diverse abstraction operations, that we shall now explore. The idea is that quantum programming can be "semantically reconstructed" a set of techniques for combining and interfacing quantum entanglement and classical abstractions. 

\subsection{Abstraction in monoidal categories}
Given a monoidal category $\CCC$ and a chosen object $A$ in it, we want to freely adjoin a variable arrow $I\stto x A$ and build the polynomial monoidal category $\CCC[\tp{x}{X}]$. Like before, $\CCC[{\tp{x}{X}}]$ can be built syntactically, as the free symmetric monoidal category over the graph spanned by $\CCC$ and $I\stto x A$, factored by the equations between the arrows of $\CCC$. Although this is not a very effective description, it does show that  the polynomial category $\CCC[{\tp{x}{X}}]$ can in this case be quite complicated\footnote{E.g., $\Rel[x]$ is not a locally small category.}. Moreover, in contrast with the cartesian (closed) case, the inclusion $\adx : \CCC\to \CCC[{\tp{x}{X}}]$ does not have an adjoint in general, and thus does not support abstraction. The task is now to extend the polynomial construction to support abstraction. We follow, refine and strengthen the results from \cite{PavlovicD:MSCS97}.

\be{defn}\label{monext}
Let $\CCC$ be a monoidal category, and $E$ a set of well typed equations between some polynomial arrows in $\CCC[{\tp{x}{X}}]$. A {\em monoidal extension\/} is the monoidal category $\CCC[{\tp{x}{X}};E] = \CCC[{\tp{x}{X}}]/E$ obtained by imposing the equations $E$ on $\CCC[{\tp{x}{X}}]$, together with all equations that make it into a monoidal category. Every monoidal extension comes with the obvious ioof $\adxE : \CCC\to \CCC[{\tp{x}{X}};E]$.

A {\em substitution functor\/} between monoidal extensions is a (strict) monoidal ioof $F:\CCC[{\tp{x}{X}};E]\to \CCC[{\tp{y}{Y}};D]$.

We denote by $\Ext{\CCC}$ the category of monoidal extensions of $\CCC$, with the substitution functors between them.
\ee{defn}

\be{defn}\label{monabs}
A {\em (monoidal) abstraction\/} over a monoidal extension $\adxE:\CCC\to \CCC[{\tp{x}{X}};E]$ is the adjunction $\abxE\dashv\adxE$ such that $\abxE (A\otimes B) = \ab(A)\otimes B$, and the unit of the adjunction $h: {\rm Id} \to \adxE\circ \abxE$ satisfies $h_{A} = x \otimes A$. We denote by $\Abs{\CCC}$ the subcategory of $\Ext{\CCC}$ spanned by the monoidal extensions that support abstraction. 
\ee{defn}

\paragraph{Notation and terminology.} Since the abstraction notation $\abxE\dashv \adxE:\CCC\to \CCC[{\tp{x}{X}};E]$ is generic, we often elide the structure and refer to an abstraction as $\CCC[{\tp{x}{X}};E]$.

\begin{thm}\label{abs1} The category $\Abs{\CCC}$ of monoidal abstractions is equivalent with the category $\CCC_\times$ of commutative comonoids in $\CCC$. Each abstraction is isomorphic with the Kleisli adjunction for the comonad induced by the corresponding comonoid. 
\end{thm}

\bprf{ (sketch)} Given a commutative comonoid $(X,\cmn,\cun)$ in $\CCC$, we construct the abstraction  $\abxE\dashv\adxE:\CCC\to \CCC[\tp{x}{X};E]$ as follows. Let 
\bear
E & = &  E_{(\cmn,\cun)}
\eear 
be the set of equations
\bear
\underset{n\ {\rm times}}{\underbrace{x\otimes x\otimes \cdots \otimes x}} & = & \cmn_n \circ x\quad \mbox{ for }n=0,1,2\ldots  
\eear
where $\cmn_n : X\to X^{\otimes n}$ is defined inductively:
\begin{gather*}
\cmn_0 = \cun \qquad 
\cmn_1 =  \id_X \qquad \cmn_2  =  \cmn\\
\cmn_{i+1} =  (\cmn \times X^{\otimes i-1})\circ \cmn_i
\end{gather*}
%\bear
%\cmn^0 & = & \cun\\
%\cmn^1 & = & \id_X\\
%\cmn^2 & = & \cmn\\
%\cmn^{i+1} & = & (\cmn \times X^{\otimes i-1})\circ \cmn^i
%\eear
This determines the extension $\adxE:\CCC\to \CCC[\tp{x}{X};E]$. Using the symmetry, it follows that every polynomial $\varphi(x)\in \CCC[\tp{x}{X};E]$ must satisfy the equation
\vspace{\baselineskip}
\newcommand{\polnm}{\varphi(x)}
\newcommand{\decomp}{\overline{\varphi}\circ (x\otimes A)}
\renewcommand{\eqls}{$\mbox{\Large =}$}
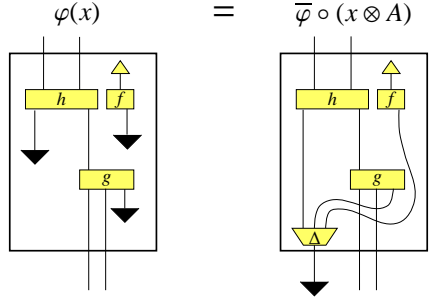
\begin{center}
\def\JPicScale{0.75}
\ifx\JPicScale\undefined\def\JPicScale{1}\fi
\psset{unit=\JPicScale mm}
\psset{linewidth=0.3,dotsep=1,hatchwidth=0.3,hatchsep=1.5,shadowsize=1,dimen=middle}
\psset{dotsize=0.7 2.5,dotscale=1 1,fillcolor=black}
\psset{arrowsize=1 2,arrowlength=1,arrowinset=0.25,tbarsize=0.7 5,bracketlength=0.15,rbracketlength=0.15}
\begin{pspicture}(0,0)(77,51)
\rput(1.75,5){}
\newrgbcolor{userFillColour}{1 1 0.4}
\pspolygon[linewidth=0.12,fillcolor=userFillColour,fillstyle=solid](5.79,34.16)(18.59,34.16)(18.59,37.58)(5.79,37.58)
\rput(12.19,35.87){$\scriptstyle h$}
\newrgbcolor{userFillColour}{1 1 0.4}
\pspolygon[linewidth=0.12,fillcolor=userFillColour,fillstyle=solid](15.39,19.91)(24.99,19.91)(24.99,23.33)(15.39,23.33)
\pspolygon[linewidth=0.12](24.99,21.62)(24.99,21.62)(24.99,21.62)(24.99,21.62)
\psline[linewidth=0.12](23.39,16.49)(23.39,19.91)
\psline[linewidth=0.12](17,2)(16.99,19.91)
\psline[linewidth=0.12](16.99,23.33)(16.99,34.16)
\rput(20.19,21.62){$\scriptstyle g$}
\newrgbcolor{userFillColour}{1 1 0.4}
\pspolygon[linewidth=0.12,fillcolor=userFillColour,fillstyle=solid](20.19,34.16)(24.99,34.16)(24.99,37.58)(20.19,37.58)
\rput(22.59,35.87){$\scriptstyle f$}
\psline[linewidth=0.12](20,2)(20,20)
\psline[linewidth=0.12](23.79,29.6)(23.79,34.16)
\psline[linewidth=0.12](8.99,37.58)(9,47)
\psline[linewidth=0.12](22.59,37.58)(22.59,40.43)
\psline[linewidth=0.12](15.39,37.58)(15.38,46.93)
\newrgbcolor{userFillColour}{1 1 0.4}
\psline[linewidth=0.12,fillcolor=userFillColour,fillstyle=solid](20.99,40.43)
(22.59,42.71)
(24.19,40.43)(20.99,40.43)
\psline[linewidth=0.12](7.44,26.75)(7.44,33.88)
\newrgbcolor{userFillColour}{1 1 0.4}
\pspolygon[linewidth=0.12,fillcolor=userFillColour,fillstyle=solid](53.79,34.16)(66.59,34.16)(66.59,37.58)(53.79,37.58)
\rput(60.19,35.87){$\scriptstyle h$}
\newrgbcolor{userFillColour}{1 1 0.4}
\pspolygon[linewidth=0.12,fillcolor=userFillColour,fillstyle=solid](63.39,19.91)(72.99,19.91)(72.99,23.33)(63.39,23.33)
\pspolygon[linewidth=0.12](72.99,21.62)(72.99,21.62)(72.99,21.62)(72.99,21.62)
\psline[linewidth=0.12](65,2)(64.99,19.91)
\psline[linewidth=0.12](64.99,23.33)(64.99,34.16)
\rput(68.19,21.62){$\scriptstyle g$}
\newrgbcolor{userFillColour}{1 1 0.4}
\pspolygon[linewidth=0.12,fillcolor=userFillColour,fillstyle=solid](68.19,34.16)(72.99,34.16)(72.99,37.58)(68.19,37.58)
\rput(70.59,35.87){$\scriptstyle f$}
\psline[linewidth=0.12](68,2)(68,20)
\psline[linewidth=0.12](57,37.56)(57,47)
\psline[linewidth=0.12](70.59,37.58)(70.59,40.43)
\psline[linewidth=0.12](63.5,37.57)(63.5,47.02)
\newrgbcolor{userFillColour}{1 1 0.4}
\psline[linewidth=0.12,fillcolor=userFillColour,fillstyle=solid](68.99,40.43)
(70.59,42.71)
(72.19,40.43)(68.99,40.43)
\psline[linewidth=0.12](55,13)(55,34)
\pspolygon[](51,44)(77,44)(77,9)(51,9)
\psecurve[linewidth=0.12,curvature=1.0 0.1 0.0](71,20)(71,20)(70,17)(60,18)(58,17)(57.25,15.38)(57,13)(57,13)(57,13)(57.25,12.88)
\psecurve[linewidth=0.15,curvature=1.0 0.1 0.0](59,13)(59,13)(59.75,16)(61.71,16.39)(72,14)(74.75,17.88)(72.88,26.62)(72.25,29.12)(72,31)(72,34)(72,34)
\psline[linewidth=0.15](57,10)(57,2)
\psline[linewidth=0.05,fillstyle=solid](9.97,26.86)
(7.47,24.36)(4.97,26.86)
\psline[linewidth=0.05,fillstyle=solid](26.36,29.54)
(23.86,27.04)(21.36,29.54)
\psline[linewidth=0.05,fillstyle=solid](25.91,16.54)
(23.41,14.04)(20.91,16.54)
\newrgbcolor{userFillColour}{1 1 0.4}
\pspolygon[linewidth=0.15,fillcolor=userFillColour,fillstyle=solid](53.05,12.98)
(61.05,12.98)
(59.05,9.98)
(55.05,9.98)(53.05,12.98)
\rput(57.03,11.39){$\scriptstyle \Delta$}
\pspolygon[](3,44)(29,44)(29,9)(3,9)
\rput(41,51){$\eqls$}
\psline[linewidth=0.05,fillstyle=solid](59.5,3.5)
(57,1)(54.5,3.5)
\rput(15,51){$\polnm$}
\rput(64,51){$\decomp$}
\end{pspicture}

\end{center}
Setting $\kappa x.\varphi(x) = \overline{\varphi}$, define
\bea
\abxE: \CCC[\tp{x}{X};E] & \to & \CCC\notag\\
A & \longmapsto & X\otimes A \notag\\
\varphi(x) & \longmapsto & \left(X\otimes \kappa x.\varphi(x)\right)\circ (\cmn \otimes A) \label{abdef}
\eea
The adjunction correspondence, with $\adxE(B) = B$, is now
\medskip
\newcommand{\leftcat}{\CCC\big(\abx(A),B\big)}
\newcommand{\rightcat}{\CCC[\tp{x}{X};E]\big(A,\adx(B)\big)}
\begin{center}
\def\JPicScale{0.75}
\ifx\JPicScale\undefined\def\JPicScale{1}\fi
\psset{unit=\JPicScale mm}
\psset{linewidth=0.3,dotsep=1,hatchwidth=0.3,hatchsep=1.5,shadowsize=1,dimen=middle}
\psset{dotsize=0.7 2.5,dotscale=1 1,fillcolor=black}
\psset{arrowsize=1 2,arrowlength=1,arrowinset=0.25,tbarsize=0.7 5,bracketlength=0.15,rbracketlength=0.15}
\begin{pspicture}(0,0)(55,27.63)
\rput(40.75,-25){}
\rput[r](10,21){$\leftcat$}
\rput[l](40,21){$\rightcat$}
\psecurve{->}(10,23)(10,23)(24,26)(40,23)(44,21)
\psecurve{->}(39.99,19.22)(39.99,19.22)(25.92,16.59)(10,20)(6.05,22.1)
\rput[l](-10,7){$\big(\kappa x.\ \varphi(x)\big)\circ \big(x\otimes A) = \varphi(x)$}
\rput[l](-8,1){$\kappa x.\ \big(f\circ (x\otimes A)\big) = f$}
\rput(25,15){$\scriptstyle \kappa x.$}
\rput(25.09,27.63){$\scriptstyle (-)\circ (x\otimes A)$}
\rput[l](55,1){$(\eta\mbox{-rule}$}
\rput[l](55,7){$(\beta\mbox{-rule}$}
\rput(25,21){$\cong$}
\end{pspicture}

\end{center}

The other way around, given an abstraction $\abx\dashv \adx : \CCC\to \CCC[\tp{x}{X};E]$, the conditions from Def.~\ref{monabs} imply that  $h(A) = x\otimes A$ and $\abxE(A) = X\otimes A$. With the transposition $\kappa x$ as above, the comonoid structure must be
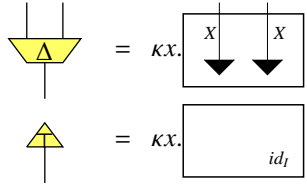
\begin{center}
\def\JPicScale{0.8}
\ifx\JPicScale\undefined\def\JPicScale{1}\fi
\psset{unit=\JPicScale mm}
\psset{linewidth=0.3,dotsep=1,hatchwidth=0.3,hatchsep=1.5,shadowsize=1,dimen=middle}
\psset{dotsize=0.7 2.5,dotscale=1 1,fillcolor=black}
\psset{arrowsize=1 2,arrowlength=1,arrowinset=0.25,tbarsize=0.7 5,bracketlength=0.15,rbracketlength=0.15}
\begin{pspicture}(0,0)(50,30)
\rput(40.75,-6){}
\psline[linewidth=0.22](4,24)(4,30)
\newrgbcolor{userFillColour}{1 1 0.4}
\pspolygon[linewidth=0.22,fillcolor=userFillColour,fillstyle=solid](1,24)
(13,24)
(10,20)
(4,20)(1,24)
\psline[linewidth=0.22](7,20)(7,14)
\psline[linewidth=0.22](7,0)(7,6)
\newrgbcolor{userFillColour}{1 1 0.4}
\psline[linewidth=0.22,fillcolor=userFillColour,fillstyle=solid](4,6)
(7,10)
(10,6)(4,6)
\psline[linewidth=0.22](10,24)(10,30)
\rput(7,7){$\top$}
\rput(6.96,21.98){$\Delta$}
\rput(20,22){$=$}
\rput(20,7){$=$}
\rput[r](30,22){$\kappa x.$}
\rput[r](30,7){$\kappa x.$}
\pspolygon[linewidth=0.25](30,28.12)(50,28.12)(50,16)(30,16)
\pspolygon[linewidth=0.25](30,13)(50,13)(50,1)(30,1)
\psline[linewidth=0.12](36,20)(36,30)
\psline[linewidth=0.05,fillstyle=solid](38.5,20.5)
(36,18)(33.5,20.5)
\psline[linewidth=0.12](44,20)(44,30)
\psline[linewidth=0.05,fillstyle=solid](46.5,20.5)
(44,18)(41.5,20.5)
\rput(41.88,3.75){}
\rput[r](35.62,25){$\scriptstyle X$}
\rput[l](45,25){$\scriptstyle X$}
\rput(46,4){$\scriptstyle id_I$}
\end{pspicture}

\end{center}

The arrow part of the claimed equivalence $\Abs{\CCC}\simeq \CCC_\times$ follows in one direction from the fact that any comonoid homomorphism $f:Y\sto X$ induces a unique ioof $F:\CCC[{\tp{x}{X}}]\to \CCC[{\tp{y}{Y}}]$, mapping $\varphi(x)$ to $F\varphi(x)  =  \varphi(f\circ y)$. Since every structure-preserving functor $F$ is easily seen to be induced by the comonoid homomorphism $f= \kappa y.\ Fx$  in this way, the bijective correspondence $\Abs{}\left(\CCC[{\tp{x}{X}}],\CCC[{\tp{y}{Y}}]\right)\cong\CCC_\times(X,Y)$ is established.

The isomorphism $\CCC[\tp{x}{X}] \cong \Kl{\CCC}{X}$, where $\Kl{\CCC}{X}$ is the Kleisli category for the comonoid $X$, is obtained by viewing the transpositions $\kappa x. (-)$ and $(-)\circ(x\otimes A)$ as functors. More precisely, this isomorphism is realized by the following ioofs:
\begin{alignat*}{4}
K: \CCC[\tp{x}{X}] & \to \Kl{\CCC}{X} \qquad & H:\Kl{\CCC}{X} & \to \CCC[\tp{x}{X}]\\
\varphi(x) &\longmapsto \kappa x.\ \varphi(x)\qquad\qquad & f &\longmapsto f\circ (x\otimes A)
\end{alignat*}
The fact that $H\circ K = \id$ is just the $\beta$-rule; the fact that $K\circ H = \id$ is the $\eta$-rule. Proving the functoriality of $K$ and $H$, and the fact that they commute with the abstraction structure $\abx\dashv\adx:\CCC\to \CCC[\tp x X;E]$ and the Kleisli adjunction $V\dashv G:\CCC\to \Kl{\CCC}{X}$  is an instructive exercise.  
%
%Towards the proof of the second statement, note that the adjunction $\abx\dashv \adx : \CCC\to \CCC[\tp{x}{X};E]$ is, by definition of the polynomial extension, initial among all adjunctions that induce on $\CCC$ the comonad corresponding to the induced comonoid $(X,\cmn,cun)$. But the Kleisli category for a comonad satisfies the same universal property: it is also initial among all the adjunctions that induce its comonad. Hence $\CCC[\tp{x}{X};E]$ must be equivalent with the Kleisli category.
\epr

\paragraph{Remarks.}(a) The upshot of the preceding theorem is that the set of equations $E$ in $\CCC[\tp{x}{X};E]$ determines the comonoid structure $(\cmn,\cun)$ over $X$; and {\em vice versa}: the comonoid structure $(\cmn,\cun)$ determines the equations $E = E_{(\cmn,\cun)}$, as in the above proof. Just like we often speak of a "comonoid $X$" and leave the actual structure $(\cmn,\cun)$ implicit, we shall often elide $E$, and write $\CCC[\tp x X]$, or even $\CCC[x]$, whenever the rest of the structure is clear from the context. We shall also blur the distinction between the comonoid $(X,\cmn, \cun)$ and the corresponding comonad, and denote both by $X$, writing $\Kl{\CCC}{X}$ for the $X$-Kleisli category, the $\EM{\CCC}{X}$ for the  $X$-Eilenberg-Moore category.

(b) The extension process can be iterated to construct $\CCC[\tp{x}{X}, \tp{y}{Y}] = \CCC[\tp{x}{X}][\tp{y}{Y}] \cong \Kl{\CCC}{X\otimes Y}$, or $\CCC[\tp{x,y}{X} \cong \Kl{\CCC}{X\otimes X}$.

%\paragraph{Remark.}  
(c) The category $\CCC_\times$ of commutative comonoids is the cofree cartesian category over the monoidal category $\CCC$ \cite{Fox}.  The equivalence of categories established in \ref{abs1} can be extended to an equivalence of 2-categories. The 2-cells of $\Abs{\CCC}$ are the monoidal natural transformations. The 2-cells of $\CCC_\times$ can be obtained by dualizing the notion of natural transformations between the monoid homomorphisms. And the monoid homomorphisms are functors between categories with one object, so the usual notion of natural transformation just needs to be internalized. The reader may find it interesting to work this out.

(d) Recall (or see \ref{MonComon}) that the tensor unit $I$ carries a canonical structure of a commutative comonoid. Adjoining a variable $I\stto y I$ leads to $\CCC[\tp y I] \cong \CCC$, because $\cmn y = y\otimes y$ and the coherence conditions imply $y = \id_I$.

\be{corollary}\label{diagx} In every extension $\CCC[\tp x X]$ that supports monoidal abstraction holds $\cmn x = x\otimes x$ and $\cun x = \id_I$.
\ee{corollary}

\bpr The first equation follows by postcomposing with $x$ the equation $\cmn = \kappa x.\ x\otimes x$, which is the definition of $\cmn$ in $\CCC[\tp x X]$, and applying the $\beta$-rule. The second one is obtained by precomposing $\cun = \kappa x.\ \id_I$ with $x$ and applyng the $\beta$-rule.
\epr

\begin{corollary} If the extension $\CCC[{\tp{x}{X}}]$ supports abstraction, then $X$ is generated by the tensor unit $I$. As a consequence, a weakly entangled vector $\eta \in \CCC[\tp x X](X\otimes X)$ is always strongly entangled.
\end{corollary}

\bpr  By definition, $I$ generates $X$ in $\CCC$ if whenever $fa=ga$ for all $a\in \CCC(X)$, then $f=g$, for any $f,g\in \CCC(X,Y)$. But the $\eta$-rule implies that $fx = gx$ implies $f=g$. Hence the first claim. Furthermore, the same fact can be used to show that condition (a) implies condition (c) in Prop.~\ref{entangprop}. E.g., going back to the proof of \ref{entangprop}, condition (c) can be obtained by composing the diagram for condition (a) and its dagger, after instantiating $a$ to $x$. Condition (c) then follows by abstracting over $x$.
\epr

\subsubsection{Substitutions}\label{Substitutions}
But what does the variable $x$ in the extension $\CCC[x]$ actually represent? What kind of vectors can be {\em substituted\/} for it?

\be{defn}
A {\em Substitution\/} for $x$ in $\CCC[{\tp{x}{X}}]$ is a monoidal functor $\CCC[{\tp{x}{X}}]\to \CCC$. 
\ee{defn}

\begin{corollary}\label{substitution-corollary} Substitutions $\CCC[{\tp{x}{X}}]\to \CCC$ are in one-to-one correspondence with the comonoid homomorphisms $I\sto X$, where $X$ is the comonoid that induces the abstraction in $\CCC[{\tp{x}{X}}]$ as in Thm.~\ref{abs1}.
\end{corollary}

\paragraph{Remark.} Only the vectors $a\in\CCC(X)$ that happen to be comonoid homomorphisms can thus be substituted for $x\in \CCC[\tp x X](X)$, leading to. In the category $\FHilb$ of finitely-dimensional Hilbert spaces, such vectors turn out to form a basis of the space $X$.

\subsubsection{Bases}
\be{defn}\label{basis}
A {\em basis vector\/} with respect to a comonoid $(X,\cmn,\cun)$ in $\CCC$ is a comonoid homomorphism from $I$, i.e. an arrow $\beta: I\sto X$ satisfying $\cmn \beta = \beta \otimes \beta$ and $ \cun \beta = \id_I$.
\newcommand{\baselt}{\scriptstyle\beta}
\newcommand{\idtty}{\scriptscriptstyle\id_I}
\newcommand{\comonoid}{\cmn}
\newcommand{\comonun}{\cun}
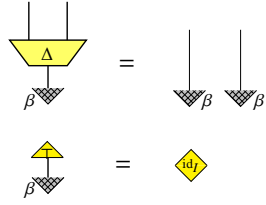
\begin{center}
\def\JPicScale{0.85}
\ifx\JPicScale\undefined\def\JPicScale{1}\fi
\psset{unit=\JPicScale mm}
\psset{linewidth=0.3,dotsep=1,hatchwidth=0.3,hatchsep=1.5,shadowsize=1,dimen=middle}
\psset{dotsize=0.7 2.5,dotscale=1 1,fillcolor=black}
\psset{arrowsize=1 2,arrowlength=1,arrowinset=0.25,tbarsize=0.7 5,bracketlength=0.15,rbracketlength=0.15}
\begin{pspicture}(0,0)(41.88,30)
\rput(40.75,-6){}
\psline[linewidth=0.22](4,24)(4,30)
\newrgbcolor{userFillColour}{1 1 0.4}
\pspolygon[linewidth=0.22,fillcolor=userFillColour,fillstyle=solid](1,24)
(13,24)
(10,20)
(4,20)(1,24)
\psline[linewidth=0.22](7,20)(7,14)
\psline[linewidth=0.22](7,0)(7,6)
\psline[linewidth=0.22](10,24)(10,30)
\rput(19.38,20.12){$=$}
\rput(18.75,4.38){$=$}
\psline[linewidth=0.12](29.12,15.62)(29.12,25.62)
\psline[linewidth=0.05,fillcolor=lightgray,fillstyle=crosshatch*,hatchwidth=0.15,hatchsep=0.55](31.62,16.12)
(29.12,13.62)(26.62,16.12)
\psline[linewidth=0.12](37.12,15.62)(37.12,25.62)
\psline[linewidth=0.05,fillcolor=lightgray,fillstyle=crosshatch*,hatchwidth=0.15,hatchsep=0.55](39.62,16.12)
(37.12,13.62)(34.62,16.12)
\rput(41.88,3.75){}
\psline[linewidth=0.05,fillcolor=lightgray,fillstyle=crosshatch*,hatchwidth=0.15,hatchsep=0.55](9.5,16.5)
(7,14)(4.5,16.5)
\psline[linewidth=0.05,fillcolor=lightgray,fillstyle=crosshatch*,hatchwidth=0.15,hatchsep=0.55](9.5,2.5)
(7,0)(4.5,2.5)
\psline[linewidth=0.15,fillcolor=yellow,fillstyle=solid](4.38,5.62)
(6.88,8.13)(9.38,5.63)
\psline[linewidth=0.15](4.38,5.62)(9.38,5.62)
\psline[linewidth=0.15,fillcolor=yellow,fillstyle=solid](31.88,4.38)
(29.38,1.88)(26.88,4.38)
\psline[linewidth=0.15,fillcolor=yellow,fillstyle=solid](26.88,4.38)
(29.38,6.88)(31.88,4.38)
\rput[l](31.25,14.38){$\baselt$}
\rput[l](39.38,14.38){$\baselt$}
\rput[r](5,14.38){$\baselt$}
\rput[r](5,0.62){$\baselt$}
\rput(29.38,4.38){$\idtty$}
\rput(6.88,21.88){$\comonoid$}
\rput(6.88,6.25){$\comonun$}
\end{pspicture}

\end{center}
The {\em basis\/} of a comonoid is the set of its basis vectors.
\ee{defn}

%\paragraph{Remark.} 
In Hopf algebra theory, our basis vectors are sometimes called {\em set-like elements}. We shall see in the next section that, for a special family of comonoids that we call classical structures, the bases tend to form categories equivalent to the category of sets. The basis vectors of a type $X$ in a monoidal category $\CCC$ are just the data that can be copied and deleted by a given comonoid structure on $X$. 
%Indeed, the purpose of this comonoid structure, and of the variable manipulations that it enables, is to propagate some indeterminate data. This capability to manipulate data that will only be determined later is the essence of programming. The basis vectors provide a mathematical  formalization of this intuition: only the data that can be copied and deleted can be substituted for variables.

\paragraph{Examples.}
Consider the monoidal category $(\Rel,\times, 1)$ of sets and relations. Every set $X$ has a standard comonoid structure $X_1 = \left(X,\cmn,\cun\right)$, induced by the cartesian structure of sets: 
\[
\cmn (x) = \{xx\} \qquad \qquad \cun (x) = \{0\}
\]
On the other hand, any monoid $(X,\mplus,\munit)$ over the same underlying set induces a nonstandard comonoid $X_2 = (X,\opr\mplus,\opr\munit)$, where $\opr r:B\sto A$ denotes the converse relation of $r:A\sto B$, and thus
\[
\opr\mplus (u) = \{vw\ |\ u = v\mplus w\} \qquad \qquad \opr{\munit} (u) = \{\munit\}
\]
These different comonoids induce different monoidal extensions $\Rel[\tp x {X};E_1]$ and $\Rel[\tp x {X};E_2]$, with different abstraction operations. Both extensions have the same objects, and even the same arrows, but these arrows  compose in different ways. Viewed in the Kleisli form, both categories consist of relations in the form $X\times A \to B$. But the composites $X\times A\tto{r;s} C$ of $X\times A\tto{r} B$ and $X\times B\tto{s}$ will respectively be
\bear
(r;s)_1(u,a,c)  &\iff & \exists b.\ \  r(u,a,b)\wedge s(u,b,c) \\
%\eear
%whereas in $\Rel_2[x]$ it is
%\be{multline*}
(r;s)_2(u,a,c)  & \iff & \exists bvw.\ r(w,a,b)\wedge  s(v,b,c) \\
&& \hspace{2.7em} \wedge\  u = v\mplus w
\eear
As a consequence, each case allows substitution of different basis vectors. With respect to the standard comonoid $X_1 = (X,\cmn,\cun)$, the basis vectors are just the singleton relations $\{u\}\in \Rel(X)$. The variable $x$ in $\Rel[\tp x {X};E_1]$ thus denotes an indeterminate element of the set $X$. On the other hand, with respect to the comonoid $X_2 = (X,\opr \mplus,\opr \munit)$, there is only one basis vector $\beta \in\Rel(X)$, which is the subset of $X$ consisting of the invertible elements with respect to the monoid $(X,\mplus,\munit)$.  The variable $x$ in $\Rel[\tp x {X};E_2]$ thus denotes this one vector $\beta \in \Rel(X)$, since there is nothing else that can be substituted for $x$.

\section{Daggers and classical structures}\label{DagAbsSec}
This section adds the dagger functor, and the dualities to the monoidal framework of abstraction (cf. \ref{dagmondefs}). The abstraction now leads to classical structures, which were introduced in \cite{PavlovicD:QMWS}, albeit without discussing their origin in the abstraction operations.

%\paragraph{Intro remarks.} Besides the monoidal structure, the category $\Rel$ also has the duality, expressed as the ioof $\ddag :\Rel^{op} \to \Rel$, mapping each relation $r:A\sto B$ to the converse relation $r^\ddag : B\sto A$. In this section we study polynomial extensions of symmetric monoidal categories with such a duality. The requirement that the abstraction operations preserve this duality naturally leads to the concept of {\em classical structure}, carried by the classical interfaces of quantum computation \cite{PavlovicD:QMWS}.

\subsection{Dagger-monoidal abstraction}
\be{defn}\label{dagext}
Let $\CCC$ be a dagger-monoidal category, and $E$ a set of equations between some parallel arrows in the dagger-monoidal polynomial category $\CCC[{\tp{x}{X}}]$. A {\em dagger-monoidal extension\/} is the dagger-monoidal category $\CCC[{\tp{x}{X}};E] = \CCC[{\tp{x}{X}}]/E$, obtained by imposing the equations $E$ on $\CCC[{\tp{x}{X}}]$, together with all equations that make it into a dagger-monoidal category. As all such constructions, it comes with the obvious ioof $\adx:\CCC\to \CCC[{\tp{x}{X}};E]$.

A {\em substitution functor\/} between the dagger-monoidal extensions is a monoidal ioof $F:\CCC[{\tp{x}{X}};E]\to \CCC[{\tp{y}{Y}};D]$ which preserves the dagger, i.e. $F(\psi^\ddag) = (F\psi)^\ddag$. 

We denote by $\dExt{\CCC}$ the category of dagger-monoidal extensions of $\CCC$, with the substitution functors between them.
\ee{defn}

\be{defn}\label{dagabs}
A {\em dagger monoidal abstraction}\/ over a dagger monoidal extension $\adxE : \CCC\to \CCC[{\tp{x}{X}};E]$ is the  adjunction $\abxE\dashv \adxE$, which satisfies the requirements of Definition \ref{monabs}, and moreover preserves the dagger, in the sense that $\abxE. \varphi(x)^\ddag = \left( \abxE. \varphi(x)\right)^\ddag$.

We denote by $\dAbs{\CCC}$ the subcategory of $\dExt{\CCC}$ where the abstraction is supported. Its objects are often called abstractions.
\ee{defn}

%\be{lemma}
%A monoidal abstraction $\abx\dashv \adxE : \CCC\to \CCC[{\tp{x}{X}};E]$, induced by a comonoid $(X,\cmn,\cun)$  as in Thm.~\ref{abs1}, preserves the dagger if and only if

%

%If an abstraction functor $\abxE$ preserves the dagger, then 
%\bear
% \kappa x. \varphi(x)^\ddag\circ (x\otimes B) & = & \left(x^\ddag \otimes B\right)\circ\left( \kappa x. \varphi(x)\right)^\ddag
%\eear
%\begin{center}
%\def\JPicScale{0.8}
%\input{PIC/DagKappa}
%\end{center}
%\ee{lemma}

%\bpr
%Unfolding $\abxE$ as in (\ref{abdef}), we see that the preservation $\abxE. \varphi(x)^\ddag = \left( \abxE. \varphi(x)\right)^\ddag$ means that
%\[
%\big(X\otimes \kappa x.\varphi(x)^\ddag \big) (\Delta \otimes B)  =  (\nabla \otimes A) \big( X\otimes \left(\kappa x. \varphi(x)\right)^\ddag \big)
%\]
%%the following diagram commutes
%%\medskip
%%\begin{center}
%%\def\JPicScale{0.75}
%%\input{PIC/DagAbs}
%%\end{center}
%%\bigskip
%As soon as $\kappa x$ is defined over $\varphi(x)^\ddag$, we can thus apply the $\beta$-rule and the functoriality of $\ddag$ to get the result
%\bear
% \kappa x. \varphi(x)^\ddag\circ (x\otimes B) & = & \varphi(x)^\ddag \\
% & = & \left(\kappa x. \varphi(x)\circ (x\otimes B)\right)^\ddag\\
% & = & \left(x^\ddag \otimes B\right)\circ\left( \kappa x. \varphi(x)\right)^\ddag
%\eear
%\epr

Thm.~\ref{abs1} established the correspondence between monoidal abstractions over $X$ and the comonoid structures carried by $X$. The next theorem extends this correspondence to dagger monoidal categories: a monoidal abstraction corresponding to a comonoid structure preserves the dagger if and only if the Kleisli category, induced by the comonoid, is (equivalent with) the dagger monoidal extension itself. 

\be{thm}\label{abs2} 
Let $\CCC$ be a dagger-monoidal category and $\adx : \CCC\to \CCC[{\tp{x}{X}};E]$ a dagger-monoidal extension. Suppose that it admits a monoidal abstraction $\abx\dashv \adx$ (as in Def.~\ref{monabs}), with the induced comonoid $(X,\cmn,\cun)$ (as in Thm.~\ref{abs1})
%, and $(X,\mnd,\unt)$ the monoid where $\mnd = \cmn^\ddag$ and $\unt = \cmn^\ddag$
. Then the following statements are equivalent:
\begin{anumerate}
\item $\abx\dashv \adx : \CCC\to \CCC[\tp{x}{X};E]$ is a dagger-abstraction, i.e.  $\abxE. \varphi(x)^\ddag = \left( \abxE. \varphi(x)\right)^\ddag$
\item $x$ is real, i.e. $x^\ast = x^\ddag$
%$\cun \circ \mnd \circ (x\otimes X) = x^\ddag$
\renewcommand{\eqls}{$\mbox{\Large =}$}
\newcommand{\comonoid}{\cmn}
\newcommand{\monoid}{\mnd}
\renewcommand{\comonunit}{\cun}
\renewcommand{\monunit}{\unt}
\renewcommand{\obj}{x^\ddag}
\newcommand{\objj}{x}
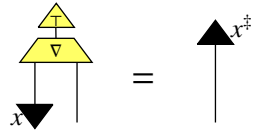
\begin{center}
\def\JPicScale{0.8}
\ifx\JPicScale\undefined\def\JPicScale{1}\fi
\psset{unit=\JPicScale mm}
\psset{linewidth=0.3,dotsep=1,hatchwidth=0.3,hatchsep=1.5,shadowsize=1,dimen=middle}
\psset{dotsize=0.7 2.5,dotscale=1 1,fillcolor=black}
\psset{arrowsize=1 2,arrowlength=1,arrowinset=0.25,tbarsize=0.7 5,bracketlength=0.15,rbracketlength=0.15}
\begin{pspicture}(0,0)(55,26.82)
\psline[linewidth=0.22](25.45,24.54)(25.5,21)
\psline[linewidth=0.22](22,7)(22,17)
\newrgbcolor{userFillColour}{1 1 0.4}
\pspolygon[linewidth=0.22,fillcolor=userFillColour,fillstyle=solid](31.5,17)
(19.5,17)
(22.5,21)
(28.5,21)(31.5,17)
\rput(25.5,19){$\monoid$}
\psline[linewidth=0.22](29,7)(29,17)
\psline[linewidth=0.22](52,21)(52,7)
\rput(40,14){$\eqls$}
\newrgbcolor{userFillColour}{1 1 0.4}
\psline[linewidth=0.22,fillcolor=userFillColour,fillstyle=solid](22.55,22.82)
(25.55,26.82)
(28.55,22.82)(22.55,22.82)
\rput(25.54,24){$\comonunit$}
\rput[l](54.38,23.12){$\obj$}
\psline[linewidth=0.22,fillstyle=solid](49,20)
(52,24)
(55,20)(49,20)
\psline[linewidth=0.22,fillstyle=solid](25,10)
(22,6)
(19,10)(25,10)
\rput[r](20,7.5){$\objj$}
\end{pspicture}

\end{center}

\item $\abx\dashv \adx : \CCC\to \CCC[\tp{x}{X};E]$ is isomorphic with the Kleisli adjunction $V\dashv G:\CCC\to \Kl{\CCC}{X}$ 
%$\CCC[\tp{x}{X};E]$ is isomorphic with the Kleisli category $\Kl{\CCC}{X}$ and the isomorphism commutes with
\end{anumerate}
%\ee{thm}
%
%\begin{prop}\label{intrinsic2} Let $\CCC$ be a dagger monoidal category $(X,\cmn,\cun)$ a comonoid in it and $(X,\mnd,unt)$ the dual monoid, i.e. $\mnd =\cmn^\ddag$, and $\unt = \cun^\ddag$. Then the following conditions are equivalent:
The following conditions provide further equivalent characterizations of (a-c), this time expressed in terms of the properties of the comonoid $(X,\cmn,\cun)$ and its dual monoid $(X,\mnd,unt)$, where $\mnd = \cmn^\ddag$ and $\unt = \cun^\ddag$.

\begin{rnumerate}
%\item $\Kl{\CCC}{X}$ is dagger monoidal, the adjunction $V\dashv G:\CCC\to \Kl{\CCC}{X}$ preserves this structure
\item $\eta = \cmn\circ \unt$ and $\varepsilon = %\eta^\ddag = 
\cun\circ \mnd$ make $X = X^\ast$  self-dual 
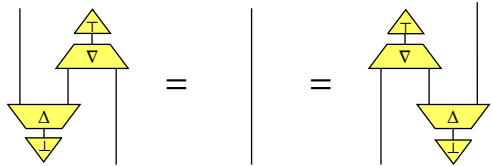
\renewcommand{\obj}{}
\renewcommand{\eqls}{$\mbox{\Large =}$}
\newcommand{\comonoid}{\cmn}
\newcommand{\monoid}{\mnd}
\renewcommand{\comonunit}{\cun}
\renewcommand{\monunit}{\unt}
\begin{center}
\def\JPicScale{0.8}
\ifx\JPicScale\undefined\def\JPicScale{1}\fi
\psset{unit=\JPicScale mm}
\psset{linewidth=0.3,dotsep=1,hatchwidth=0.3,hatchsep=1.5,shadowsize=1,dimen=middle}
\psset{dotsize=0.7 2.5,dotscale=1 1,fillcolor=black}
\psset{arrowsize=1 2,arrowlength=1,arrowinset=0.25,tbarsize=0.7 5,bracketlength=0.15,rbracketlength=0.15}
\begin{pspicture}(0,0)(91.5,28)
\psline[linewidth=0.22](13.5,11)(13.5,27)
\psline[linewidth=0.22](25.45,24.54)(25.5,21)
\psline[linewidth=0.22](21.5,11)(21.5,17)
\newrgbcolor{userFillColour}{1 1 0.4}
\pspolygon[linewidth=0.22,fillcolor=userFillColour,fillstyle=solid](11.5,11)
(23.5,11)
(20.5,7)
(14.5,7)(11.5,11)
\psline[linewidth=0.22](17.5,7)(17.5,3.02)
\newrgbcolor{userFillColour}{1 1 0.4}
\pspolygon[linewidth=0.22,fillcolor=userFillColour,fillstyle=solid](31.5,17)
(19.5,17)
(22.5,21)
(28.5,21)(31.5,17)
\rput(25.5,19){$\monoid$}
\psline[linewidth=0.22](29.5,1)(29.5,17)
\rput(17.5,9){$\comonoid$}
\psline[linewidth=0.22](52,27)(52,1)
\rput(39.5,14){$\eqls$}
\psline[linewidth=0.22](89.51,10.92)(89.5,28)
\psline[linewidth=0.22](77.5,24.71)(77.5,21)
\psline[linewidth=0.22](73.5,1)(73.5,17)
\newrgbcolor{userFillColour}{1 1 0.4}
\pspolygon[linewidth=0.22,fillcolor=userFillColour,fillstyle=solid](79.5,11)
(91.5,11)
(88.5,7)
(82.5,7)(79.5,11)
\psline[linewidth=0.22](85.5,7)(85.45,2.13)
\newrgbcolor{userFillColour}{1 1 0.4}
\pspolygon[linewidth=0.22,fillcolor=userFillColour,fillstyle=solid](83.5,17)
(71.5,17)
(74.5,21)
(80.5,21)(83.5,17)
\rput(77.5,19){$\monoid$}
\psline[linewidth=0.22](81.5,11)(81.5,17)
\rput(85.5,9){$\comonoid$}
\rput(63.5,14){$\eqls$}
\newrgbcolor{userFillColour}{1 1 0.4}
\psline[linewidth=0.22,fillcolor=userFillColour,fillstyle=solid](22.55,22.82)
(25.55,26.82)
(28.55,22.82)(22.55,22.82)
\newrgbcolor{userFillColour}{1 1 0.4}
\psline[linewidth=0.22,fillcolor=userFillColour,fillstyle=solid](74.62,22.77)
(77.62,26.77)
(80.62,22.77)(74.62,22.77)
\newrgbcolor{userFillColour}{1 1 0.4}
\psline[linewidth=0.22,fillcolor=userFillColour,fillstyle=solid](20.45,5.45)
(17.45,1.45)
(14.45,5.45)(20.45,5.45)
\rput(17.59,4.18){$\monunit$}
\newrgbcolor{userFillColour}{1 1 0.4}
\psline[linewidth=0.22,fillcolor=userFillColour,fillstyle=solid](88.55,5.09)
(85.55,1.09)
(82.55,5.09)(88.55,5.09)
\rput(85.54,3.82){$\monunit$}
\rput(77.59,23.82){$\comonunit$}
\rput(25.54,24){$\comonunit$}
\rput[l](52.86,5.61){$\obj$}
\end{pspicture}

\end{center}

\item $(X\otimes \mnd)\circ (\eta\otimes X) = \cmn = (\mnd\otimes X)\circ(X\otimes \eta)$
\begin{center}
\def\JPicScale{0.8}
\ifx\JPicScale\undefined\def\JPicScale{1}\fi
\psset{unit=\JPicScale mm}
\psset{linewidth=0.3,dotsep=1,hatchwidth=0.3,hatchsep=1.5,shadowsize=1,dimen=middle}
\psset{dotsize=0.7 2.5,dotscale=1 1,fillcolor=black}
\psset{arrowsize=1 2,arrowlength=1,arrowinset=0.25,tbarsize=0.7 5,bracketlength=0.15,rbracketlength=0.15}
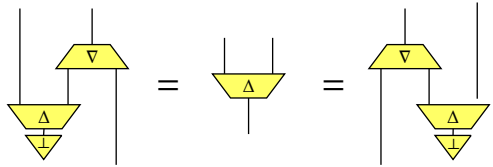
\begin{pspicture}(0,0)(91.5,28)
\rput(58.25,14.12){}
\psline[linewidth=0.22](13.5,11)(13.5,27)
\psline[linewidth=0.22](25.5,27)(25.5,21)
\psline[linewidth=0.22](47.5,16.12)(47.5,22.12)
\newrgbcolor{userFillColour}{1 1 0.4}
\pspolygon[linewidth=0.22,fillcolor=userFillColour,fillstyle=solid](45.5,16.12)
(57.5,16.12)
(54.5,12.12)
(48.5,12.12)(45.5,16.12)
\psline[linewidth=0.22](21.5,11)(21.5,17)
\newrgbcolor{userFillColour}{1 1 0.4}
\pspolygon[linewidth=0.22,fillcolor=userFillColour,fillstyle=solid](11.5,11)
(23.5,11)
(20.5,7)
(14.5,7)(11.5,11)
\psline[linewidth=0.22](17.5,7)(17.32,3.93)
\psline[linewidth=0.22](55.5,16.12)(55.5,22.12)
\newrgbcolor{userFillColour}{1 1 0.4}
\pspolygon[linewidth=0.22,fillcolor=userFillColour,fillstyle=solid](31.5,17)
(19.5,17)
(22.5,21)
(28.5,21)(31.5,17)
\rput(25.5,19){$\monoid$}
\psline[linewidth=0.22](29.5,1)(29.5,17)
\rput(17.5,9){$\comonoid$}
\psline[linewidth=0.22](51.5,12.12)(51.5,6.12)
\rput(51.5,14.12){$\comonoid$}
\rput(38.12,13.75){$\eqls$}
\psline[linewidth=0.22](89.5,12)(89.5,28)
\psline[linewidth=0.22](77.5,27)(77.5,21)
\psline[linewidth=0.22](73.5,1)(73.5,17)
\newrgbcolor{userFillColour}{1 1 0.4}
\pspolygon[linewidth=0.22,fillcolor=userFillColour,fillstyle=solid](79.5,11)
(91.5,11)
(88.5,7)
(82.5,7)(79.5,11)
\psline[linewidth=0.22](85.5,7)(85.45,4.11)
\newrgbcolor{userFillColour}{1 1 0.4}
\pspolygon[linewidth=0.22,fillcolor=userFillColour,fillstyle=solid](83.5,17)
(71.5,17)
(74.5,21)
(80.5,21)(83.5,17)
\rput(77.5,19){$\monoid$}
\psline[linewidth=0.22](81.5,11)(81.5,17)
\rput(85.5,9){$\comonoid$}
\rput(65.62,13.75){$\eqls$}
\newrgbcolor{userFillColour}{1 1 0.4}
\psline[linewidth=0.22,fillcolor=userFillColour,fillstyle=solid](20.45,5.89)
(17.45,1.89)
(14.45,5.89)(20.45,5.89)
\rput(17.45,4.67){$\monunit$}
\newrgbcolor{userFillColour}{1 1 0.4}
\psline[linewidth=0.22,fillcolor=userFillColour,fillstyle=solid](88.48,5.89)
(85.48,1.89)
(82.48,5.89)(88.48,5.89)
\rput(85.57,4.58){$\monunit$}
\end{pspicture}

\end{center}

\item $(X\otimes \mnd)\circ (\cmn\otimes X) = \cmn \circ \mnd = (\mnd\otimes X)\circ(X\otimes \cmn)$
\begin{center}
\def\JPicScale{0.8}
\ifx\JPicScale\undefined\def\JPicScale{1}\fi
\psset{unit=\JPicScale mm}
\psset{linewidth=0.3,dotsep=1,hatchwidth=0.3,hatchsep=1.5,shadowsize=1,dimen=middle}
\psset{dotsize=0.7 2.5,dotscale=1 1,fillcolor=black}
\psset{arrowsize=1 2,arrowlength=1,arrowinset=0.25,tbarsize=0.7 5,bracketlength=0.15,rbracketlength=0.15}
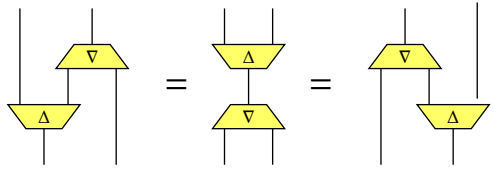
\begin{pspicture}(0,0)(91.5,28)
\rput(58.25,19){}
\psline[linewidth=0.22](13.5,11)(13.5,27)
\psline[linewidth=0.22](25.5,27)(25.5,21)
\psline[linewidth=0.22](47.5,21)(47.5,27)
\newrgbcolor{userFillColour}{1 1 0.4}
\pspolygon[linewidth=0.22,fillcolor=userFillColour,fillstyle=solid](45.5,21)
(57.5,21)
(54.5,17)
(48.5,17)(45.5,21)
\psline[linewidth=0.22](21.5,11)(21.5,17)
\newrgbcolor{userFillColour}{1 1 0.4}
\pspolygon[linewidth=0.22,fillcolor=userFillColour,fillstyle=solid](11.5,11)
(23.5,11)
(20.5,7)
(14.5,7)(11.5,11)
\psline[linewidth=0.22](17.5,7)(17.5,1)
\psline[linewidth=0.22](55.5,21)(55.5,27)
\newrgbcolor{userFillColour}{1 1 0.4}
\pspolygon[linewidth=0.22,fillcolor=userFillColour,fillstyle=solid](31.5,17)
(19.5,17)
(22.5,21)
(28.5,21)(31.5,17)
\rput(25.5,19){$\monoid$}
\psline[linewidth=0.22](29.5,1)(29.5,17)
\rput(17.5,9){$\comonoid$}
\psline[linewidth=0.22](51.5,17)(51.5,11)
\psline[linewidth=0.22](47.5,1)(47.5,7)
\newrgbcolor{userFillColour}{1 1 0.4}
\pspolygon[linewidth=0.22,fillcolor=userFillColour,fillstyle=solid](57.5,7)
(45.5,7)
(48.5,11)
(54.5,11)(57.5,7)
\rput(51.5,9){$\monoid$}
\psline[linewidth=0.22](55.5,1)(55.5,7)
\rput(51.5,19){$\comonoid$}
\rput(39.5,14){$\eqls$}
\psline[linewidth=0.22](89.5,12)(89.5,28)
\psline[linewidth=0.22](77.5,27)(77.5,21)
\psline[linewidth=0.22](73.5,1)(73.5,17)
\newrgbcolor{userFillColour}{1 1 0.4}
\pspolygon[linewidth=0.22,fillcolor=userFillColour,fillstyle=solid](79.5,11)
(91.5,11)
(88.5,7)
(82.5,7)(79.5,11)
\psline[linewidth=0.22](85.5,7)(85.5,1)
\newrgbcolor{userFillColour}{1 1 0.4}
\pspolygon[linewidth=0.22,fillcolor=userFillColour,fillstyle=solid](83.5,17)
(71.5,17)
(74.5,21)
(80.5,21)(83.5,17)
\rput(77.5,19){$\monoid$}
\psline[linewidth=0.22](81.5,11)(81.5,17)
\rput(85.5,9){$\comonoid$}
\rput(63.5,14){$\eqls$}
\end{pspicture}

\end{center}
\end{rnumerate}
%\ee{prop}
\ee{thm}

\paragraph{Remark.} Condition (iii) is the {\em Frobenius condition}, analyzed in \cite{Carboni-Walters,Carboni,KockJ:book,PavlovicD:QMWS}. Condition (ii) is Lawvere's earlier version of the same \cite{Lawvere:Ordinal}. In each of the last three conditions, the commutativity assumption makes one of the equations redundant. The equivalence of (i-iii), however, holds without this commutativity.

\bpr {\bf (a$\Rightarrow$b)} Using the definition \eqref{abdef} of $\abx$, condition (a) implies that
$\nabla\ = \left(\abx. x\right)^\ddag \ = \ \abx. x^\ddag\ = \ (X\otimes \kappa x.x^\ddag) \circ (\Delta\otimes X)$, or graphically
\newcommand{\monoid}{\mnd}
\newcommand{\comonoid}{\cmn}
\begin{center}
\def\JPicScale{0.8}
\ifx\JPicScale\undefined\def\JPicScale{1}\fi
\psset{unit=\JPicScale mm}
\psset{linewidth=0.3,dotsep=1,hatchwidth=0.3,hatchsep=1.5,shadowsize=1,dimen=middle}
\psset{dotsize=0.7 2.5,dotscale=1 1,fillcolor=black}
\psset{arrowsize=1 2,arrowlength=1,arrowinset=0.25,tbarsize=0.7 5,bracketlength=0.15,rbracketlength=0.15}
\begin{pspicture}(0,0)(47.5,17.5)
\psline[linewidth=0.22](2.5,1.88)(2.5,6)
\newrgbcolor{userFillColour}{1 1 0.4}
\pspolygon[linewidth=0.22,fillcolor=userFillColour,fillstyle=solid](12.12,6)
(0.12,6)
(3.12,10)
(9.12,10)(12.12,6)
\rput(6.12,8){$\monoid$}
\psline[linewidth=0.22](10,1.88)(10,6)
\rput(20.62,8){$\eqls$}
\newrgbcolor{userFillColour}{1 1 0.4}
\pspolygon[linewidth=0.22,fillcolor=userFillColour,fillstyle=solid](28.63,10)
(40.62,10)
(37.62,6)
(31.62,6)(28.63,10)
\psline[linewidth=0.22](34.38,3.12)(34.38,6)
\psline[linewidth=0.22](30.62,10)(30.62,17.5)
\psline[linewidth=0.22](38.62,10)(38.62,16)
\newrgbcolor{userFillColour}{1 1 0.4}
\pspolygon[linewidth=0.2,fillcolor=userFillColour,fillstyle=solid](35.62,16.88)(47.5,16.88)(47.5,11.88)(35.62,11.88)
\psline[linewidth=0.2](5.62,10)(5.62,17.5)
\psline[linewidth=0.2](44.38,3.12)(44.38,14.38)
\pscustom[linewidth=0.22,fillstyle=solid]{\psline(42.07,13.12)(44.47,15.62)
%\psbezier(44.47,15.62)(44.47,15.62)(47.07,13.12)(46.88,13.12)
\psline(46.88,13.12)(42.07,13.12)
}
\rput(34.38,8.12){$\comonoid$}
\end{pspicture}

\end{center}
from which (b) follows by precomposing both sides with $(x\otimes X)$ and postcomposing with $\top$.
%, we get
%\begin{multline*}
%x^\ast = \varepsilon \circ (x\otimes X)\ =\ \top \circ \nabla \circ (x\otimes X)\ = \\ (\kappa x.x^\ddag)\circ (\top\otimes X\otimes X)\circ (\Delta \otimes X) \circ (x\otimes X) = (\kappa x.x^\ddag)\circ(x\otimes X)\ = \ x^\ddag
%\end{multline*}

\noindent{\bf (b$\Rightarrow$i)} Dualizing (b) gives $x = x_\ast = x^{\ddag\ast}$, i.e.
\renewcommand{\monoid}{\cmn}
\renewcommand{\comonunit}{\unt}
\renewcommand{\obj}{x}
\newcommand{\objj}{x^\ddag}
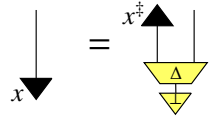
\begin{center}
\def\JPicScale{0.7}
\ifx\JPicScale\undefined\def\JPicScale{1}\fi
\psset{unit=\JPicScale mm}
\psset{linewidth=0.3,dotsep=1,hatchwidth=0.3,hatchsep=1.5,shadowsize=1,dimen=middle}
\psset{dotsize=0.7 2.5,dotscale=1 1,fillcolor=black}
\psset{arrowsize=1 2,arrowlength=1,arrowinset=0.25,tbarsize=0.7 5,bracketlength=0.15,rbracketlength=0.15}
\begin{pspicture}(0,0)(35.5,19)
\psline[linewidth=0.22](29.55,0.46)(29.5,4)
\psline[linewidth=0.22](33,18)(33,8)
\newrgbcolor{userFillColour}{1 1 0.4}
\pspolygon[linewidth=0.22,fillcolor=userFillColour,fillstyle=solid](23.5,8)
(35.5,8)
(32.5,4)
(26.5,4)(23.5,8)
\rput(29.5,6){$\monoid$}
\psline[linewidth=0.22](26,18)(26,8)
\psline[linewidth=0.22](3,4)(3,18)
\rput(15,11){$\eqls$}
\newrgbcolor{userFillColour}{1 1 0.4}
\psline[linewidth=0.22,fillcolor=userFillColour,fillstyle=solid](32.45,2.18)
(29.45,-1.82)
(26.45,2.18)(32.45,2.18)
\rput(29.46,1){$\comonunit$}
\rput[r](0.62,1.88){$\obj$}
\psline[linewidth=0.22,fillstyle=solid](6,5)
(3,1)
(-0,5)(6,5)
\psline[linewidth=0.22,fillstyle=solid](23,15)
(26,19)
(29,15)(23,15)
\rput[r](23.75,17.5){$\objj$}
\end{pspicture}

\end{center}
Combining (b) and its dual gives
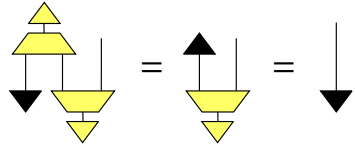
\begin{center}
\def\JPicScale{0.7}
\ifx\JPicScale\undefined\def\JPicScale{1}\fi
\psset{unit=\JPicScale mm}
\psset{linewidth=0.3,dotsep=1,hatchwidth=0.3,hatchsep=1.5,shadowsize=1,dimen=middle}
\psset{dotsize=0.7 2.5,dotscale=1 1,fillcolor=black}
\psset{arrowsize=1 2,arrowlength=1,arrowinset=0.25,tbarsize=0.7 5,bracketlength=0.15,rbracketlength=0.15}
\begin{pspicture}(0,0)(62,26.82)
\psline[linewidth=0.22](3.45,24.54)(3.5,21)
\psline[linewidth=0.22](0,7)(0,17)
\newrgbcolor{userFillColour}{1 1 0.4}
\pspolygon[linewidth=0.22,fillcolor=userFillColour,fillstyle=solid](9.5,17)
(-2.5,17)
(0.5,21)
(6.5,21)(9.5,17)
\psline[linewidth=0.22](7,9)(7,17)
\rput(24,14){$\eqls$}
\newrgbcolor{userFillColour}{1 1 0.4}
\psline[linewidth=0.22,fillcolor=userFillColour,fillstyle=solid](0.55,22.82)
(3.55,26.82)
(6.55,22.82)(0.55,22.82)
\psline[linewidth=0.22,fillstyle=solid](3,10)
(0,6)
(-3,10)(3,10)
\psline[linewidth=0.22](36.55,2.46)(36.5,6)
\psline[linewidth=0.22](40,20)(40,10)
\newrgbcolor{userFillColour}{1 1 0.4}
\pspolygon[linewidth=0.22,fillcolor=userFillColour,fillstyle=solid](30.5,10)
(42.5,10)
(39.5,6)
(33.5,6)(30.5,10)
\psline[linewidth=0.22](33,20)(33,10)
\psline[linewidth=0.22](59,9)(59,23)
\rput(49,14){$\eqls$}
\newrgbcolor{userFillColour}{1 1 0.4}
\psline[linewidth=0.22,fillcolor=userFillColour,fillstyle=solid](39.45,4.18)
(36.45,0.18)
(33.45,4.18)(39.45,4.18)
\psline[linewidth=0.22,fillstyle=solid](62,10)
(59,6)
(56,10)(62,10)
\psline[linewidth=0.22,fillstyle=solid](30,17)
(33,21)
(36,17)(30,17)
\psline[linewidth=0.22](10.92,2.46)(10.87,6)
\psline[linewidth=0.22](14.38,20)(14.37,10)
\newrgbcolor{userFillColour}{1 1 0.4}
\pspolygon[linewidth=0.22,fillcolor=userFillColour,fillstyle=solid](4.88,10)
(16.88,10)
(13.87,6)
(7.88,6)(4.88,10)
\newrgbcolor{userFillColour}{1 1 0.4}
\psline[linewidth=0.22,fillcolor=userFillColour,fillstyle=solid](13.82,4.18)
(10.82,0.18)
(7.82,4.18)(13.82,4.18)
\end{pspicture}

\end{center}
from which (i) follows, because the $\eta$-rule implies that 
$f \circ (x\otimes A) = g\circ (x\otimes A) \Longrightarrow  f=g$ 

%\epr
%
%To prove \ref{abs2}(b)$\Rightarrow$\ref{intrinsic2}(b), we need the following

%\be{lemma}\label{lemmaB}
%If $\CCC[\tp{x}{X}; E]$ supports abstraction, then $x:I\sto X$ is a tensor-stable epimorphism, i.e., for all $f,g\in \CCC(X\otimes A,B)$ holds 
%\bear
%f \circ (x\otimes A) = g\circ (x\otimes A)& \Longrightarrow & f=g
%\eear 
%\ee{lemma}

%\bpr $f =\kappa x.\ f \circ (x\otimes A)= \kappa x.\ g\circ (x\otimes A) =g$ follows from the $\eta$-rule.
%\epr
%
%{\bf (i$\Rightarrow$ii)}
%Lemma \ref{lemmaB} implies that the first equation of (c) follows from
%\newcommand{\comonoid}{\cmn}
%\newcommand{\monoid}{\mnd}
%%\newcommand{\comonunit}{\cun}
%%\newcommand{\monunit}{\unt}
%%\newcommand{\obj}{x}
%\begin{center}
%\def\JPicScale{0.6}
%\input{PIC/Frob2}
%\end{center}
%But this equation obviously follows from (b) and its $\ddag$-dual. 
%\epr

%\bprf{ of \ref{intrinsic2}(b)$\Rightarrow$\ref{abs2}(b)}
%On one hand, by definition, the dual of any vector $\varphi$ is $\varphi^\ast  =  \varepsilon\circ (\varphi\otimes X)$. On the other hand, the (c$\Rightarrow$a)-part of proposition \ref{entangprop} says that \ref{intrinsic2}(b) implies $\varepsilon\circ (x\otimes X)   =  x^\ddag$. Hence $x^\ast = x^\ddag$.
%\epr

\noindent{\bf (i$\Rightarrow$ii)} On one hand, if $X$ is self-dual, then $X\otimes X$ is self-dual too, because
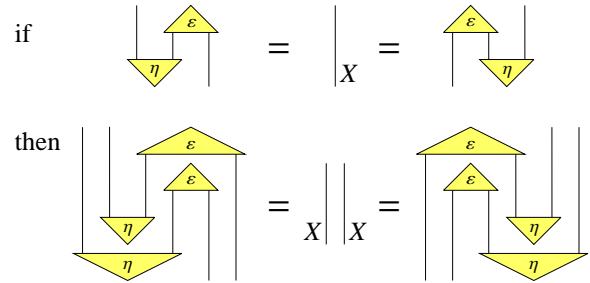
\newcommand{\lablb}{then}
\newcommand{\labla}{if}
\renewcommand{\comonunit}{\scriptstyle \varepsilon}
\renewcommand{\monunit}{\scriptstyle \eta}
\renewcommand{\obj}{X}
\begin{center}
\def\JPicScale{.6}
\ifx\JPicScale\undefined\def\JPicScale{1}\fi
\psset{unit=\JPicScale mm}
\psset{linewidth=0.3,dotsep=1,hatchwidth=0.3,hatchsep=1.5,shadowsize=1,dimen=middle}
\psset{dotsize=0.7 2.5,dotscale=1 1,fillcolor=black}
\psset{arrowsize=1 2,arrowlength=1,arrowinset=0.25,tbarsize=0.7 5,bracketlength=0.15,rbracketlength=0.15}
\begin{pspicture}(0,0)(125,61)
\psline[linewidth=0.22](25,49)(25,61)
\psline[linewidth=0.22](33,49)(33,55)
\psline[linewidth=0.22](41,43)(41,55)
\psline[linewidth=0.22](69,61)(69,43)
\rput(56.5,51){$\eqls$}
\psline[linewidth=0.22](111.01,48.92)(111,61)
\psline[linewidth=0.22](95,43)(95,55)
\psline[linewidth=0.22](103,49)(103,55)
\rput(81,51){$\eqls$}
\newrgbcolor{userFillColour}{1 1 0.4}
\psline[linewidth=0.22,fillcolor=userFillColour,fillstyle=solid](93,55)
(99,61)
(105,55)(93,55)
\rput(99.07,57.29){$\comonunit$}
\rput[l](70,46){$\obj$}
\newrgbcolor{userFillColour}{1 1 0.4}
\psline[linewidth=0.22,fillcolor=userFillColour,fillstyle=solid](31,55)
(37,61)
(43,55)(31,55)
\rput(37.25,57.11){$\comonunit$}
\newrgbcolor{userFillColour}{1 1 0.4}
\psline[linewidth=0.22,fillcolor=userFillColour,fillstyle=solid](113,49)
(107,43)
(101,49)(113,49)
\rput(107.11,46.39){$\monunit$}
\newrgbcolor{userFillColour}{1 1 0.4}
\psline[linewidth=0.22,fillcolor=userFillColour,fillstyle=solid](35,49)
(29,43)
(23,49)(35,49)
\rput(29.04,46.57){$\monunit$}
\rput[l](-2,31){\lablb}
\rput[l](-2,55){\labla}
\psline[linewidth=0.22](67,26)(67,8)
\rput(56.5,16){$\eqls$}
\psline[linewidth=0.22](117.01,13.92)(117,34)
\psline[linewidth=0.22](95,0)(95,20)
\psline[linewidth=0.22](103,6)(103,20)
\rput(81,16){$\eqls$}
\newrgbcolor{userFillColour}{1 1 0.4}
\psline[linewidth=0.22,fillcolor=userFillColour,fillstyle=solid](93,20)
(99,26)
(105,20)(93,20)
\rput(99.07,22.29){$\comonunit$}
\rput[l](62,11){$\obj$}
\newrgbcolor{userFillColour}{1 1 0.4}
\psline[linewidth=0.22,fillcolor=userFillColour,fillstyle=solid](119,14)
(113,8)
(107,14)(119,14)
\rput(113.11,11.39){$\monunit$}
\psline[linewidth=0.22](123,6)(123,34)
\newrgbcolor{userFillColour}{1 1 0.4}
\psline[linewidth=0.22,fillcolor=userFillColour,fillstyle=solid](87,28)
(99,34)
(111,28)(87,28)
\newrgbcolor{userFillColour}{1 1 0.4}
\psline[linewidth=0.22,fillcolor=userFillColour,fillstyle=solid](125,6)
(113,0)
(101,6)(125,6)
\psline[linewidth=0.22](109.01,13.92)(109,28)
\psline[linewidth=0.22](89,0)(89,28)
\rput(99,30){$\comonunit$}
\rput(113,3){$\monunit$}
\psline[linewidth=0.22](19,14)(18.99,34.08)
\psline[linewidth=0.22](33,6)(33,20)
\psline[linewidth=0.22](41,0)(41,20)
\newrgbcolor{userFillColour}{1 1 0.4}
\psline[linewidth=0.22,fillcolor=userFillColour,fillstyle=solid](31,20)
(37,26)
(43,20)(31,20)
\rput(37.07,22.29){$\comonunit$}
\newrgbcolor{userFillColour}{1 1 0.4}
\psline[linewidth=0.22,fillcolor=userFillColour,fillstyle=solid](29,14)
(23,8)
(17,14)(29,14)
\rput(23.11,11.39){$\monunit$}
\psline[linewidth=0.22](13,6)(13,34)
\newrgbcolor{userFillColour}{1 1 0.4}
\psline[linewidth=0.22,fillcolor=userFillColour,fillstyle=solid](25,28)
(37,34)
(49,28)(25,28)
\newrgbcolor{userFillColour}{1 1 0.4}
\psline[linewidth=0.22,fillcolor=userFillColour,fillstyle=solid](35,6)
(23,0)
(11,6)(35,6)
\psline[linewidth=0.22](47,0)(47,28)
\psline[linewidth=0.22](27,14)(27,28)
\rput(37,30){$\comonunit$}
\rput(23,3){$\monunit$}
\psline[linewidth=0.22](71,26)(71,8)
\rput[l](72,11){$\obj$}
\end{pspicture}

\end{center}

On the other hand, (i) also implies that $\mnd^\ddag = \mnd^\ast$, and since $\cmn = \mnd^\ddag$ holds by definition, we have
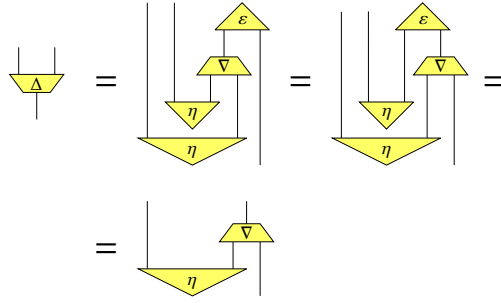
\renewcommand{\comonoid}{\cmn}
\renewcommand{\monoid}{\mnd}
\begin{center}
\def\JPicScale{.6}
\ifx\JPicScale\undefined\def\JPicScale{1}\fi
\psset{unit=\JPicScale mm}
\psset{linewidth=0.3,dotsep=1,hatchwidth=0.3,hatchsep=1.5,shadowsize=1,dimen=middle}
\psset{dotsize=0.7 2.5,dotscale=1 1,fillcolor=black}
\psset{arrowsize=1 2,arrowlength=1,arrowinset=0.25,tbarsize=0.7 5,bracketlength=0.15,rbracketlength=0.15}
\begin{pspicture}(0,0)(108,60)
\rput(65,42){$\eqls$}
\psline[linewidth=0.22](37,38)(37,60)
\psline[linewidth=0.22](51,30)(51,44)
\newrgbcolor{userFillColour}{1 1 0.4}
\psline[linewidth=0.22,fillcolor=userFillColour,fillstyle=solid](47,38)
(41,32)
(35,38)(47,38)
\rput(41.11,35.39){$\monunit$}
\psline[linewidth=0.22](31,30)(31,60)
\newrgbcolor{userFillColour}{1 1 0.4}
\psline[linewidth=0.22,fillcolor=userFillColour,fillstyle=solid](53,30)
(41,24)
(29,30)(53,30)
\psline[linewidth=0.22](45,38)(45,44)
\rput(41,27){$\monunit$}
\rput(13.25,42.12){}
\psline[linewidth=0.22](2.5,44.12)(2.5,50.12)
\newrgbcolor{userFillColour}{1 1 0.4}
\pspolygon[linewidth=0.22,fillcolor=userFillColour,fillstyle=solid](0.5,44.12)
(12.5,44.12)
(9.5,40.12)
(3.5,40.12)(0.5,44.12)
\psline[linewidth=0.22](10.5,44.12)(10.5,50.12)
\psline[linewidth=0.22](6.5,40.12)(6.5,34.12)
\rput(6.5,42.12){$\comonoid$}
\rput(22,42){$\eqls$}
\psline[linewidth=0.22](48,54)(48,48)
\newrgbcolor{userFillColour}{1 1 0.4}
\pspolygon[linewidth=0.22,fillcolor=userFillColour,fillstyle=solid](54,44)
(42,44)
(45,48)
(51,48)(54,44)
\rput(48,46){$\monoid$}
\psline[linewidth=0.22](56,24)(56,54)
\newrgbcolor{userFillColour}{1 1 0.4}
\psline[linewidth=0.22,fillcolor=userFillColour,fillstyle=solid](46,54)
(52,60)
(58,54)(46,54)
\rput(52.25,56.11){$\comonunit$}
\psline[linewidth=0.22](80,38)(79.99,58.08)
\psline[linewidth=0.22](99,24)(99,44)
\newrgbcolor{userFillColour}{1 1 0.4}
\psline[linewidth=0.22,fillcolor=userFillColour,fillstyle=solid](90,38)
(84,32)
(78,38)(90,38)
\rput(84.11,35.39){$\monunit$}
\psline[linewidth=0.22](74,30)(74,58)
\newrgbcolor{userFillColour}{1 1 0.4}
\psline[linewidth=0.22,fillcolor=userFillColour,fillstyle=solid](96,30)
(84,24)
(72,30)(96,30)
\psline[linewidth=0.22](93,30)(93,44)
\rput(84,27){$\monunit$}
\psline[linewidth=0.22](88,54)(88,38)
\newrgbcolor{userFillColour}{1 1 0.4}
\pspolygon[linewidth=0.22,fillcolor=userFillColour,fillstyle=solid](102,44)
(90,44)
(93,48)
(99,48)(102,44)
\rput(96,46){$\monoid$}
\psline[linewidth=0.22](96,48)(96,54)
\newrgbcolor{userFillColour}{1 1 0.4}
\psline[linewidth=0.22,fillcolor=userFillColour,fillstyle=solid](86,54)
(92,60)
(98,54)(86,54)
\rput(92.25,56.11){$\comonunit$}
\rput(108,42){$\eqls$}
\rput(22,5){$\eqls$}
\psline[linewidth=0.22](56,-5)(56,7)
\psline[linewidth=0.22](31,1)(31,16)
\newrgbcolor{userFillColour}{1 1 0.4}
\psline[linewidth=0.22,fillcolor=userFillColour,fillstyle=solid](53,1)
(41,-5)
(29,1)(53,1)
\psline[linewidth=0.22](50,1)(50,7)
\rput(41,-2){$\monunit$}
\newrgbcolor{userFillColour}{1 1 0.4}
\pspolygon[linewidth=0.22,fillcolor=userFillColour,fillstyle=solid](59,7)
(47,7)
(50,11)
(56,11)(59,7)
\rput(53,9){$\monoid$}
\psline[linewidth=0.22](53,11)(53,16)
\end{pspicture}

\end{center}

\noindent{\bf (ii$\Rightarrow$iii)} Using (ii) to expand $\cmn$ at the first step, and to collapse it at the last step, we get
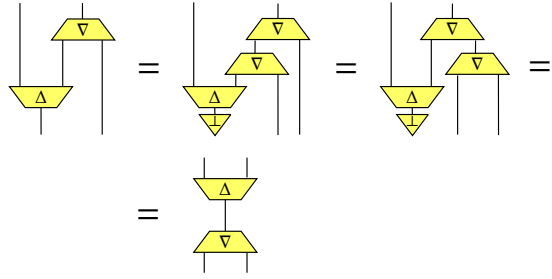
\renewcommand{\comonoid}{\cmn}
\renewcommand{\monoid}{\mnd}
\renewcommand{\comonunit}{\cun}
\renewcommand{\monunit}{\unt}
%\vspace{2\baselineskip}
%\begin{figure}[htbp]
\begin{center}
\def\JPicScale{0.7}
\ifx\JPicScale\undefined\def\JPicScale{1}\fi
\psset{unit=\JPicScale mm}
\psset{linewidth=0.3,dotsep=1,hatchwidth=0.3,hatchsep=1.5,shadowsize=1,dimen=middle}
\psset{dotsize=0.7 2.5,dotscale=1 1,fillcolor=black}
\psset{arrowsize=1 2,arrowlength=1,arrowinset=0.25,tbarsize=0.7 5,bracketlength=0.15,rbracketlength=0.15}
\begin{pspicture}(0,0)(100.62,51.38)
\rput(47.62,15.88){}
\psline[linewidth=0.22](1.88,35.38)(1.88,51.25)
\psline[linewidth=0.22](13.75,51.25)(13.75,48.12)
\psline[linewidth=0.22](36.88,17.88)(36.88,21.88)
\newrgbcolor{userFillColour}{1 1 0.4}
\pspolygon[linewidth=0.22,fillcolor=userFillColour,fillstyle=solid](34.88,17.88)
(46.88,17.88)
(43.88,13.88)
(37.88,13.88)(34.88,17.88)
\psline[linewidth=0.22](10,35)(10,44.38)
\newrgbcolor{userFillColour}{1 1 0.4}
\pspolygon[linewidth=0.22,fillcolor=userFillColour,fillstyle=solid](-0.12,35.38)
(11.88,35.38)
(8.88,31.38)
(2.88,31.38)(-0.12,35.38)
\psline[linewidth=0.22](5.88,31.38)(5.89,26.25)
\psline[linewidth=0.22](45,18.12)(45,21.88)
\newrgbcolor{userFillColour}{1 1 0.4}
\pspolygon[linewidth=0.22,fillcolor=userFillColour,fillstyle=solid](19.88,44.23)
(7.88,44.23)
(10.88,48.23)
(16.88,48.23)(19.88,44.23)
\rput(13.88,46.23){$\monoid$}
\psline[linewidth=0.22](17.5,26.25)(17.5,44.2)
\rput(5.88,33.38){$\comonoid$}
\psline[linewidth=0.22](40.88,13.88)(40.88,7.88)
\psline[linewidth=0.22](36.88,0)(36.88,3.88)
\newrgbcolor{userFillColour}{1 1 0.4}
\pspolygon[linewidth=0.22,fillcolor=userFillColour,fillstyle=solid](46.88,3.88)
(34.88,3.88)
(37.88,7.88)
(43.88,7.88)(46.88,3.88)
\rput(40.88,5.88){$\monoid$}
\psline[linewidth=0.22](45,0)(45,3.75)
\rput(40.88,15.88){$\comonoid$}
\rput(26.38,38.38){$\eqls$}
\rput(26.25,10.62){$\eqls$}
\psline[linewidth=0.22](34.88,35.38)(34.88,51.38)
\psline[linewidth=0.22](42.86,35.27)(42.88,37.71)
\newrgbcolor{userFillColour}{1 1 0.4}
\pspolygon[linewidth=0.22,fillcolor=userFillColour,fillstyle=solid](32.88,35.38)
(44.88,35.38)
(41.88,31.38)
(35.88,31.38)(32.88,35.38)
\psline[linewidth=0.22](38.88,31.38)(38.93,28.21)
\newrgbcolor{userFillColour}{1 1 0.4}
\pspolygon[linewidth=0.22,fillcolor=userFillColour,fillstyle=solid](52.88,37.71)
(40.88,37.71)
(43.88,41.71)
(49.88,41.71)(52.88,37.71)
\rput(46.88,39.71){$\monoid$}
\psline[linewidth=0.22](50.89,26.16)(50.88,37.71)
\rput(38.88,33.38){$\comonoid$}
\psline[linewidth=0.22](50.89,51.16)(50.88,48.32)
\newrgbcolor{userFillColour}{1 1 0.4}
\pspolygon[linewidth=0.22,fillcolor=userFillColour,fillstyle=solid](56.88,44.32)
(44.88,44.32)
(47.88,48.32)
(53.88,48.32)(56.88,44.32)
\rput(50.88,46.32){$\monoid$}
\psline[linewidth=0.22](55,26.25)(55,44.38)
\newrgbcolor{userFillColour}{1 1 0.4}
\psline[linewidth=0.22,fillcolor=userFillColour,fillstyle=solid](41.93,30.07)
(38.93,26.07)
(35.93,30.07)(41.93,30.07)
\rput(38.93,28.75){$\monunit$}
\psline[linewidth=0.22](46.52,41.7)(46.54,44.14)
\rput(63.88,38.38){$\eqls$}
\psline[linewidth=0.22](72.38,35.38)(72.38,51.38)
\psline[linewidth=0.22](80,35)(80,44.38)
\newrgbcolor{userFillColour}{1 1 0.4}
\pspolygon[linewidth=0.22,fillcolor=userFillColour,fillstyle=solid](70.38,35.38)
(82.38,35.38)
(79.38,31.38)
(73.38,31.38)(70.38,35.38)
\psline[linewidth=0.22](76.38,31.38)(76.43,28.21)
\newrgbcolor{userFillColour}{1 1 0.4}
\pspolygon[linewidth=0.22,fillcolor=userFillColour,fillstyle=solid](94.75,37.71)
(82.75,37.71)
(85.75,41.71)
(91.75,41.71)(94.75,37.71)
\rput(88.75,39.71){$\monoid$}
\psline[linewidth=0.22](92.77,26.16)(92.75,37.71)
\rput(76.38,33.38){$\comonoid$}
\psline[linewidth=0.22](84.02,51.16)(84,48.32)
\newrgbcolor{userFillColour}{1 1 0.4}
\pspolygon[linewidth=0.22,fillcolor=userFillColour,fillstyle=solid](90,44.32)
(78,44.32)
(81,48.32)
(87,48.32)(90,44.32)
\rput(84,46.32){$\monoid$}
\psline[linewidth=0.22](85,26.25)(85,37.5)
\newrgbcolor{userFillColour}{1 1 0.4}
\psline[linewidth=0.22,fillcolor=userFillColour,fillstyle=solid](79.43,30.07)
(76.43,26.07)
(73.43,30.07)(79.43,30.07)
\rput(76.43,28.75){$\monunit$}
\psline[linewidth=0.22](88.39,41.7)(88.41,44.14)
\rput(100.62,38.75){$\eqls$}
\end{pspicture}

%\caption{default}
%\label{default}
\end{center}
%\end{figure}

\noindent{\bf (iii$\Rightarrow$i)} follows in a way obvious from the diagrams, by precomposing the first equation of (iii) with $\unt\otimes X$ and postcomposing it with $X\otimes \cun$; and by precomposing the second equation with $X\otimes \unt$ and postcomposing it with $\cun\otimes X$.

\noindent{\bf (i$\Rightarrow$c)} Using the self-duality of $X$, the dagger on $\Kl{\CCC}{X}$ is defined by
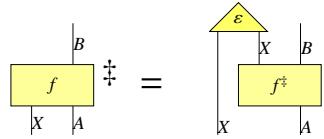
\newcommand{\aaa}{\scriptstyle A}
\newcommand{\bbb}{\scriptstyle B}
\newcommand{\xxx}{\scriptstyle X}
\newcommand{\funct}{\scriptstyle f}
\newcommand{\gadder}{\mbox{\large $\ddag$}}
\newcommand{\fundag}{\scriptstyle f^\ddag}
\newcommand{\dwn}{\scriptstyle \varepsilon}
\begin{center}
\def\JPicScale{0.55}
\ifx\JPicScale\undefined\def\JPicScale{1}\fi
\psset{unit=\JPicScale mm}
\psset{linewidth=0.3,dotsep=1,hatchwidth=0.3,hatchsep=1.5,shadowsize=1,dimen=middle}
\psset{dotsize=0.7 2.5,dotscale=1 1,fillcolor=black}
\psset{arrowsize=1 2,arrowlength=1,arrowinset=0.25,tbarsize=0.7 5,bracketlength=0.15,rbracketlength=0.15}
\begin{pspicture}(0,0)(75,35)
\rput(65.75,-6){}
\psline[linewidth=0.15](70,20)(70,30)
\psline[linewidth=0.15](60,29)(60,20)
\psline[linewidth=0.15](70,10)(70,3)
\newrgbcolor{userFillColour}{1 1 0.6}
\pspolygon[linewidth=0.1,fillcolor=userFillColour,fillstyle=solid](55,20)(75,20)(75,10)(55,10)
\psline[linewidth=0.15](15,20)(15,30)
\psline[linewidth=0.15](5,10)(5,3)
\psline[linewidth=0.15](15,10)(15,3)
\newrgbcolor{userFillColour}{1 1 0.6}
\pspolygon[linewidth=0.1,fillcolor=userFillColour,fillstyle=solid](0,20)(20,20)(20,10)(0,10)
\rput(7.5,15){}
\rput(34,15){$\eqls$}
\rput(10,15){$\funct$}
\rput[l](22,18){$\gadder$}
\newrgbcolor{userFillColour}{1 1 0.4}
\psline[linewidth=0.22,fillcolor=userFillColour,fillstyle=solid](48,28)
(55,35)
(62,28)(48,28)
\psline[linewidth=0.15](50,28)(50,3)
\rput(55,31){$\dwn$}
\rput(65,15){$\fundag$}
\rput[l](15,6){$\aaa$}
\rput[l](15,25){$\bbb$}
\rput[l](5,6){$\xxx$}
\rput[l](50,5){$\xxx$}
\rput[l](60,24){$\xxx$}
\rput[l](70,5){$\aaa$}
\rput[l](70,24){$\bbb$}
\end{pspicture}

\end{center}
Since this implies $\kappa x.\ \varphi(x)^\ddag = \left(\kappa x.\ \varphi(x)\right)^\ddag$, it follows that the isomorphism $\CCC[\tp{x}{X}]\cong \Kl{\CCC}{X}$, defined in the proof of Thm.~\ref{abs1}, preserves the dagger.

\noindent{\bf (c$\Rightarrow$a)} Since the dagger preservation under the isomorphism $\CCC[\tp{x}{X}]\cong \Kl{\CCC}{X}$ means that the dagger in $\Kl\CCC X$ must be as above, it follows

By \eqref{abdef}, the left-hand side is $\abx.\ \varphi(x)^\ddag$, whereas the right-hand side is $\left(\abx.\ \varphi(x)\right)^\ddag$. Hence (a).
\epr

\be{defn}
A {\em Frobenius algebra\/} in a monoidal category $\CCC$ is a structure $(X,\mnd,\cmn,\unt,\cun)$ such that 
\begin{itemize}
\item $(X,\mnd,\unt)$ is a monoid,
\item $(X,\cmn,\cun)$ is a comonoid, and
\item the equivalent conditions (i-iii) of Thm.~\ref{abs2} are satisfied.
\end{itemize}
A {\em dagger-Frobenius algebra\/} in a dagger-monoidal category $\CCC$ is a Frobenius algebra where $\mnd = \cmn^\ddag$ and $\unt = \cun^\ddag$.
\ee{defn}

Thm.~\ref{abs2} can now be summarized as follows.

\begin{corollary}
The category of dagger-monoidal abstractions $\dAbs{\CCC}$ is equivalent with the category $\CCC_{\cmn}$ of commutative dagger-Frobenius algebras and comonoid homomorphisms in $\CCC$.
\end{corollary}

\paragraph{Summary.} The upshot of Thm.~\ref{abs2} is thus that a monoidal extension $\CCC[\tp x X]$, induced by a commutative comonoid $X$ which also happens to be a dagger-Frobenius algebra, is necessarily a dagger-monoidal extension. The immediate corollary is the following.

\begin{corollary} \label{dagger-substitution-corollary} 
The substitutions $\CCC[\tp x X]\to \CCC$ of the basis vectors with respect to a Frobenius algebra $X$ preserve not only the tensors and their unit, but also the daggers.
\end{corollary}

Furthermore, since the basis vectors of the Frobenius algebra $X$ are substituted for the variable $x$, which must be real, it is natural to expect, and easy to prove that

\begin{corollary}\label{basis-real} The basis vectors with respect to a dagger-Frobenius algebra are always real.
\end{corollary}

\paragraph{Remark.} This last statement may sound curious. There are many complex vectors in a complex Hilbert space, and each of them may participate some basis. However, after a change of basis they may become real; and some vectors that were real will cease to be real. The notion of reality depends on the choice of basis. However, just like people, the basis vectors themselves always satisfy their own notion of reality: they are in the form $\beta_1 = (1,0,0,\ldots, 0), \beta_2 = (0,1,0,\ldots,0),\ldots, \beta_n = (0,0,\ldots,0,1)$.

\subsection{Classical structures}
It turns out that Frobenius algebras with additional properties provide a purely algebraic characterization of the choice of a basis, e.g. in a Hilbert space. More generally, in an abstract quantum universe, we can thus distinguish classical data types, by means of algebraic operations. We begin by describing the additional property needed for this. 

\be{lemma}\label{special}
Let $\CCC[\tp{x,y}{X}]$ be a dagger-monoidal extension induced by the Frobenius algebra $(X,\mnd,\cmn,\unt,\cun)$. Then the following conditions are equivalent:
\begin{anumerate}
\item $\mnd \circ \cmn = \id_X$ 
\item $\mnd(x\otimes x) = x$
\item $<x|y>^2 = <x|y>$
\end{anumerate}
and they imply
\begin{anumerate}
\setcounter{countalpha}{3}
\item $<x|x> = \id_I$
\end{anumerate}
The equivalence of (a) and (b) is also valid for monoidal categories, with no dagger.
\ee{lemma}

\bpr {\bf (a$\Rightarrow$b)} $\mnd(x\otimes x) = \mnd\cmn x = x$, using Cor.~\ref{diagx}.

\noindent {\bf (b$\Rightarrow$c)} $<x|y> = x^\ddag \circ y = x^\ddag \circ \mnd \circ (y\otimes y) = x^\ddag \circ \cmn^\ddag \circ (y\otimes y) = (x^\ddag \otimes x^\ddag)(y\otimes y) = (x^\ddag\circ y)\otimes (x^\ddag \circ y) = <x|y>^2$, i.e.
\newcommand{\monoid}{\mnd}
\newcommand{\xxx}{\scriptstyle y}
\newcommand{\yyy}{\scriptstyle x^\ddag}
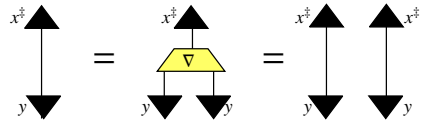
\begin{center}
\def\JPicScale{.75}
\ifx\JPicScale\undefined\def\JPicScale{1}\fi
\psset{unit=\JPicScale mm}
\psset{linewidth=0.3,dotsep=1,hatchwidth=0.3,hatchsep=1.5,shadowsize=1,dimen=middle}
\psset{dotsize=0.7 2.5,dotscale=1 1,fillcolor=black}
\psset{arrowsize=1 2,arrowlength=1,arrowinset=0.25,tbarsize=0.7 5,bracketlength=0.15,rbracketlength=0.15}
\begin{pspicture}(0,0)(71.25,21.25)
\psline[linewidth=0.22](7,21)(7,1)
\rput(18.13,11.25){$\eqls$}
\psline[linewidth=0.22](33.75,18.75)(33.75,12.5)
\psline[linewidth=0.22](28.75,4.38)(28.75,10)
\newrgbcolor{userFillColour}{1 1 0.4}
\pspolygon[linewidth=0.22,fillcolor=userFillColour,fillstyle=solid](39.5,9.38)
(27.5,9.38)
(30.5,13.38)
(36.5,13.38)(39.5,9.38)
\rput(33.12,11.25){$\monoid$}
\psline[linewidth=0.22](37.5,5)(37.5,9.38)
\psline[linewidth=0.22,fillstyle=solid](31.88,5)
(28.88,1)
(25.88,5)(31.88,5)
\psline[linewidth=0.22,fillstyle=solid](10.38,5)
(7.38,1)
(4.38,5)(10.38,5)
\psline[linewidth=0.22,fillstyle=solid](4,17)
(7,21)
(10,17)(4,17)
\psline[linewidth=0.22,fillstyle=solid](40.5,5)
(37.5,1)
(34.5,5)(40.5,5)
\psline[linewidth=0.22](57.5,18.75)(57.5,1.75)
\psline[linewidth=0.22,fillstyle=solid](60.38,5)
(57.38,1)
(54.38,5)(60.38,5)
\rput(48.13,11.25){$\eqls$}
\psline[linewidth=0.22,fillstyle=solid](30.5,17)
(33.5,21)
(36.5,17)(30.5,17)
\psline[linewidth=0.22,fillstyle=solid](54.5,17.25)
(57.5,21.25)
(60.5,17.25)(54.5,17.25)
\psline[linewidth=0.22](68.12,18.75)(68.12,1.75)
\psline[linewidth=0.22,fillstyle=solid](71.01,5)
(68.01,1)
(65.01,5)(71.01,5)
\psline[linewidth=0.22,fillstyle=solid](65.12,17.25)
(68.12,21.25)
(71.12,17.25)(65.12,17.25)
\rput[r](4.38,2.5){$\xxx$}
\rput[r](4.38,19.38){$\yyy$}
\rput[r](26.25,2.5){$\xxx$}
\rput[r](31.25,19.38){$\yyy$}
\rput[r](55,2.5){$\xxx$}
\rput[r](55,19.38){$\yyy$}
\rput[l](71.25,2.5){$\xxx$}
\rput[l](71.25,19.38){$\yyy$}
\rput[l](39.38,2.5){$\xxx$}
\end{pspicture}

\end{center}

\noindent {\bf (c$\Rightarrow$a)} $x^\ddag \circ \mnd\circ \cmn \circ y = (x^\ddag \otimes x^\ddag)(y\otimes y) = (x^\ddag\circ y)\otimes (x^\ddag \circ y) = <x|y>^2 = <x|y> = x^\ddag \circ y$, and then use the $\eta$-rule.

\noindent {\bf (b$\Rightarrow$d)} Since by Thm.~\ref{abs2} $x^\ddag = x^\ast$, and by Cor.~\ref{diagx} $\cun x = \id_I$, we have$<x|x> = x^\ddag x = \cun x = \id_I$.
\renewcommand{\comonunit}{\cun}
\renewcommand{\xxx}{\scriptstyle x}
\renewcommand{\yyy}{\scriptstyle x^\ddag}
\begin{center}
\def\JPicScale{.75}
\ifx\JPicScale\undefined\def\JPicScale{1}\fi
\psset{unit=\JPicScale mm}
\psset{linewidth=0.3,dotsep=1,hatchwidth=0.3,hatchsep=1.5,shadowsize=1,dimen=middle}
\psset{dotsize=0.7 2.5,dotscale=1 1,fillcolor=black}
\psset{arrowsize=1 2,arrowlength=1,arrowinset=0.25,tbarsize=0.7 5,bracketlength=0.15,rbracketlength=0.15}
\begin{pspicture}(0,0)(61.01,21.25)
\psline[linewidth=0.22](7,21)(7,1)
\rput(18.76,11.25){$\eqls$}
\psline[linewidth=0.22](34.12,19)(34.12,13)
\psline[linewidth=0.22](29.62,1)(29.38,11.88)
\newrgbcolor{userFillColour}{1 1 0.4}
\pspolygon[linewidth=0.22,fillcolor=userFillColour,fillstyle=solid](40.12,9.88)
(28.12,9.88)
(31.12,13.88)
(37.12,13.88)(40.12,9.88)
\rput(33.75,11.75){$\monoid$}
\psline[linewidth=0.22](38.12,5)(38.12,10)
\newrgbcolor{userFillColour}{1 1 0.4}
\psline[linewidth=0.22,fillcolor=userFillColour,fillstyle=solid](31.24,16.88)
(34.25,20.88)
(37.24,16.88)(31.24,16.88)
\rput(34.38,18.12){$\comonunit$}
\psline[linewidth=0.22,fillstyle=solid](32.5,5)
(29.5,1)
(26.5,5)(32.5,5)
\psline[linewidth=0.22,fillstyle=solid](10.38,5)
(7.38,1)
(4.38,5)(10.38,5)
\psline[linewidth=0.22,fillstyle=solid](4,17)
(7,21)
(10,17)(4,17)
\psline[linewidth=0.22,fillstyle=solid](41.12,5)
(38.12,1)
(35.12,5)(41.12,5)
\psline[linewidth=0.22](57.63,21)(57.63,1)
\psline[linewidth=0.22,fillstyle=solid](61.01,5)
(58.01,1)
(55.01,5)(61.01,5)
\rput(48.76,11.25){$\eqls$}
\newrgbcolor{userFillColour}{1 1 0.4}
\psline[linewidth=0.22,fillcolor=userFillColour,fillstyle=solid](54.5,17.25)
(57.5,21.25)
(60.5,17.25)(54.5,17.25)
\rput[r](5,3){$\xxx$}
\rput[r](27.12,3){$\xxx$}
\rput[r](55.25,3){$\xxx$}
\rput[l](40.12,3){$\xxx$}
\rput[r](4.38,19.38){$\yyy$}
\rput(57.5,18.75){$\comonunit$}
\end{pspicture}

\end{center}
\epr

\be{defn}
A {\em classical structure\/} is a commutative dagger-Frobenius algebra satifsying \ref{special}(a). A {\em classical extension\/} of $\CCC$ is a dagger-monoidal extension $\CCC[\tp {x} X]$ induced by a classical structure, i.e. satisfying \ref{special}(b-c).
\ee{defn}

\paragraph{Remark.} Lemma~\ref{special}(b) and Thm.~\ref{abs2} together say that a monoidal extension $\CCC[\tp x X]$ of a dagger monoidal category $\CCC$ is a classical extension if and only the variable $x$ is real and idempotent, i.e. $x=x_\ast = x\starr x$, where $a\starr b = \mnd(a\otimes b) $ is the convolution, mentioned in \ref{Convolution-Representation}. Lemma~\ref{special}(c) says that the idempotence of $x$ is equivalent with the idempotence of the inner product $<x|y>$ of any two variables of type $X$. (Idempotence with respect to which monoid? Recall from Sec.~\ref{Convolution-Representation} that the convolution, the composition, and the tensor of scalars all induce the same monoid, since $s\starr t = s\circ t = s\otimes t$ holds for all $s,t\in \CCC(I)$.)

Note that, by the $\eta$-rule, $<x|y> = <x|z>\Rightarrow y=z$. It follows that the monoid of scalars in a polynomial extension $\CCC[\tp{x,y,z} X]$ must have freshly adjoined elements, if $x\neq y \neq z$. Another interesting point is that the implication $<x|a> = <x|b>\Rightarrow a = b$, valid in $\CCC[\tp{x} X]$, is preserved under the substitutions {\em jointly}, provided that the basis vectors generate $X$: if $<\beta |a> = <\beta |b>$ holds for all basis vectors $\beta$, then $a=b$. Elaborating this, one could formulate the suitable soundness and completeness notions and for reasoning with polynomials and classical structures, but we shall not pursue this thread.

\begin{corollary}
The category $\bsAbs{\CCC}\subseteq \dAbs{\CCC}$ of classical abstractions of $\CCC$ is equivalent with the category $\CCC_{\bs}$ of classical structures and comonoid homomorphisms in $\CCC$.
\end{corollary}

%\paragraph{Remark.} 
Note that the category $\CCC_\bs$ is a cartesian subcategory of the category $\CCC_\times$ of commutative comonoids. While the forgetful functor $\CCC_{\times} \to \CCC$ was couniversal for all monoidal functors from cartesian categories  to $\CCC$, the forgetful functor $\CCC_{\bs} \to \CCC$  is couniversal for the  conservative functors among them. The exactness properties of $\CCC_{\bs}$, induced by the various properties of $\CCC$,  were analyzed in \cite{Carboni}. If $\CCC$ is compact \cite{Kelly-Laplaza} and right exact with biproducts, then $\CCC_{\bs}$ turns out to be a pretopos. In any case, if $\CCC$ represents a quantum universe, $\CCC_\bs$ can be thought of as the category of classical data.

\subsubsection{Orthonormality of bases}
Definition \ref{basis} stipulated an abstract notion of a basis with respect to a comonoid. The notion of a classical structure now characterizes just those comonoids whose bases are {\em orthonormal}, in the sense of the following

\be{defn}
A vector $a\in \CCC(A)$ is {\em normalized\/} if $<a | b> = \id_I$. A pair of vectors $a,b \in \CCC(A)$ is {\em orthogonal\/} if $<a|b>^2 = <a |b>$. A set of vectors is {\em orthonormal\/} when each element is normalized, and each pair orthogonal.
\ee{defn}

Lemma~\ref{special} and Cor.~\ref{substitution-corollary} imply that

\be{prop}
The basis set of every classical structure is orthonormal.
\ee{prop}

\subsubsection{Succinct classical structures}
The following lemma shows that being a classical structure is a property of a comonoid (or of a monoid), rather than additional structure.
\be{lemma}\label{uniq}
The monoid and the comonoid part of a classical structure determine each other: e.g., $(X,\mnd,\cmn_1,\unt,\cun_1)$ and $(X,\mnd,\cmn_2,\unt,\cun_2)$ are classical structures, then $\cmn_1 = \cmn_2$ and $\cun_1=\cun_2$.
\ee{lemma}
Since $(X,\mnd,\cmn,\unt,\cun)$ is completely determined by $(X,\mnd,\unt)$ (and by $(X,\cmn,\cun)$), it is justified to speak succinctly of the classical structure $(X,\mnd,\unt)$ (and of the classical structure $(X,\cmn,\cun)$).

\bpr It is enough to prove $\cmn_1 \circ \mnd = \cmn_2 \circ \mnd$, because this and $\mnd \circ \cmn_1 = \id_X$ give
\[
\cmn_1 \ = \ \cmn_1\circ \mnd \circ \cmn_1 \ = \ \cmn_2\circ \mnd \circ \cmn_1 \ = \ \cmn_2
\]
Here is a diagrammatic proof $\cmn_1 \circ \mnd = \cmn_2 \circ \mnd$:
\newcommand{\comonoid}{\scriptscriptstyle 2}
\newcommand{\monoid}{\scriptscriptstyle 1}
\begin{center}
\def\JPicScale{.65}
\ifx\JPicScale\undefined\def\JPicScale{1}\fi
\psset{unit=\JPicScale mm}
\psset{linewidth=0.3,dotsep=1,hatchwidth=0.3,hatchsep=1.5,shadowsize=1,dimen=middle}
\psset{dotsize=0.7 2.5,dotscale=1 1,fillcolor=black}
\psset{arrowsize=1 2,arrowlength=1,arrowinset=0.25,tbarsize=0.7 5,bracketlength=0.15,rbracketlength=0.15}
\begin{pspicture}(0,0)(157.74,75.13)
\rput(13.62,58.13){}
\psline[linewidth=0.22](106.5,10.63)(106.5,26.5)
\psline[linewidth=0.22](118.38,26.5)(118.38,23.37)
\psline[linewidth=0.22](2.88,65.13)(2.88,69.13)
\newrgbcolor{userFillColour}{1 1 0.4}
\pspolygon[linewidth=0.22,fillcolor=userFillColour,fillstyle=solid](0.88,65.13)
(12.88,65.13)
(9.88,61.13)
(3.88,61.13)(0.88,65.13)
\psline[linewidth=0.22](114.62,10.25)(114.62,19.63)
\newrgbcolor{userFillColour}{1 1 0.4}
\pspolygon[linewidth=0.22,fillcolor=userFillColour,fillstyle=solid](104.5,10.63)
(116.5,10.63)
(113.5,6.63)
(107.5,6.63)(104.5,10.63)
\psline[linewidth=0.22](110.5,6.63)(110.51,1.5)
\psline[linewidth=0.22](11,65.37)(11,69.13)
\newrgbcolor{userFillColour}{1 1 0.4}
\pspolygon[linewidth=0.22,fillcolor=userFillColour,fillstyle=solid](124.5,19.48)
(112.5,19.48)
(115.5,23.48)
(121.5,23.48)(124.5,19.48)
\psline[linewidth=0.22](122.12,1.5)(122.12,19.45)
\rput(110.5,8.63){$\comonoid$}
\psline[linewidth=0.22](6.87,61.12)(6.88,53.75)
\psline[linewidth=0.22](2.88,47.88)(2.88,51.75)
\newrgbcolor{userFillColour}{1 1 0.4}
\pspolygon[linewidth=0.22,fillcolor=userFillColour,fillstyle=solid](12.88,51.75)
(0.88,51.75)
(3.88,55.75)
(9.88,55.75)(12.88,51.75)
\psline[linewidth=0.22](11,47.88)(11,51.62)
\rput(21.25,59.38){$\eqls$}
\rput(56,16.25){$\eqls$}
\psline[linewidth=0.22](27.88,13.75)(27.88,29.75)
\psline[linewidth=0.22](34.11,13.19)(34.12,16.88)
\newrgbcolor{userFillColour}{1 1 0.4}
\pspolygon[linewidth=0.22,fillcolor=userFillColour,fillstyle=solid](25.12,14)
(37.12,14)
(34.12,10)
(28.12,10)(25.12,14)
\psline[linewidth=0.22](31,10)(31,5.62)
\newrgbcolor{userFillColour}{1 1 0.4}
\pspolygon[linewidth=0.22,fillcolor=userFillColour,fillstyle=solid](43.25,16.46)
(31.25,16.46)
(34.25,20.46)
(40.25,20.46)(43.25,16.46)
\rput(31,12.5){$\monoid$}
\psline[linewidth=0.22](40.38,6.88)(40.37,16.55)
\psline[linewidth=0.22](41.26,29.91)(41.25,27.07)
\newrgbcolor{userFillColour}{1 1 0.4}
\pspolygon[linewidth=0.22,fillcolor=userFillColour,fillstyle=solid](47.25,23.07)
(35.25,23.07)
(38.25,27.07)
(44.25,27.07)(47.25,23.07)
\psline[linewidth=0.22](45.38,1.25)(45.38,23.13)
\psline[linewidth=0.22](36.9,20.45)(36.91,22.89)
\rput(95,58.75){$\eqls$}
\rput(44.87,58.13){}
\psline[linewidth=0.22](34.13,57)(34.13,61)
\newrgbcolor{userFillColour}{1 1 0.4}
\pspolygon[linewidth=0.22,fillcolor=userFillColour,fillstyle=solid](32.13,57)
(44.13,57)
(41.13,53)
(35.13,53)(32.13,57)
\psline[linewidth=0.22](42.25,57.25)(42.25,61)
\psline[linewidth=0.22](38.12,53)(38.12,50)
\psline[linewidth=0.22](34.13,42.25)(34.13,46.13)
\newrgbcolor{userFillColour}{1 1 0.4}
\pspolygon[linewidth=0.22,fillcolor=userFillColour,fillstyle=solid](44.13,46.13)
(32.13,46.13)
(35.13,50.13)
(41.13,50.13)(44.13,46.13)
\psline[linewidth=0.22](42.25,42.25)(42.25,46)
\rput(38.13,55){$\comonoid$}
\rput(44.87,72.26){}
\psline[linewidth=0.22](34.13,71.13)(34.13,75.13)
\newrgbcolor{userFillColour}{1 1 0.4}
\pspolygon[linewidth=0.22,fillcolor=userFillColour,fillstyle=solid](32.13,71.13)
(44.13,71.13)
(41.13,67.13)
(35.13,67.13)(32.13,71.13)
\psline[linewidth=0.22](42.25,71.37)(42.25,75.13)
\psline[linewidth=0.22](38.12,67.12)(38.12,64.12)
\newrgbcolor{userFillColour}{1 1 0.4}
\pspolygon[linewidth=0.22,fillcolor=userFillColour,fillstyle=solid](44.13,60.25)
(32.13,60.25)
(35.13,64.26)
(41.13,64.26)(44.13,60.25)
\rput(38,69.25){$\monoid$}
\psline[linewidth=0.22](35.38,4.38)(35.38,0)
\rput(55,58.75){$\eqls$}
\psline[linewidth=0.22](106.13,56)(106.13,72)
\psline[linewidth=0.22](113.75,55.62)(113.75,65)
\newrgbcolor{userFillColour}{1 1 0.4}
\pspolygon[linewidth=0.22,fillcolor=userFillColour,fillstyle=solid](104.13,56)
(116.13,56)
(113.13,52)
(107.13,52)(104.13,56)
\psline[linewidth=0.22](110.13,52)(110,48.5)
\newrgbcolor{userFillColour}{1 1 0.4}
\pspolygon[linewidth=0.22,fillcolor=userFillColour,fillstyle=solid](128.5,58.34)
(116.5,58.34)
(119.5,62.34)
(125.5,62.34)(128.5,58.34)
\psline[linewidth=0.22](126.25,44.38)(126.23,58.43)
\psline[linewidth=0.22](117.77,71.78)(117.75,68.94)
\newrgbcolor{userFillColour}{1 1 0.4}
\pspolygon[linewidth=0.22,fillcolor=userFillColour,fillstyle=solid](123.75,64.94)
(111.75,64.94)
(114.75,68.94)
(120.75,68.94)(123.75,64.94)
\rput(110,53.75){$\monoid$}
\psline[linewidth=0.22](118.12,48.12)(118.12,58.12)
\psline[linewidth=0.22](122.14,62.32)(122.16,64.76)
\newrgbcolor{userFillColour}{1 1 0.4}
\pspolygon[linewidth=0.22,fillcolor=userFillColour,fillstyle=solid](108.12,50)
(120.12,50)
(117.12,46)
(111.12,46)(108.12,50)
\psline[linewidth=0.22](114.5,46)(114.38,43.75)
\psline[linewidth=0.22](66.25,64.38)(66.25,73.12)
\psline[linewidth=0.22](73.75,63.75)(73.75,66.26)
\newrgbcolor{userFillColour}{1 1 0.4}
\pspolygon[linewidth=0.22,fillcolor=userFillColour,fillstyle=solid](64.13,64.13)
(76.13,64.13)
(73.13,60.13)
(67.13,60.13)(64.13,64.13)
\psline[linewidth=0.22](70,60)(70,49.75)
\rput(70,62.5){$\monoid$}
\psline[linewidth=0.22](85.64,45.32)(85.62,54.38)
\rput(114.38,48.12){$\comonoid$}
\psline[linewidth=0.22](77.77,73.04)(77.75,70.2)
\newrgbcolor{userFillColour}{1 1 0.4}
\pspolygon[linewidth=0.22,fillcolor=userFillColour,fillstyle=solid](83.75,66.2)
(71.75,66.2)
(74.75,70.2)
(80.75,70.2)(83.75,66.2)
\rput(6.88,63.12){$\monoid$}
\psline[linewidth=0.22](78.12,50)(78.12,53.75)
\psline[linewidth=0.22](81.85,55.69)(81.88,66.25)
\newrgbcolor{userFillColour}{1 1 0.4}
\pspolygon[linewidth=0.22,fillcolor=userFillColour,fillstyle=solid](68.12,50)
(80.12,50)
(77.12,46)
(71.12,46)(68.12,50)
\psline[linewidth=0.22](74.5,46)(74.38,43.75)
\newrgbcolor{userFillColour}{1 1 0.4}
\pspolygon[linewidth=0.22,fillcolor=userFillColour,fillstyle=solid](88.12,52.5)
(76.12,52.5)
(79.12,56.5)
(85.12,56.5)(88.12,52.5)
\rput(74.38,48.12){$\comonoid$}
\psline[linewidth=0.22](68.5,6.25)(68.5,29.12)
\psline[linewidth=0.22](74.11,12.56)(74.12,16.25)
\psline[linewidth=0.22](77.25,9.38)(77.25,5)
\newrgbcolor{userFillColour}{1 1 0.4}
\pspolygon[linewidth=0.22,fillcolor=userFillColour,fillstyle=solid](83.25,15.84)
(71.25,15.84)
(74.25,19.84)
(80.25,19.84)(83.25,15.84)
\psline[linewidth=0.22](80.38,12.5)(80.37,15.92)
\psline[linewidth=0.22](81.26,29.28)(81.25,26.44)
\newrgbcolor{userFillColour}{1 1 0.4}
\pspolygon[linewidth=0.22,fillcolor=userFillColour,fillstyle=solid](87.26,22.44)
(75.25,22.44)
(78.25,26.44)
(84.25,26.44)(87.26,22.44)
\psline[linewidth=0.22](85.38,0.62)(85.38,22.51)
\psline[linewidth=0.22](76.9,19.83)(76.91,22.26)
\psline[linewidth=0.22](72.88,3.75)(72.88,0)
\newrgbcolor{userFillColour}{1 1 0.4}
\pspolygon[linewidth=0.22,fillcolor=userFillColour,fillstyle=solid](71.12,13.38)
(83.12,13.38)
(80.12,9.38)
(74.12,9.38)(71.12,13.38)
\rput(77.25,11.25){$\monoid$}
\newrgbcolor{userFillColour}{1 1 0.4}
\pspolygon[linewidth=0.22,fillcolor=userFillColour,fillstyle=solid](66.75,6.5)
(78.75,6.5)
(75.75,2.5)
(69.75,2.5)(66.75,6.5)
\rput(72.88,4.38){$\comonoid$}
\newrgbcolor{userFillColour}{1 1 0.4}
\pspolygon[linewidth=0.22,fillcolor=userFillColour,fillstyle=solid](29.62,7.5)
(41.62,7.5)
(38.62,3.5)
(32.62,3.5)(29.62,7.5)
\rput(35.38,5.62){$\comonoid$}
\rput(95.38,16.25){$\eqls$}
\rput(157.74,15.01){}
\psline[linewidth=0.22](147,22)(147,26)
\newrgbcolor{userFillColour}{1 1 0.4}
\pspolygon[linewidth=0.22,fillcolor=userFillColour,fillstyle=solid](145,22)
(157,22)
(154,18)
(148,18)(145,22)
\psline[linewidth=0.22](155.12,22.25)(155.12,26)
\psline[linewidth=0.22](151,18)(151,10.62)
\psline[linewidth=0.22](147,4.75)(147,8.63)
\newrgbcolor{userFillColour}{1 1 0.4}
\pspolygon[linewidth=0.22,fillcolor=userFillColour,fillstyle=solid](157,8.63)
(145,8.63)
(148,12.63)
(154,12.63)(157,8.63)
\psline[linewidth=0.22](155.12,4.75)(155.12,8.5)
\rput(135,16){$\eqls$}
\rput(137.5,58.12){$\eqls$}
\rput(151,20){$\comonoid$}
\rput(20,16){$\eqls$}
\end{pspicture}

\end{center}
\epr

\subsubsection{Classifying classical structures}\label{Classifying}
\be{prop}{\rm \cite{PavlovicD:MSCS08}} In the category $(\FHilb,\otimes,\CCc,\ddag)$ of finitely-dimensional complex Hilbert spaces and linear maps, the classical structures correspond to the orthonormal bases in the usual sense. $\FHilb_\bs$ is equivalent with the category $\FSet$ of finite sets and functions.
\ee{prop}

\be{prop}{\rm \cite{PavlovicD:QI09}} In the category $\left(\Rel,\times,1,\opr{(-)}\right)$ of sets and relations, the classical structures are just the biproducts (disjoint unions) of abelian groups. $\Rel_\bs$ is equivalent with the category $\Set$ of sets and functions.
\end{prop}

Each classical structure $X$ in $\Rel$ decomposes as a disjoint union $X = \sum_{j\in J} X_j$ where each restriction $(X_j,\mnd_j,\unt_j)$ of $(X,\mnd,\unt)$ is an abelian group. A classical structure on $X$ thus consists of (1) a partition $X = \sum_{j\in J} X_j$ and (2) an abelian group structure on each $X_j$. These partitions and group structures, and even the size of $X$ are,  however, indistinguishable by the morphisms of $\Rel_\bs$, because any two classical structures with the same number $J$ of components are isomorphic. 

\paragraph{Bases in $\Rel$.} 
The basis induced by the classical structure $X = \sum_{j\in J} X_j$ is in the form $\BBB(X) = \{X_j\}_{j\in J}$. While the bases with the same number of elements are indistinguishable in $\Rel_\bs$, they are the crucial resource for quantum computation in $\Rel$. The bases induced by the {\em rectangular\/} structures $({\rect}_n,\cmn,\cun)$, will be particularly useful, where
\bear
{\rect}_n & = &  \sum_n \ZZz_n = \{ij \ |\  0\leq i,j\leq n-1\}\\
\cmn (ij) & = & \left\{\left<ik,i\ell \right>\ |\  j = k+\ell \right\}
\\
\cun & = & \left\{i0\ |\  0\leq i\leq n-1\right\}\\
\BBB({\rect}_n) & = & \left\{\beta_i = \{ij\}\ |\  0\leq i,j\leq n-1 \right\}
\eear

\subsection{Bases for Simon's algorithm}
Any bitstring function $f:\ZZz_2^m\sto \ZZz_2^n$, considered in Simon's algorithm, can be viewed as a morphism $f\in \pw\FSet(m, n)$ in the category of finite powersets and all functions between them. It is easy to see that this is a cartesian closed category, with $+$ as the cartesian product\footnote{$\pw\FSet$ is opposite to the Kleisli category for the $\wp\wp$-monad. Along the discrete Stone duality, $\pw\FSet$ is thus dual to the category of free finite atomic Boolean algebras. Since Boolean algebras are primal, every function between them can be expressed as a polynomial.}. The program transformation from the function $f$ to the corresponding Hilbert space unitary $U_f$ is formalized as follows
{\small \[
\prooftree
\prooftree
f(x) = f\circ x \in \FSet_\wp [\tp x m](n)
\justifies
f'(x,y) = <x,y\oplus f(x)> \in \FSet_\wp [\tp{x,y}{m+n}](m+n)
\endprooftree
\justifies
U_f|x,y> = \BBb^{\otimes f'(x,y)} \in \FHilb\left[\tp{|x,y>}{\BBb^{\otimes (m+n)}}\right]\left(\BBb^{\otimes (m+n)}\right)
\endprooftree
\]}
where $\BBb = \CCc^2$. The unitary $U_f$ is thus the image of $f'$ along the functor 
{\small
\[ 
\BBb^{\otimes (-)}\ :\  \FSet_\wp [\tp{x,y}{m+n}]  \to  \FHilb\left[\tp{|x,y>}{\BBb^{\otimes (m+n)}}\right] 
\]}
which maps finite sets to the tensor powers of $\BBb$. Since $\BBb^{\otimes m} = \CCc^{\left(2^m\right)}$, any function $f:2^m \sto 2^n$ in $\pw\Set$ is mapped to a linear operator $\BBb^{\otimes f}: \BBb^{\otimes m}\to \BBb^{\otimes n}$ in $\FHilb$, represented by the matrix $F = \left(F_{ij}\right)_{2^n\times 2^m}$ where $F_{ij} = 1$ whenever $f(j) = i$, otherwise $F_{ij} = 0$. This determines a functor $\FSet_\wp \to \FHilb$. It is extended to a substitution $\FSet_\wp\left[\tp{x,y}{m+n}\right]\to \FHilb\left[\tp{|x,y>}{\BBb^{\otimes (m+n)}}\right]$ by stipulating that the variables $x,y$ are mapped to the variables $|x,y>$.

The function $f\in \pw{\FSet} (m,n)$ has a simpler, though nonstandard interpretation in the dagger-{\em pre\/}monoidal\footnote{The tensor $m\otimes n = m\times n$ is functorial in each argument, but it is not a bifunctor. See \cite{Power-Robinson} for a discussion about such structures. This has no repercussions for us, since the definition of the functor ${\rect}^{\otimes(-)}$, spelled out explicitly below, makes no use of the arrow part of $\otimes$.} category $(\pw\Rel,\otimes, 1,\ddag)$, where  $\pw\Rel(m,n) = \Rel(2^m, 2^n)$ and $m\otimes n = m\times n$. 
%The reader may wish to work out the arrow part of this tensor. 
The dagger is still just the relational converse.
%of $r\in \pw(m,n)$ is defined by $r^\ddag (b,a) = r(\neg a,\neg b)$, for $a\subseteq m$ and $b\subseteq n$. 
Like before, we define
{\small
\[ 
{\rect}^{\otimes(-)}\ :\  \pw{\FSet} [\tp{x,y}{m+n}]  \to  \pw{\Rel}\left[\tp{|x,y>}{{\rect}^{\otimes(m+n)}}\right] 
\]}
this time over the rectangular structure
\begin{gather*}
{\rect} = {\rect}_2  =  \{00,01,10,11\}\\
\cmn(i0)  =  \left\{<i0,i0>,<i1,i1>\right\}\quad \cmn(i1)  =  \left\{<i0,i1>,<i1,i0>\right\}
\\
\cun  =  \left\{00,10\right\}\\
\BBB({\rect})  =  \left\{\beta_0 = \{00,01\},\beta_1 =\{10,11\} \right\}
\end{gather*}
Note that this comonoid structure lifts from $(\Rel,\times,1)$ to $(\pw\Rel,\otimes,1)$ because ${\rect}\otimes {\rect} = 2^2 \otimes 2^2 = 2^{2\times 2} = 2^{2+2} = 2^2\times 2^2 = {\rect}\times {\rect}$. It furthermore lifts to any ${\rect}^{\otimes m}$, since the commutative (co)monoid structures always extend to the tensor powers. 

Since the underlying set of ${\rect}^{\otimes m}$ is $2^{\left(2^m\right)}$, any function $f:2^m \sto 2^n$ in $\pw\Set$, is mapped to a relation ${\rect}^{\otimes f}: {\rect}^{\otimes m}\to {\rect}^{\otimes n}$ in $\pw\Rel$, represented by the matrix $F = \left(F_{ij}\right)_{2^n\times 2^m}$ where $F_{ij} = 1$ whenever $f(j) = i$, otherwise $F_{ij} = 0$. The functor is extended into a substitution $\pw\Set[\tp{x,y}{m+n}] \to \pw\Rel\left[\tp{|x,y>}{{\rect}^{\otimes (m+n)}}\right]$ like before. Mapping  the polynomial $f'(x,y)$, constructed above, along this functor, we get a polynomial unitary relation $\Upsilon_f|x,y> = {\rect}^{\otimes f'(x,y)}$ on ${\rect}^{\otimes (m+n)}$ in $\pw\Rel\left[\tp{|x,y>}{{\rect}^{\otimes (m+n)}}\right]$. This polynomial can be viewed as a family of unitary relations indexed over the basis of ${\rect}^{\otimes (m+n)}$; and each member of the family is a permutation on ${\rect}^{\otimes (m+n)} = 2^{\left(2^{m+n}\right)}$.

\section{Complementarity}\label{Unbiased}
\subsection{Complementary classical structures}
\be{defn}\label{unbiased}
A vector $a \in \CCC(X)$ is\/ {\em unbiased} (or complementary) with respect to a classical structure $(X,\cmn,\cun)$ if $\cmn a\in \CCC(X\otimes X)$ is strongly entangled (in the sense of Sec.~\ref{entangled}).  Two classical structures are complementary if every every basis vector with respect to one is complementary with respect to the other one, and {\em vice versa}.
\ee{defn}

\paragraph{Remark.} In the framework of Hilbert spaces, this definition is equivalent to the standard notion of  complementary bases, used for describing the quantum uncertainty relations \cite{Kraus,Wootters}. Coecke, Duncan and Edwards \cite{Coecke-Duncan,Coecke-Edwards} have characterized complementary vectors in terms of their representations (cf.~Sec.~\ref{Convolution-Representation}~\eqref{Cayley}).  The first part of the following proposition says that our definition is equivalent to theirs.

\be{prop}\label{uu}
With respect to a classical structure $X$, the representative $\yon b \in \CCC(X,X)$ of $b\in \CCC(X)$ is
\begin{anumerate}
\item unitary if and only if $b$ is unbiased; 
\item a pure projector if $b$ is a basis vector.
\end{anumerate}
The converse of (b) holds whenever the basis vectors generate $X$.
\ee{prop}

Recall from Sec.~\ref{dagmondefs} that the usual definitions of projectors and unitaries lift to dagger-categories: a unitary is an endomorphism $u$ such that  $u^\ddag = u^{-1}$, whereas a projector $p$ satisfies $p = p^\ddag = p\circ p$. For a pure projector over $X$ we moreover require $\Tr(p) = \varepsilon \circ (X\otimes p) \circ \eta = \id_I$. The assumption that a set of vectors $\Gamma \subseteq \CCC(X)$ generates an object $X$ means that for any $f\neq g \in \CCC(X,Y)$ there must be a basis vector $a\in \Gamma$ such that $fa\neq ga$.

\bprf{ of \ref{uu}}{\bf (a)} Since $\mnd$ is commutative, by the definition of $\yon b$ in \eqref{Cayley}, $\yon b^\ddag = (\mnd(b\otimes X))^\ddag = \left(X\otimes b^\ddag\right)\cmn$. The composites $\yon b \circ \yon b^\ddag$ and $\yon b^\ddag \circ \yon b$ can thus be viewed as the left-hand side and the right-hand side of the following diagram.
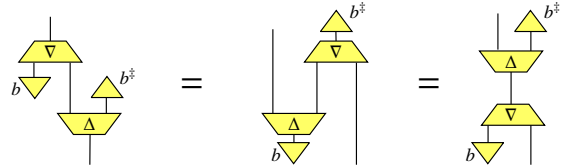
\renewcommand{\eqls}{$\mbox{\Large =}$}
\newcommand{\comonoid}{\cmn}
\newcommand{\monoid}{\mnd}
\renewcommand{\comonunit}{\scriptstyle b^\ddag}
\renewcommand{\monunit}{\scriptstyle b}
\begin{center}
\def\JPicScale{0.7}
\ifx\JPicScale\undefined\def\JPicScale{1}\fi
\psset{unit=\JPicScale mm}
\psset{linewidth=0.3,dotsep=1,hatchwidth=0.3,hatchsep=1.5,shadowsize=1,dimen=middle}
\psset{dotsize=0.7 2.5,dotscale=1 1,fillcolor=black}
\psset{arrowsize=1 2,arrowlength=1,arrowinset=0.25,tbarsize=0.7 5,bracketlength=0.15,rbracketlength=0.15}
\begin{pspicture}(0,0)(102.3,29.38)
\psline[linewidth=0.22](50.38,9.75)(50.38,25.75)
\psline[linewidth=0.22](62.32,27.04)(62.38,23.5)
\psline[linewidth=0.22](58.75,9.38)(58.75,20)
\newrgbcolor{userFillColour}{1 1 0.4}
\pspolygon[linewidth=0.22,fillcolor=userFillColour,fillstyle=solid](48.38,9.75)
(60.38,9.75)
(57.38,5.75)
(51.38,5.75)(48.38,9.75)
\psline[linewidth=0.22](54.38,5.75)(54.38,1.77)
\newrgbcolor{userFillColour}{1 1 0.4}
\pspolygon[linewidth=0.22,fillcolor=userFillColour,fillstyle=solid](68.38,19.5)
(56.38,19.5)
(59.38,23.5)
(65.38,23.5)(68.38,19.5)
\rput(62.38,21.5){$\monoid$}
\psline[linewidth=0.22](66.25,0)(66.25,19.38)
\rput(54.38,7.75){$\comonoid$}
\rput(35,15){$\eqls$}
\psline[linewidth=0.22](18.75,7.5)(18.75,14.38)
\psline[linewidth=0.22](8.12,28.12)(8.12,23.5)
\psline[linewidth=0.22](5,13.75)(5,19.38)
\psline[linewidth=0.22](15.67,6.74)(15.62,0)
\newrgbcolor{userFillColour}{1 1 0.4}
\pspolygon[linewidth=0.22,fillcolor=userFillColour,fillstyle=solid](14.12,19.5)
(2.12,19.5)
(5.12,23.5)
(11.12,23.5)(14.12,19.5)
\rput(8.12,21.5){$\monoid$}
\psline[linewidth=0.22](11.88,9.38)(11.88,19.38)
\newrgbcolor{userFillColour}{1 1 0.4}
\psline[linewidth=0.22,fillcolor=userFillColour,fillstyle=solid](59.42,25.32)
(62.42,29.32)
(65.42,25.32)(59.42,25.32)
\newrgbcolor{userFillColour}{1 1 0.4}
\psline[linewidth=0.22,fillcolor=userFillColour,fillstyle=solid](15.87,12.77)
(18.87,16.77)
(21.87,12.77)(15.87,12.77)
\newrgbcolor{userFillColour}{1 1 0.4}
\psline[linewidth=0.22,fillcolor=userFillColour,fillstyle=solid](57.32,4.2)
(54.32,0.2)
(51.32,4.2)(57.32,4.2)
\rput[r](51.88,1.88){$\monunit$}
\newrgbcolor{userFillColour}{1 1 0.4}
\psline[linewidth=0.22,fillcolor=userFillColour,fillstyle=solid](8.17,16.59)
(5.17,12.59)
(2.17,16.59)(8.17,16.59)
\rput[r](2.5,14.38){$\monunit$}
\rput[l](21.25,16.88){$\comonunit$}
\rput[l](65,29.38){$\comonunit$}
\newrgbcolor{userFillColour}{1 1 0.4}
\pspolygon[linewidth=0.22,fillcolor=userFillColour,fillstyle=solid](9.5,9.75)
(21.5,9.75)
(18.5,5.75)
(12.5,5.75)(9.5,9.75)
\rput(15.5,7.75){$\comonoid$}
\psline[linewidth=0.22](93.12,21.88)(93.12,28.75)
\psline[linewidth=0.22](99.38,26.88)(99.38,21.25)
\psline[linewidth=0.22](95.62,9.38)(95.62,20)
\newrgbcolor{userFillColour}{1 1 0.4}
\pspolygon[linewidth=0.22,fillcolor=userFillColour,fillstyle=solid](89.62,21.62)
(101.62,21.62)
(98.62,17.62)
(92.62,17.62)(89.62,21.62)
\psline[linewidth=0.22](91.25,7.5)(91.25,1.77)
\newrgbcolor{userFillColour}{1 1 0.4}
\pspolygon[linewidth=0.22,fillcolor=userFillColour,fillstyle=solid](102,7.5)
(90,7.5)
(93,11.5)
(99,11.5)(102,7.5)
\rput(95.62,9.38){$\monoid$}
\psline[linewidth=0.22](99.38,0)(99.38,7.5)
\rput(95.62,19.62){$\comonoid$}
\rput(80,15){$\eqls$}
\newrgbcolor{userFillColour}{1 1 0.4}
\psline[linewidth=0.22,fillcolor=userFillColour,fillstyle=solid](96.3,25.32)
(99.3,29.32)
(102.3,25.32)(96.3,25.32)
\newrgbcolor{userFillColour}{1 1 0.4}
\psline[linewidth=0.22,fillcolor=userFillColour,fillstyle=solid](94.2,4.2)
(91.2,0.2)
(88.2,4.2)(94.2,4.2)
\rput[r](88.75,1.88){$\monunit$}
\rput[l](101.88,29.38){$\comonunit$}
\end{pspicture}

\end{center}
Both side diagrams can be transformed into the middle one by applying the Frobenius condition \ref{abs2}(iii). Thus 
\[ \yon b \circ \yon b^\ddag = \id_X\quad \iff \quad (X\otimes b^\ddag\mnd)(\cmn b \otimes X) = \id_X \quad \iff \quad \yon b^\ddag \circ \yon b = \id_X
\]
But by Defn.~\ref{strong-entanglement}, the middle equation just says that $\cmn b$ is strongly entangled, i.e. that $b$ is unbiased. Hence the claim.

{\bf (b)} To begin from the easiest, first note that $\Tr(\yon b) = \id_I\  \iff \ \cun b = \id_I$, because $\Tr(\yon b) = \cun b$:  
\newcommand{\beee}{\scriptstyle b}
\renewcommand{\monunit}{\scriptscriptstyle \bot}
\renewcommand{\comonunit}{\scriptscriptstyle \top}
\begin{center}
\def\JPicScale{0.7}
\ifx\JPicScale\undefined\def\JPicScale{1}\fi
\psset{unit=\JPicScale mm}
\psset{linewidth=0.3,dotsep=1,hatchwidth=0.3,hatchsep=1.5,shadowsize=1,dimen=middle}
\psset{dotsize=0.7 2.5,dotscale=1 1,fillcolor=black}
\psset{arrowsize=1 2,arrowlength=1,arrowinset=0.25,tbarsize=0.7 5,bracketlength=0.15,rbracketlength=0.15}
\begin{pspicture}(0,0)(119.74,26.88)
\psline[linewidth=0.22](19.38,19.38)(19.38,9.25)
\psline[linewidth=0.22](13.75,23.12)(13.75,25)
\psline[linewidth=0.22](8.62,19.75)(8.62,15.75)
\newrgbcolor{userFillColour}{1 1 0.4}
\psline[linewidth=0.22,fillcolor=userFillColour,fillstyle=solid](11.25,24.38)
(13.75,26.88)
(16.25,24.38)(11.25,24.38)
\psline[linewidth=0.22](116.25,21.88)(116.25,3.75)
\newrgbcolor{userFillColour}{1 1 0.4}
\psline[linewidth=0.22,fillcolor=userFillColour,fillstyle=solid](119.62,5.12)
(116.62,1.12)
(113.62,5.12)(119.62,5.12)
\newrgbcolor{userFillColour}{1 1 0.4}
\psline[linewidth=0.22,fillcolor=userFillColour,fillstyle=solid](113.62,20.5)
(116.62,24.5)
(119.62,20.5)(113.62,20.5)
\rput(13.75,25){$\comonunit$}
\rput(116.62,21.5){$\comonunit$}
\rput[l](119.74,2.38){$\beee$}
\newrgbcolor{userFillColour}{1 1 0.4}
\pspolygon[linewidth=0.22,fillcolor=userFillColour,fillstyle=solid](14.62,12.75)
(2.62,12.75)
(5.62,16.75)
(11.62,16.75)(14.62,12.75)
\psline[linewidth=0.22](4.62,12.75)(4.62,3.75)
\newrgbcolor{userFillColour}{1 1 0.4}
\rput(8.62,14.75){$\monoid$}
\psline[linewidth=0.22](12.62,12.75)(12.62,8.75)
\newrgbcolor{userFillColour}{1 1 0.4}
\pspolygon[linewidth=0.22,fillcolor=userFillColour,fillstyle=solid](21.25,18.75)
(6.25,18.75)
(10.62,23.12)
(16.88,23.12)(21.25,18.75)
\newrgbcolor{userFillColour}{1 1 0.4}
\rput(13.75,20.62){$\monoid$}
\newrgbcolor{userFillColour}{1 1 0.4}
\psline[linewidth=0.22,fillcolor=userFillColour,fillstyle=solid](7.88,4.5)
(4.88,0.5)
(1.88,4.5)(7.88,4.5)
\rput[r](2.12,2.38){$\beee$}
\psline[linewidth=0.22](16.25,5.25)(16.25,2.5)
\newrgbcolor{userFillColour}{1 1 0.4}
\psline[linewidth=0.22,fillcolor=userFillColour,fillstyle=solid](19,3.75)
(16,0.75)
(13,3.75)(18.75,3.75)
\rput(16.25,2.5){$\monunit$}
\newrgbcolor{userFillColour}{1 1 0.4}
\pspolygon[linewidth=0.22,fillcolor=userFillColour,fillstyle=solid](10.38,9.12)
(22.38,9.12)
(19.38,5.12)
(13.38,5.12)(10.38,9.12)
\rput(16.38,7.12){$\comonoid$}
\psline[linewidth=0.22](58.12,13.12)(58.12,9.25)
\psline[linewidth=0.22](49.38,23.12)(49.38,25)
\psline[linewidth=0.22](54.38,19.62)(54.38,15.62)
\newrgbcolor{userFillColour}{1 1 0.4}
\psline[linewidth=0.22,fillcolor=userFillColour,fillstyle=solid](46.88,24.38)
(49.38,26.88)
(51.88,24.38)(46.88,24.38)
\rput(49.38,25){$\comonunit$}
\newrgbcolor{userFillColour}{1 1 0.4}
\pspolygon[linewidth=0.22,fillcolor=userFillColour,fillstyle=solid](60.88,12.75)
(48.88,12.75)
(51.88,16.75)
(57.88,16.75)(60.88,12.75)
\psline[linewidth=0.22](43.75,19.38)(43.75,3.12)
\newrgbcolor{userFillColour}{1 1 0.4}
\rput(54.88,14.75){$\monoid$}
\psline[linewidth=0.22](51.38,12.75)(51.38,8.75)
\newrgbcolor{userFillColour}{1 1 0.4}
\pspolygon[linewidth=0.22,fillcolor=userFillColour,fillstyle=solid](56.88,18.75)
(41.88,18.75)
(46.25,23.12)
(52.5,23.12)(56.88,18.75)
\newrgbcolor{userFillColour}{1 1 0.4}
\rput(49.38,20.62){$\monoid$}
\newrgbcolor{userFillColour}{1 1 0.4}
\psline[linewidth=0.22,fillcolor=userFillColour,fillstyle=solid](46.62,4.5)
(43.62,0.5)
(40.62,4.5)(46.62,4.5)
\rput[r](40.88,2.38){$\beee$}
\psline[linewidth=0.22](55,5.25)(55,2.5)
\newrgbcolor{userFillColour}{1 1 0.4}
\psline[linewidth=0.22,fillcolor=userFillColour,fillstyle=solid](57.75,3.75)
(54.75,0.75)
(51.75,3.75)(57.5,3.75)
\rput(55,2.5){$\monunit$}
\newrgbcolor{userFillColour}{1 1 0.4}
\pspolygon[linewidth=0.22,fillcolor=userFillColour,fillstyle=solid](49.13,9.12)
(61.13,9.12)
(58.13,5.12)
(52.13,5.12)(49.13,9.12)
\rput(55.13,7.12){$\comonoid$}
\rput(32.5,14.38){$\eqls$}
\rput(104.38,14.38){$\eqls$}
\psline[linewidth=0.22](87.5,23.12)(87.5,25)
\psline[linewidth=0.22](92.5,19.62)(92.5,3.12)
\newrgbcolor{userFillColour}{1 1 0.4}
\psline[linewidth=0.22,fillcolor=userFillColour,fillstyle=solid](85,24.38)
(87.5,26.88)
(90,24.38)(85,24.38)
\rput(87.5,25){$\comonunit$}
\psline[linewidth=0.22](81.88,19.38)(81.88,3.12)
\newrgbcolor{userFillColour}{1 1 0.4}
\pspolygon[linewidth=0.22,fillcolor=userFillColour,fillstyle=solid](95,18.75)
(80,18.75)
(84.38,23.12)
(90.62,23.12)(95,18.75)
\newrgbcolor{userFillColour}{1 1 0.4}
\rput(87.5,20.62){$\monoid$}
\newrgbcolor{userFillColour}{1 1 0.4}
\psline[linewidth=0.22,fillcolor=userFillColour,fillstyle=solid](84.75,4.5)
(81.75,0.5)
(78.75,4.5)(84.75,4.5)
\rput[r](79,2.38){$\beee$}
\rput(70.62,14.38){$\eqls$}
\newrgbcolor{userFillColour}{1 1 0.4}
\psline[linewidth=0.22,fillcolor=userFillColour,fillstyle=solid](95.62,4.38)
(92.62,0.38)
(89.62,4.38)(95.62,4.38)
\rput(92.5,2.5){$\monunit$}
\end{pspicture}

\end{center}

Secondly, we want to show that $\yon b = \yon b^\ddag\  \iff\  b^\ast = b^\ddag$, i.e.
\renewcommand{\comonunit}{\scriptstyle b}
\renewcommand{\beee}{\scriptstyle b^\ddag}
\newcommand{\eqlnt}{\mbox{\large $\iff$}}
\renewcommand{\monunit}{\scriptscriptstyle \top}
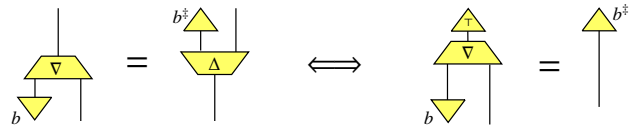
\begin{center}
\def\JPicScale{0.75}
\ifx\JPicScale\undefined\def\JPicScale{1}\fi
\psset{unit=\JPicScale mm}
\psset{linewidth=0.3,dotsep=1,hatchwidth=0.3,hatchsep=1.5,shadowsize=1,dimen=middle}
\psset{dotsize=0.7 2.5,dotscale=1 1,fillcolor=black}
\psset{arrowsize=1 2,arrowlength=1,arrowinset=0.25,tbarsize=0.7 5,bracketlength=0.15,rbracketlength=0.15}
\begin{pspicture}(0,0)(109.88,20)
\psline[linewidth=0.22](83.58,17.66)(83.62,14.12)
\psline[linewidth=0.22](80,0)(80,10.62)
\newrgbcolor{userFillColour}{1 1 0.4}
\pspolygon[linewidth=0.22,fillcolor=userFillColour,fillstyle=solid](89.62,10.12)
(77.62,10.12)
(80.62,14.12)
(86.62,14.12)(89.62,10.12)
\rput(83.62,12.12){$\monoid$}
\psline[linewidth=0.22](87.5,-0.62)(87.5,10)
\newrgbcolor{userFillColour}{1 1 0.4}
\psline[linewidth=0.22,fillcolor=userFillColour,fillstyle=solid](80.68,15.94)
(83.68,19.94)
(86.68,15.94)(80.68,15.94)
\rput[r](4.38,0.62){$\comonunit$}
\psline[linewidth=0.22](36.25,12.5)(36.25,19.38)
\psline[linewidth=0.22](42.5,20)(42.5,11.88)
\psline[linewidth=0.22](38.75,0)(38.75,10.62)
\newrgbcolor{userFillColour}{1 1 0.4}
\pspolygon[linewidth=0.22,fillcolor=userFillColour,fillstyle=solid](32.75,12.25)
(44.75,12.25)
(41.75,8.25)
(35.75,8.25)(32.75,12.25)
\psline[linewidth=0.22](6.88,7.5)(6.88,1.77)
\psline[linewidth=0.22](15,0)(15,7.5)
\rput(38.75,10.25){$\comonoid$}
\rput(25,10){$\eqls$}
\newrgbcolor{userFillColour}{1 1 0.4}
\psline[linewidth=0.22,fillcolor=userFillColour,fillstyle=solid](33.25,16)
(36.25,20)
(39.25,16)(33.25,16)
\newrgbcolor{userFillColour}{1 1 0.4}
\psline[linewidth=0.22,fillcolor=userFillColour,fillstyle=solid](9.83,4.2)
(6.83,0.2)
(3.83,4.2)(9.83,4.2)
\rput(83.75,17.5){$\monunit$}
\psline[linewidth=0.22](11,9.38)(11,20)
\newrgbcolor{userFillColour}{1 1 0.4}
\pspolygon[linewidth=0.22,fillcolor=userFillColour,fillstyle=solid](17,7.5)
(5,7.5)
(8,11.5)
(14,11.5)(17,7.5)
\rput(10.62,9.38){$\monoid$}
\newrgbcolor{userFillColour}{1 1 0.4}
\psline[linewidth=0.22,fillcolor=userFillColour,fillstyle=solid](83.12,3.75)
(80.12,-0.25)
(77.12,3.75)(83.12,3.75)
\rput[r](77.5,0.62){$\comonunit$}
\psline[linewidth=0.22](106.88,2.5)(106.88,19.38)
\rput(98.12,9.38){$\eqls$}
\newrgbcolor{userFillColour}{1 1 0.4}
\psline[linewidth=0.22,fillcolor=userFillColour,fillstyle=solid](103.88,16)
(106.88,20)
(109.88,16)(103.88,16)
\rput[l](109.38,19.38){$\beee$}
\rput[r](34.38,18.75){$\beee$}
\rput(60,10){$\eqlnt$}
\end{pspicture}

\end{center}
The right-hand equation says that $b$ is real, which is a property of every basis vector, according Cor.~\ref{basis-real}. The implication from left to right is obtained by postcomposing both sides of the left-hand equation with $\cun$. The implication from right to left is obtained by tensoring by $X$ on the right both sides of the right-hand equation, and then precomposing them with $\cmn$. The left-hand equation is then obtained using \ref{abs2}(ii).

To complete the proof, we show that $\cmn b = b\otimes b$ implies $\yon b\circ \yon b = \yon b$, by the following diagram:
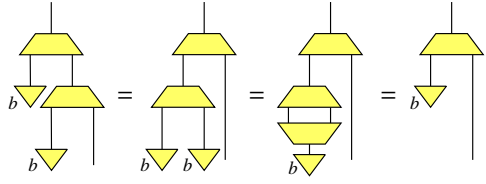
\begin{center}
\def\JPicScale{0.7}
\ifx\JPicScale\undefined\def\JPicScale{1}\fi
\psset{unit=\JPicScale mm}
\psset{linewidth=0.3,dotsep=1,hatchwidth=0.3,hatchsep=1.5,shadowsize=1,dimen=middle}
\psset{dotsize=0.7 2.5,dotscale=1 1,fillcolor=black}
\psset{arrowsize=1 2,arrowlength=1,arrowinset=0.25,tbarsize=0.7 5,bracketlength=0.15,rbracketlength=0.15}
\begin{pspicture}(0,0)(89,31)
\psline[linewidth=0.22](40,21)(40,1)
\newrgbcolor{userFillColour}{1 1 0.4}
\pspolygon[linewidth=0.22,fillcolor=userFillColour,fillstyle=solid](42,21)
(30,21)
(33,25)
(39,25)(42,21)
\psline[linewidth=0.22](36,25)(36,31)
\psline[linewidth=0.22](32,21)(32,15)
\psline[linewidth=0.22](36,11)(36,2)
\newrgbcolor{userFillColour}{1 1 0.4}
\pspolygon[linewidth=0.22,fillcolor=userFillColour,fillstyle=solid](38,11)
(26,11)
(29,15)
(35,15)(38,11)
\psline[linewidth=0.22](28,11)(28,2)
\rput(21,13){$=$}
\psline[linewidth=0.22](2.96,20.98)(3,15)
\newrgbcolor{userFillColour}{1 1 0.4}
\pspolygon[linewidth=0.22,fillcolor=userFillColour,fillstyle=solid](12.96,20.98)
(0.96,20.98)
(3.96,24.98)
(9.96,24.98)(12.96,20.98)
\psline[linewidth=0.22](6.96,24.98)(6.96,30.98)
\psline[linewidth=0.22](10.96,20.98)(10.96,14.98)
\psline[linewidth=0.22](15,11)(15,0)
\newrgbcolor{userFillColour}{1 1 0.4}
\pspolygon[linewidth=0.22,fillcolor=userFillColour,fillstyle=solid](16.96,10.98)
(4.96,10.98)
(7.96,14.98)
(13.96,14.98)(16.96,10.98)
\psline[linewidth=0.22](6.96,10.98)(7,3)
\psline[linewidth=0.22](64,21)(64,1)
\newrgbcolor{userFillColour}{1 1 0.4}
\pspolygon[linewidth=0.22,fillcolor=userFillColour,fillstyle=solid](66,21)
(54,21)
(57,25)
(63,25)(66,21)
\psline[linewidth=0.22](60,25)(60,31)
\psline[linewidth=0.22](56,21)(56,15)
\psline[linewidth=0.22](60,11)(60,7)
\newrgbcolor{userFillColour}{1 1 0.4}
\pspolygon[linewidth=0.22,fillcolor=userFillColour,fillstyle=solid](62,11)
(50,11)
(53,15)
(59,15)(62,11)
\psline[linewidth=0.22](52,11)(52,7)
\newrgbcolor{userFillColour}{1 1 0.4}
\pspolygon[linewidth=0.22,fillcolor=userFillColour,fillstyle=solid](50,8)
(62,8)
(59,4)
(53,4)(50,8)
\psline[linewidth=0.22](56,4)(56,1)
\newrgbcolor{userFillColour}{1 1 0.4}
\psline[linewidth=0.22,fillcolor=userFillColour,fillstyle=solid](59,2)
(56,-2)
(53,2)(59,2)
\newrgbcolor{userFillColour}{1 1 0.4}
\psline[linewidth=0.22,fillcolor=userFillColour,fillstyle=solid](31,3)
(28,-1)
(25,3)(31,3)
\newrgbcolor{userFillColour}{1 1 0.4}
\psline[linewidth=0.22,fillcolor=userFillColour,fillstyle=solid](39,3)
(36,-1)
(33,3)(39,3)
\newrgbcolor{userFillColour}{1 1 0.4}
\psline[linewidth=0.22,fillcolor=userFillColour,fillstyle=solid](10,3)
(7,-1)
(4,3)(10,3)
\newrgbcolor{userFillColour}{1 1 0.4}
\psline[linewidth=0.22,fillcolor=userFillColour,fillstyle=solid](6,15)
(3,11)
(0,15)(6,15)
\rput(46,13){$=$}
\psline[linewidth=0.22](87,21)(87,1)
\newrgbcolor{userFillColour}{1 1 0.4}
\pspolygon[linewidth=0.22,fillcolor=userFillColour,fillstyle=solid](89,21)
(77,21)
(80,25)
(86,25)(89,21)
\psline[linewidth=0.22](83,25)(83,31)
\psline[linewidth=0.22](79,21)(79,15)
\newrgbcolor{userFillColour}{1 1 0.4}
\psline[linewidth=0.22,fillcolor=userFillColour,fillstyle=solid](82,15)
(79,11)
(76,15)(82,15)
\rput(71,13){$=$}
\rput[r](53.75,-0.62){$\comonunit$}
\rput[r](25.62,0){$\comonunit$}
\rput[r](4.38,0){$\comonunit$}
\rput[r](0.62,11.88){$\comonunit$}
\rput[r](33.75,0){$\comonunit$}
\rput[r](76.88,11.88){$\comonunit$}
\end{pspicture}

\end{center}
%
%The converse of (b) follows from Lemma~\ref{diagonals}.
\epr

%\begin{lemma}\label{diagonals}
%For  every classical structure $X$ which is generated by its basis, and every vector $b\in \CCC(X)$ holds
%\bear
%\mnd (b\otimes b) = b & \iff &\cmn b = b\otimes b
%\eear
%\end{lemma}

%\bpr
%One direction is trivial: if $\cmn b = b\otimes b$, then obviously $b = \mnd\cmn b = \mnd(b\otimes b)$. 

%The other way around, if $\mnd(b\otimes b) = b$, we show that 
%\bea\label{xdag}
%(x^\ddag \otimes y^\ddag) \cmn b & = &  (x^\ddag \otimes y^\ddag) (b\otimes b)
%\eea
%\newcommand{\monoid}{\mnd}
%\newcommand{\comonoid}{\cmn}
%\renewcommand{\monunit}{\scriptstyle b}
%\renewcommand{\comonunit}{\scriptstyle x^\ddag}
%\begin{center}
%\def\JPicScale{0.65}
%\input{PIC/nab}
%\end{center}
%holds for every substitution of a pair of basis vectors for $x$ and $y$.

%

%where we use $\mnd(b\otimes b) = b$ at the first step, and at the last two steps the fact that $\cmn x = x\otimes x$ (Cor.~\ref{diagx}), i.e. $x^\ddag \circ \mnd = x^\ddag\otimes x^\ddag$. The result follows by abstracting $x$ from \eqref{xdag}, and applying the $\eta$-rule.

%\epr

\subsection{Transforms}
A given basis of a Hilbert space can be mapped into a complementary one using a Fourier transform. This is done in all HSP-algorithms: the basis vectors are entangled into one complementary vector, and the unitary $U_f$ is then evaluated over that vector, thus computing all values of $f$ in one sweep. 

In order to complete the implementation of Simon's algorithm in $\pw\Rel$, we need a pair of complementary bases for ${\rect}^{\otimes (m+n)}$. As mentioned above, the classical structures of $\rect$ lift from $\Rel$ to $\pw\Rel$. And in $\Rel$ in general, for a given classical structure $X = \sum_{j\leq m} X^1_j$ in $\Rel$, a complementary vector is a set $\gamma \subseteq X$ such that $\gamma_j = \gamma \cap X^1_j$ is a singleton for every $j\leq m$. Another classical structure $X = \sum_{k\leq n} X^2_k$ over the same set is thus complementary if and only if $X^1_j\cap X^2_k$ is a singleton for all $j\leq m, k\leq n$. Since $X^1$ and $X^2$ are partitions, it follows that all $\# X^1_j = n$ and all $\#X^2_k = m$. So $X$ must decompose to $m$ groups of order $n$, and to $n$ groups of order $n$. In order to have an invertible transform from one basis to another, we need $m=n$. Unless we are interested in the various forms of entanglement engendered by the various group structures, we can thus restrict attention to rectangular structures from sec.~\ref{Classifying}. A simple transform mapping the basis vectors of $\rect_\ell$ into a complementary basis is
\bear
H_\ell\ :\ \rect_\ell & \to & \rect_\ell\\
ij & \longmapsto & ji
\eear
Using $H=H_2$ to transform $H^{\otimes m}: {\rect}^{\otimes m}\to {\rect}^{\otimes m}$ we can now produce the superposition of all the basis vectors, representing the inputs of the function $f:\ZZz_2^m\sto \ZZz_2^n$ from Simon's algorithm. The other way around, the $H$-image of any basis vector is the superposition of the complementary basis of ${\rect}^{\otimes m}$. We can thus define the unitary polynomial $(H^{\otimes m}\otimes \id)\circ \Upsilon_f |x,y>\circ  (H^{\otimes m}\otimes \id)$ on ${\rect}^{\otimes (m+n)}$ in $\pw\Rel\left[|x,y>:{\rect}^{\otimes (m+n)}\right]$
and evaluate it on the vector $|0,0>=\unt \in \pw\Rel\left( {\rect}^{\otimes (m+n)}\right)$, to get the outcome $S|x,y> \in \pw\Rel\left[|x,y>:{\rect}^{\otimes (m+n)}\right]\left( {\rect}^{\otimes (m+n)}\right)$. To complete the execution of Simon's algorithm in $\pw\Rel$, we just need to measure this outcome.

\section{Measurements}\label{Measurement}
So far, we have seen that the classical data in a quantum universe, represented by a dagger-monoidal category $\CCC$, can be characterized as just those data that can be annotated by the variables in $\CCC[x,y,\ldots]$, i.e. those data that support the abstraction operation $\kappa x$. Quantum programs are thus viewed as polynomial arrows $\varphi(x,y,\ldots)\in \CCC[x,y,\ldots]$. In this respect, quantum programs are similar to classical programs: they specify that some operations should be applied to some input data, always classical, denoted by the variables. Semantics of computation is captured through abstractions and substitutions. Program execution, in particular, corresponds to substituting some input data for the variables, and evaluating the resulting expressions.

In classical computation, such evaluations yield the outputs. In quantum computation, however, there is more: the outputs need to be {\em measured}. The view of quantum programs as polynomials in dagger-monoidal categories needs to be refined to capture measurements. In the simplest case, a measurement will turn out to be just a projector in $\CCC[\tp x X]$. 

\be{defn}
A morphism $X\otimes A\tto\alpha A$ in $\CCC$ on is an \/{\em $X$-action}  $A$ if $\alpha\circ (X\otimes\alpha) = \alpha \circ \mnd$. An $X$-action is\/ {\em normal} if moreover $\alpha\circ (\unt \times A) = \id_A$.
\newcommand{\objX}{{\scriptstyle X}}
\newcommand{\objA}{{\scriptstyle A}}
\newcommand{\comonoid}{\cmn}
\newcommand{\monoid}{\mnd}
\renewcommand{\monunit}{\scriptscriptstyle \bot}
\newcommand{\coalgebra}{{\scriptstyle \alpha^\ddag}}
\newcommand{\algebra}{{\scriptstyle \alpha}}
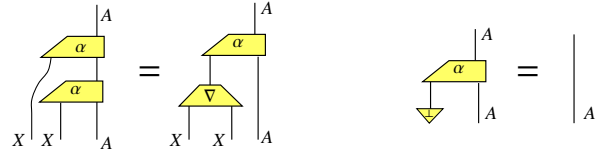
\begin{center}
\def\JPicScale{0.7}
\ifx\JPicScale\undefined\def\JPicScale{1}\fi
\psset{unit=\JPicScale mm}
\psset{linewidth=0.3,dotsep=1,hatchwidth=0.3,hatchsep=1.5,shadowsize=1,dimen=middle}
\psset{dotsize=0.7 2.5,dotscale=1 1,fillcolor=black}
\psset{arrowsize=1 2,arrowlength=1,arrowinset=0.25,tbarsize=0.7 5,bracketlength=0.15,rbracketlength=0.15}
\begin{pspicture}(0,0)(107.5,26.63)
\psline[linewidth=0.22](45,26.63)(45,20.63)
\psline[linewidth=0.22](36.5,10.63)(36.5,16.63)
\newrgbcolor{userFillColour}{1 1 0.4}
\pspolygon[linewidth=0.22,fillcolor=userFillColour,fillstyle=solid](47,16.63)
(35,16.63)
(40,20.63)
(47,20.63)(47,16.63)
\psline[linewidth=0.22](45.62,0.62)(45.62,16.38)
\rput(25,13.12){$\eqls$}
\rput(42,18.63){}
\rput(41.88,19){$\algebra$}
\psline[linewidth=0.22](15,0.62)(15,26.25)
\rput(11.5,18.38){}
\psecurve[linewidth=0.2,curvature=1.0 0.1 0.0](6.38,16.88)(6.38,16.88)(5.75,13.75)(3.88,11.26)(2.62,8.12)(2.62,5.62)(2.62,0.62)(2.62,-0.62)
\rput[l](46.25,1.25){$\objA$}
\rput[l](15.75,24.38){$\objA$}
\rput[l](15.75,-0){$\objA$}
\rput[r](1.38,0.62){$\objX$}
\rput[r](6.88,0.62){$\objX$}
\newrgbcolor{userFillColour}{1 1 0.4}
\pspolygon[linewidth=0.22,fillcolor=userFillColour,fillstyle=solid](16.38,16.25)
(4.38,16.25)
(9.38,20.25)
(16.38,20.25)(16.38,16.25)
\rput(12,18.12){$\algebra$}
\newrgbcolor{userFillColour}{1 1 0.4}
\pspolygon[linewidth=0.22,fillcolor=userFillColour,fillstyle=solid](16.12,7.76)
(4.12,7.76)
(9.12,11.76)
(16.12,11.76)(16.12,7.76)
\psline[linewidth=0.22](8.12,0.62)(8.12,7.63)
\rput(10.12,9.76){}
\rput(11,10.12){$\algebra$}
\psline[linewidth=0.22](32.5,0.62)(32.5,7.5)
\newrgbcolor{userFillColour}{1 1 0.4}
\pspolygon[linewidth=0.22,fillcolor=userFillColour,fillstyle=solid](42.61,7)
(30.61,7)
(34.5,11)
(38.5,11)(42.61,7)
\psline[linewidth=0.22](40.62,0.62)(40.61,7)
\rput(36.5,9){$\monoid$}
\rput[l](46.25,25.62){$\objA$}
\rput[r](31.25,0.62){$\objX$}
\rput[r](39.38,0.62){$\objX$}
\psline[linewidth=0.22](86.87,21.63)(86.87,15.63)
\psline[linewidth=0.22](78.37,5.63)(78.37,11.63)
\newrgbcolor{userFillColour}{1 1 0.4}
\pspolygon[linewidth=0.22,fillcolor=userFillColour,fillstyle=solid](88.87,11.63)
(76.87,11.63)
(81.87,15.63)
(88.87,15.63)(88.87,11.63)
\psline[linewidth=0.22](87.5,3.75)(87.5,11.38)
\rput(83.87,13.63){}
\rput(83.75,14){$\algebra$}
\rput[l](88.75,5.62){$\objA$}
\rput[l](88.12,20.62){$\objA$}
\rput(96.88,13.12){$\eqls$}
\rput[l](107.5,5.62){$\objA$}
\psline[linewidth=0.22](105.62,20.62)(105.62,3.75)
\newrgbcolor{userFillColour}{1 1 0.4}
\pscustom[linewidth=0.2,fillcolor=userFillColour,fillstyle=solid]{\psline(80.75,6.5)(75.75,6.5)
\psline(75.75,6.5)(78.13,3.75)
\psbezier(78.13,3.75)(78.13,3.75)(78.13,3.75)
\psline(78.13,3.75)(80.75,6.5)
\closepath}
\rput(78.12,5.62){$\monunit$}
\end{pspicture}

\end{center}

An $X$-equivariant homomorphism from $X\otimes A \tto \alpha A$ to $X\otimes B\tto \beta B$ is an arrow $f\in \CCC(A,B)$ such that $f\circ \alpha = \beta \circ(X\otimes f)$. The category of $X$-actions and $X$-equivariant homomorphisms is denoted $\Ac \CCC X$.

The full subcategory of \/{\em normal} $X$-actions is $\EM \CCC X \hookrightarrow \Ac\CCC X$.
\ee{defn}

\paragraph{Remark.} Normal $X$-actions are the Eilenberg-Moore algebras for the monad $X\otimes (-):\CCC\to \CCC$. Equivalently, they are also actions of the monoid $X$, and this terminology tends to lead to less confusion.

\begin{lemma}\label{meas-lemma} Let $(X,\cmn,\cun)$ be a classical structure, $\alpha(x):A\to A$ an endomorphism in $\CCC[\tp{x}{X}]$ and $\alpha = \kappa x.\ \alpha(x) : X\otimes A \to A$ its abstraction. 
\begin{anumerate}
\item The following conditions are equivalent:
\begin{rnumerate}
\item $\alpha(x)= \alpha(x)\circ \alpha(x)$, i.e. $\alpha(x)$ is idempotent
\item $\alpha\circ (X\otimes \alpha) = \alpha \circ \mnd$, i.e. $\alpha$ is an $X$-action
\item $\alpha\circ (X\otimes \alpha) \circ (\cmn\otimes A) = \alpha$, i.e. $\alpha$ is idempotent as an endomorphism on $A$ in $\Kl\CCC X$.
\end{rnumerate}

\item On the other hand, the following conditions are also equivalent:
\begin{rnumerate}
\item $\alpha(x) = \alpha(x)^\ddag$, i.e. $\alpha(x)$ is self-adjoint 
\item $\alpha = (\varepsilon\otimes A)\circ \alpha^\ddag$
\newcommand{\objX}{{\scriptstyle X}}
\newcommand{\objA}{{\scriptstyle A}}
\newcommand{\monoid}{\varepsilon}
\newcommand{\coalgebra}{{\scriptstyle \alpha^\ddag}}
\newcommand{\algebra}{{\scriptstyle \alpha}}
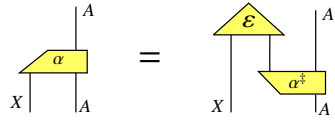
\begin{center}
\def\JPicScale{0.75}
\ifx\JPicScale\undefined\def\JPicScale{1}\fi
\psset{unit=\JPicScale mm}
\psset{linewidth=0.3,dotsep=1,hatchwidth=0.3,hatchsep=1.5,shadowsize=1,dimen=middle}
\psset{dotsize=0.7 2.5,dotscale=1 1,fillcolor=black}
\psset{arrowsize=1 2,arrowlength=1,arrowinset=0.25,tbarsize=0.7 5,bracketlength=0.15,rbracketlength=0.15}
\begin{pspicture}(0,0)(66.12,23.12)
\psline[linewidth=0.22](13.75,3.75)(13.75,11.25)
\psline[linewidth=0.22](21.88,3.75)(21.88,22.5)
\psline[linewidth=0.22](64.38,10.88)(64.38,21.88)
\psline[linewidth=0.22](49.38,3.75)(49.38,17.5)
\newrgbcolor{userFillColour}{1 1 0.4}
\pspolygon[linewidth=0.22,fillcolor=userFillColour,fillstyle=solid](54.24,11)
(66.12,11)
(66.12,7)
(58.12,7)(54.24,11)
\psline[linewidth=0.22](64.38,6.88)(64.38,3.75)
\psline[linewidth=0.22](56.24,11)(56.25,18.12)
\rput(35,13.12){$\eqls$}
\rput(18.88,19){}
\rput(61.25,9.38){$\coalgebra$}
\rput[l](22.5,5){$\objA$}
\rput[l](23.12,21.88){$\objA$}
\rput[l](65.62,21.25){$\objA$}
\rput[l](65,4.38){$\objA$}
\rput[r](12.5,5.62){$\objX$}
\rput[r](48.12,5){$\objX$}
\newrgbcolor{userFillColour}{1 1 0.4}
\psline[linewidth=0.22,fillcolor=userFillColour,fillstyle=solid](46.25,17.5)
(52.5,23.12)
(58.75,17.5)(46.25,17.5)
\rput(52.5,20){$\monoid$}
\newrgbcolor{userFillColour}{1 1 0.4}
\pspolygon[linewidth=0.22,fillcolor=userFillColour,fillstyle=solid](23.88,10.75)
(11.88,10.75)
(16.88,14.75)
(23.88,14.75)(23.88,10.75)
\rput(18.76,13.13){$\algebra$}
\end{pspicture}

\end{center}

\item $ (X\otimes \alpha)\circ (\cmn \otimes A) = (\mnd \otimes A)\circ (X\otimes \alpha^\ddag)$
\nopagebreak
\newcommand{\comonoid}{\cmn}
\renewcommand{\monoid}{\mnd}
\begin{center}
\def\JPicScale{0.75}
\ifx\JPicScale\undefined\def\JPicScale{1}\fi
\psset{unit=\JPicScale mm}
\psset{linewidth=0.3,dotsep=1,hatchwidth=0.3,hatchsep=1.5,shadowsize=1,dimen=middle}
\psset{dotsize=0.7 2.5,dotscale=1 1,fillcolor=black}
\psset{arrowsize=1 2,arrowlength=1,arrowinset=0.25,tbarsize=0.7 5,bracketlength=0.15,rbracketlength=0.15}
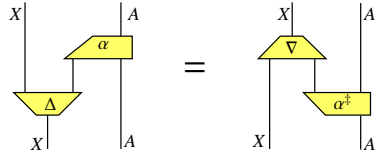
\begin{pspicture}(0,0)(66.12,27)
\psline[linewidth=0.22](4.88,11)(4.88,27)
\psline[linewidth=0.22](21.88,27)(21.88,21)
\psline[linewidth=0.22](13.38,11)(13.38,17)
\newrgbcolor{userFillColour}{1 1 0.4}
\pspolygon[linewidth=0.22,fillcolor=userFillColour,fillstyle=solid](2.88,11)
(14.88,11)
(10.88,7)
(6.88,7)(2.88,11)
\psline[linewidth=0.22](8.88,7)(8.88,1)
\newrgbcolor{userFillColour}{1 1 0.4}
\pspolygon[linewidth=0.22,fillcolor=userFillColour,fillstyle=solid](23.88,17)
(11.88,17)
(16.88,21)
(23.88,21)(23.88,17)
\psline[linewidth=0.22](21.88,1)(21.88,17)
\rput(9.38,9){$\comonoid$}
\psline[linewidth=0.22](64.38,10.88)(64.38,26.88)
\psline[linewidth=0.22](52.24,27)(52.24,21)
\psline[linewidth=0.22](48.24,1)(48.24,17)
\newrgbcolor{userFillColour}{1 1 0.4}
\pspolygon[linewidth=0.22,fillcolor=userFillColour,fillstyle=solid](54.24,11)
(66.12,11)
(66.12,7)
(58.12,7)(54.24,11)
\psline[linewidth=0.22](64.38,6.88)(64.38,0.88)
\newrgbcolor{userFillColour}{1 1 0.4}
\pspolygon[linewidth=0.22,fillcolor=userFillColour,fillstyle=solid](58.24,17)
(46.24,17)
(50.12,21)
(54.12,21)(58.24,17)
\psline[linewidth=0.22](56.24,11)(56.24,17)
\rput(35,15){$\eqls$}
\rput(18.88,19){}
\rput(52.12,19){$\monoid$}
\rput(18.76,19.38){$\algebra$}
\rput(61.25,9.38){$\coalgebra$}
\rput[l](22.5,2.5){$\objA$}
\rput[l](23.12,25){$\objA$}
\rput[l](65,25){$\objA$}
\rput[l](65,1.88){$\objA$}
\rput[r](8.12,1.88){$\objX$}
\rput[r](47.5,2.5){$\objX$}
\rput[r](4.38,25){$\objX$}
\rput[r](51.88,25){$\objX$}
\end{pspicture}

\end{center}
\end{rnumerate}
\end{anumerate}
\end{lemma}

The {\bf proofs} of the above equivalences are easy exercises with classical structure. The equivalence (b)(ii$\Leftrightarrow$iii) can be viewed, and proven, in analogy with Thm.~\ref{abs2}(ii$\Leftrightarrow$iii).

%\subsection{Category of measurements}
%The category of measurements $\Mes \CCC {\tp{x}{X}}$ is the smallest completion and cocompletion of $\CCC[\tp{x}{X}]$ which is dagger monoidal, and such that the functor $\CCC[\tp{x}{X}]\to \Mes \CCC {\tp{x}{X}}$ preserves the structure. It is clear that $\Mes \CCC {\tp{x}{X}}$ must at least split the idempotents in $\CCC[\tp{x}{X}]$. But it can only split the self-adjoint idempotents, or else it will not have the dagger, at least not consistent with the one in $\CCC[\tp{x}{X}]$.

%Recall that a self-adjoint idempotent, i.e. an endomorphism $p:A\to A$ such that $p\circ p = p = p^\ddag$ is called a {\em projector}.

\be{defn}\label{Frobcoalg-def} Let $X$ be a classical structure in $\CCC$. An $X$-{\em measurement\/} over $A\in \CCC$ is a projector $\alpha(x):A\to A$ in $\CCC[\tp x X]$, i.e. a self-adjoint idempotent $\alpha(x) = \alpha(x)^\ddag = \alpha(x)\circ \alpha(x)$.

A homomorphism $f:\alpha(x)\to \beta(x)$, where $\alpha(x)$ is an $X$-measurement over $A$ and $\beta(x)$ is an $X$-measurement over $B$, is an arrow $f\in \CCC(A,B)$ such that
%\bear
$f\circ \alpha(x)  =  \beta(x) \circ f$.
%\eear
The category of measurements in the classical structure $(X,\cmn,\cun)$ is denoted by $\Mes{\CCC}{\tp{x}{X}}$.
\ee{defn}

\paragraph{Remark.} Substituting a basis vector $\varphi\in \BBB(X)$ into a measurement $\alpha(x)\in \CCC[\tp{x}{X}](A,A)$ yields a projector $\alpha(\varphi)\in \CCC(A,A)$. The intuition is that this projector corresponds to an the outcome of the measurement $\alpha$. 

It is easy to see that $\Mes \CCC {\tp{x}{X}}$ is a dagger-monoidal category. The following two propositions show that this notion of a measurement is equivalent with the one from \cite{PavlovicD:QMWS}.

\be{thm}\label{Frobcoalg} Let $X$ be a classical structure, and $\alpha(x):A\to A$ an endomorphism in $\CCC[\tp{x}{X}]$. Then $\mbox{(a)}\iff\mbox{(b)}\Longleftarrow\mbox{(c)}$.

\be{anumerate}
\item $\alpha(x):A\to A$ is a measurement
\item $\alpha  = \kappa x.\ \alpha(x): XA\to A$ is an $X$-action such that $\alpha\circ (x\otimes A)  =  (x^\ddag \otimes A)\circ \alpha^\ddag$
\newcommand{\objX}{{\scriptstyle X}}
\newcommand{\objA}{{\scriptstyle A}}
\newcommand{\comonoid}{\cmn}
\newcommand{\monoid}{\mnd}
\newcommand{\coalgebra}{{\scriptstyle \alpha^\ddag}}
\newcommand{\algebra}{{\scriptstyle \alpha}}
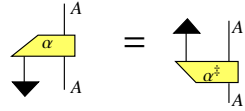
\begin{center}
\def\JPicScale{0.7}
\ifx\JPicScale\undefined\def\JPicScale{1}\fi
\psset{unit=\JPicScale mm}
\psset{linewidth=0.3,dotsep=1,hatchwidth=0.3,hatchsep=1.5,shadowsize=1,dimen=middle}
\psset{dotsize=0.7 2.5,dotscale=1 1,fillcolor=black}
\psset{arrowsize=1 2,arrowlength=1,arrowinset=0.25,tbarsize=0.7 5,bracketlength=0.15,rbracketlength=0.15}
\begin{pspicture}(0,0)(48.62,13.88)
\psline[linewidth=0.22](15.62,13.88)(15.62,7.88)
\psline[linewidth=0.22](8,-2)(8,4)
\newrgbcolor{userFillColour}{1 1 0.4}
\pspolygon[linewidth=0.22,fillcolor=userFillColour,fillstyle=solid](17.62,3.88)
(5.62,3.88)
(10.62,7.88)
(17.62,7.88)(17.62,3.88)
\psline[linewidth=0.22](15.62,-2.5)(15.62,3.88)
\psline[linewidth=0.22](46.88,2.75)(46.88,10.62)
\newrgbcolor{userFillColour}{1 1 0.4}
\pspolygon[linewidth=0.22,fillcolor=userFillColour,fillstyle=solid](36.75,2.88)
(48.62,2.88)
(48.62,-1.12)
(40.62,-1.12)(36.75,2.88)
\psline[linewidth=0.22](46.88,-1.25)(46.88,-5)
\psline[linewidth=0.22](38.75,2.88)(38.75,8.88)
\rput(29,6){$\eqls$}
\rput(12.62,5.88){}
\rput(12.5,6.25){$\algebra$}
\rput(43.75,0.62){$\coalgebra$}
\pscustom[linewidth=0.2,fillstyle=solid]{\psline(36.38,8.25)(41.38,8.25)
\psline(41.38,8.25)(39,11)
\psbezier(39,11)(39,11)(39,11)
\psline(39,11)(36.38,8.25)
\closepath}
\rput[l](16.88,-1.88){$\objA$}
\rput[l](16.88,13.12){$\objA$}
\rput[l](47.5,9.38){$\objA$}
\rput[l](47.5,-4.38){$\objA$}
\pscustom[linewidth=0.2,fillstyle=solid]{\psline(10.75,-1)(5.75,-1)
\psline(5.75,-1)(8.13,-3.75)
\psbezier(8.13,-3.75)(8.13,-3.75)(8.13,-3.75)
\psline(8.13,-3.75)(10.75,-1)
\closepath}
\end{pspicture}

\end{center}

\item $\alpha$ is an $X$-action satisfying the following equivalent conditions
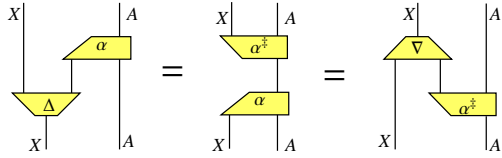
\begin{rnumerate}
\item $(X\otimes \alpha)\circ (\cmn \otimes A)\ =\  \alpha^\ddag \circ \alpha\  = \   (\mnd \otimes A)\circ (X\otimes \alpha^\ddag)$
%\vspace{-.5\baselineskip}
%\newcommand{\eqls}{$\mbox{\Large =}$}
%\newcommand{\objX}{{\scriptstyle X}}
%\newcommand{\objA}{{\scriptstyle A}}
%\newcommand{\comonoid}{\cmn}
%\newcommand{\monoid}{\mnd}
%\newcommand{\coalgebra}{{\scriptstyle \alpha^\ddag}}
%\newcommand{\algebra}{{\scriptstyle \alpha}}
\begin{center}
\def\JPicScale{0.75}
\ifx\JPicScale\undefined\def\JPicScale{1}\fi
\psset{unit=\JPicScale mm}
\psset{linewidth=0.3,dotsep=1,hatchwidth=0.3,hatchsep=1.5,shadowsize=1,dimen=middle}
\psset{dotsize=0.7 2.5,dotscale=1 1,fillcolor=black}
\psset{arrowsize=1 2,arrowlength=1,arrowinset=0.25,tbarsize=0.7 5,bracketlength=0.15,rbracketlength=0.15}
\begin{pspicture}(0,0)(88.62,27.13)
\psline[linewidth=0.22](4.88,11)(4.88,27)
\psline[linewidth=0.22](21.88,27)(21.88,21)
\psline[linewidth=0.22](13.38,11)(13.38,17)
\newrgbcolor{userFillColour}{1 1 0.4}
\pspolygon[linewidth=0.22,fillcolor=userFillColour,fillstyle=solid](2.88,11)
(14.88,11)
(10.88,7)
(6.88,7)(2.88,11)
\psline[linewidth=0.22](8.88,7)(8.88,1)
\newrgbcolor{userFillColour}{1 1 0.4}
\pspolygon[linewidth=0.22,fillcolor=userFillColour,fillstyle=solid](23.88,17)
(11.88,17)
(16.88,21)
(23.88,21)(23.88,17)
\psline[linewidth=0.22](21.88,1)(21.88,17)
\rput(9.38,9){$\comonoid$}
\psline[linewidth=0.22](86.88,10.88)(86.88,26.88)
\psline[linewidth=0.22](74.74,27)(74.74,21)
\psline[linewidth=0.22](70.74,1)(70.74,17)
\newrgbcolor{userFillColour}{1 1 0.4}
\pspolygon[linewidth=0.22,fillcolor=userFillColour,fillstyle=solid](76.74,11)
(88.62,11)
(88.62,7)
(80.62,7)(76.74,11)
\psline[linewidth=0.22](86.88,6.88)(86.88,0.88)
\newrgbcolor{userFillColour}{1 1 0.4}
\pspolygon[linewidth=0.22,fillcolor=userFillColour,fillstyle=solid](80.74,17)
(68.74,17)
(72.62,21)
(76.62,21)(80.74,17)
\psline[linewidth=0.22](78.74,11)(78.74,17)
\rput(31.25,14.38){$\eqls$}
\rput(18.88,19){}
\rput(74.62,19){$\monoid$}
\rput(18.76,19.38){$\algebra$}
\rput(83.74,8.75){$\coalgebra$}
\rput[l](22.5,2.5){$\objA$}
\rput[l](23.12,25){$\objA$}
\rput[l](87.5,25){$\objA$}
\rput[l](87.5,1.88){$\objA$}
\rput[r](8.12,1.88){$\objX$}
\rput[r](70,2.5){$\objX$}
\rput[r](4.38,25){$\objX$}
\rput[r](74.38,25){$\objX$}
\psline[linewidth=0.22](49.87,17.13)(49.87,11.13)
\psline[linewidth=0.22](41.37,1.13)(41.37,7.13)
\newrgbcolor{userFillColour}{1 1 0.4}
\pspolygon[linewidth=0.22,fillcolor=userFillColour,fillstyle=solid](51.87,7.13)
(39.87,7.13)
(44.87,11.13)
(51.87,11.13)(51.87,7.13)
\psline[linewidth=0.22](49.87,0.75)(49.87,7.13)
\psline[linewidth=0.22](49.87,21)(49.87,27)
\newrgbcolor{userFillColour}{1 1 0.4}
\pspolygon[linewidth=0.22,fillcolor=userFillColour,fillstyle=solid](39.75,21.13)
(51.63,21.13)
(51.63,17.13)
(43.63,17.13)(39.75,21.13)
\psline[linewidth=0.22](41.75,21.13)(41.75,27.13)
\rput(46.87,9.13){}
\rput(46.75,9.5){$\algebra$}
\rput(46.75,19.5){$\coalgebra$}
\rput[l](51.25,1.88){$\objA$}
\rput[l](51.25,24.38){$\objA$}
\rput[r](40.62,1.88){$\objX$}
\rput[r](41.25,24.38){$\objX$}
\rput(60,13.75){$\eqls$}
\end{pspicture}

\end{center}
%i.e. $\alpha$ is an $X$-coalgebra homomorphism $\cmn \tto{\alpha} \alpha^\ddag$, whereas $\alpha^\ddag$ is an $X$-algebra homomorphism  $\alpha\tto{\alpha^\ddag} \mnd$
%

%
%\item  
%\bear (X\otimes \alpha)\circ (\cmn \otimes A) & = &  \alpha^\ddag \circ \alpha
%\eear
%\vspace{-.5\baselineskip}
%\begin{center}
%\def\JPicScale{0.7}
%\input{PIC/Mes-b2}
%\end{center}
%\item $\alpha$ is a retract of $\mnd$ and $\alpha^\ddag$ is a retract of $\cmn$
\item $\alpha^\ddag\circ\alpha = (X\otimes \alpha)\circ (c\otimes A)\circ (X\otimes \alpha^\ddag)$
\begin{center}
\def\JPicScale{0.75}
\ifx\JPicScale\undefined\def\JPicScale{1}\fi
\psset{unit=\JPicScale mm}
\psset{linewidth=0.3,dotsep=1,hatchwidth=0.3,hatchsep=1.5,shadowsize=1,dimen=middle}
\psset{dotsize=0.7 2.5,dotscale=1 1,fillcolor=black}
\psset{arrowsize=1 2,arrowlength=1,arrowinset=0.25,tbarsize=0.7 5,bracketlength=0.15,rbracketlength=0.15}
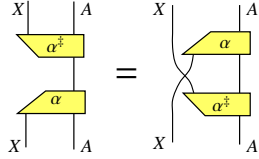
\begin{pspicture}(0,0)(41.88,28.88)
\psline[linewidth=0.22](10.62,18.88)(10.62,12.88)
\psline[linewidth=0.22](2.12,2.88)(2.12,8.88)
\newrgbcolor{userFillColour}{1 1 0.4}
\pspolygon[linewidth=0.22,fillcolor=userFillColour,fillstyle=solid](12.62,8.88)
(0.62,8.88)
(5.62,12.88)
(12.62,12.88)(12.62,8.88)
\psline[linewidth=0.22](10.62,2.5)(10.62,8.88)
\psline[linewidth=0.22](10.62,22.75)(10.62,28.75)
\newrgbcolor{userFillColour}{1 1 0.4}
\pspolygon[linewidth=0.22,fillcolor=userFillColour,fillstyle=solid](0.5,22.88)
(12.38,22.88)
(12.38,18.88)
(4.38,18.88)(0.5,22.88)
\psline[linewidth=0.22](2.5,22.88)(2.5,28.88)
\rput(20,15.62){$\eqls$}
\rput(7.62,10.88){}
\rput(7.5,11.25){$\algebra$}
\rput(7.5,21.25){$\coalgebra$}
\psline[linewidth=0.22](40,28.12)(40,23.5)
\psline[linewidth=0.22](40,12.5)(40,19.5)
\psline[linewidth=0.22](40,3.38)(40,9.38)
\rput(37,21.5){}
\psecurve[linewidth=0.2,curvature=1.0 0.1 0.0](31.88,12.5)(31.88,12.5)(31.25,14.38)(28.75,17.5)(28.12,20.62)(28.12,23.12)(28.12,28.12)(28.12,29.38)
\psecurve[linewidth=0.2,curvature=1.0 0.1 0.0](31.88,20)(31.88,20)(31.25,16.88)(29.38,14.38)(28.12,11.25)(28.12,8.75)(28.12,3.75)(28.12,2.5)
\newrgbcolor{userFillColour}{1 1 0.4}
\pspolygon[linewidth=0.22,fillcolor=userFillColour,fillstyle=solid](30,12.5)
(41.88,12.5)
(41.88,8.5)
(33.88,8.5)(30,12.5)
\rput(36.88,10.62){$\coalgebra$}
\rput[l](11.88,3.12){$\objA$}
\rput[l](11.88,27.5){$\objA$}
\rput[l](41.25,27.5){$\objA$}
\rput[l](41.25,3.12){$\objA$}
\rput[r](1.25,3.12){$\objX$}
\rput[r](26.88,3.75){$\objX$}
\rput[r](1.88,27.5){$\objX$}
\rput[r](26.88,27.5){$\objX$}
\newrgbcolor{userFillColour}{1 1 0.4}
\pspolygon[linewidth=0.22,fillcolor=userFillColour,fillstyle=solid](41.88,19.38)
(29.88,19.38)
(34.88,23.38)
(41.88,23.38)(41.88,19.38)
\rput(37.5,21.25){$\algebra$}
\end{pspicture}

\end{center}
\end{rnumerate}
\ee{anumerate}

The converse $\mbox{(c)}\Longrightarrow\mbox{(a)}\wedge\mbox{(b)}$ holds if the $X$-action $\alpha$ is normal. When this is the case, then also 
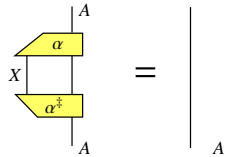
\bear
\alpha \circ \alpha^\ddag & = & \id_A
\eear
\vspace{-1\baselineskip}
\newcommand{\objX}{{\scriptstyle X}}
\newcommand{\objA}{{\scriptstyle A}}
\newcommand{\coalgebra}{{\scriptstyle \alpha^\ddag}}
\newcommand{\algebra}{{\scriptstyle \alpha}}
\begin{center}
\def\JPicScale{0.75}
\ifx\JPicScale\undefined\def\JPicScale{1}\fi
\psset{unit=\JPicScale mm}
\psset{linewidth=0.3,dotsep=1,hatchwidth=0.3,hatchsep=1.5,shadowsize=1,dimen=middle}
\psset{dotsize=0.7 2.5,dotscale=1 1,fillcolor=black}
\psset{arrowsize=1 2,arrowlength=1,arrowinset=0.25,tbarsize=0.7 5,bracketlength=0.15,rbracketlength=0.15}
\begin{pspicture}(0,0)(38,25.12)
\psline[linewidth=0.22](5,18)(5,9)
\rput(26,13){$\eqls$}
\psline[linewidth=0.22](13,25.12)(13,20.5)
\psline[linewidth=0.22](13,9.5)(13,16.5)
\psline[linewidth=0.22](13,0.38)(13,6.38)
\rput(10,18.5){}
\newrgbcolor{userFillColour}{1 1 0.4}
\pspolygon[linewidth=0.22,fillcolor=userFillColour,fillstyle=solid](3,9.5)
(14.88,9.5)
(14.88,5.5)
(6.88,5.5)(3,9.5)
\rput(9.88,7.62){$\coalgebra$}
\rput[l](14.25,24.5){$\objA$}
\rput[l](14.25,0.12){$\objA$}
\rput[r](4,13){$\objX$}
\newrgbcolor{userFillColour}{1 1 0.4}
\pspolygon[linewidth=0.22,fillcolor=userFillColour,fillstyle=solid](14.88,16.38)
(2.88,16.38)
(7.88,20.38)
(14.88,20.38)(14.88,16.38)
\rput(10.5,18.25){$\algebra$}
\rput[l](38,0){$\objA$}
\psline[linewidth=0.22](34,25)(34,0)
\end{pspicture}

\end{center}
\end{thm}

\paragraph{Remarks.} The two equations in Thm.~\ref{Frobcoalg}(i) imply each other by applying the dagger. They also imply that 
\begin{itemize}
\item $X\otimes A\tto{\alpha} A$ is a retract of $X\otimes X\tto\mnd X$ in the category of $X$-actions, along the restriction $\alpha^\ddag:\alpha \rightarrowtail \mnd$, and that 
\item $A\tto{\alpha^\ddag} X\otimes A$ is a retract of $X\tto\cmn X\otimes X$ in the category of $X$-coactions, along the retraction $\alpha:\cmn\twoheadrightarrow \alpha^\ddag$. 
\end{itemize}

The Frobenius condition is the special case of both (i) and (ii), since $\cmn$ and $\mnd$ are just special actions.

\bpr {\bf (a$\iff$ b)} follows directly from Lemma~\ref{meas-lemma}. Part (a) of the lemma says that $\alpha(x)$ is idempotent if and only if  $\alpha$ is an $X$-action. Part (b) says that $\alpha(x)$ is self-adjoint if and only if $\alpha = (\varepsilon \otimes A)\circ \alpha^\ddag$, which is equivalent to $\alpha\circ(x\otimes ) = (x^\ddag\otimes A)\circ \alpha^\ddag$ by the $\eta$-rule, using Thm.~\ref{abs2}(b).

{\bf (a$\Longrightarrow$ii)} is proved as follows:
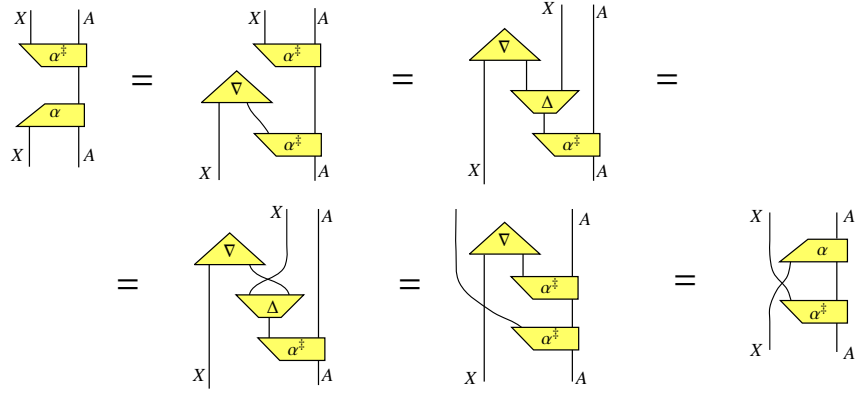
\newcommand{\objX}{{\scriptstyle X}}
\newcommand{\objA}{{\scriptstyle A}}
\newcommand{\comonoid}{\cmn}
\newcommand{\monoid}{\mnd}
\newcommand{\coalgebra}{{\scriptstyle \alpha^\ddag}}
\newcommand{\algebra}{{\scriptstyle \alpha}}
\begin{center}
\def\JPicScale{0.75}
\ifx\JPicScale\undefined\def\JPicScale{1}\fi
\psset{unit=\JPicScale mm}
\psset{linewidth=0.3,dotsep=1,hatchwidth=0.3,hatchsep=1.5,shadowsize=1,dimen=middle}
\psset{dotsize=0.7 2.5,dotscale=1 1,fillcolor=black}
\psset{arrowsize=1 2,arrowlength=1,arrowinset=0.25,tbarsize=0.7 5,bracketlength=0.15,rbracketlength=0.15}
\begin{pspicture}(0,0)(149.38,66.88)
\psline[linewidth=0.22](4.38,38.12)(4.38,45.62)
\psline[linewidth=0.22](13.12,38)(13.12,66)
\newrgbcolor{userFillColour}{1 1 0.4}
\pspolygon[linewidth=0.22,fillcolor=userFillColour,fillstyle=solid](2.62,59.88)
(14.51,59.88)
(14.51,55.88)
(6.5,55.88)(2.62,59.88)
\psline[linewidth=0.22](4.62,59.88)(4.62,65.88)
\rput(9.75,47.88){}
\rput(9.62,58.25){$\coalgebra$}
\psline[linewidth=0.22](147.5,30)(147.5,25.38)
\psline[linewidth=0.22](147.5,14.38)(147.5,21.38)
\psline[linewidth=0.22](147.5,5.25)(147.5,11.25)
\rput(144.5,23.38){}
\psecurve[linewidth=0.2,curvature=1.0 0.1 0.0](139.38,14.38)(139.38,14.38)(138.75,16.25)(136.25,19.38)(135.62,22.5)(135.62,25)(135.62,30)(135.62,31.25)
\psecurve[linewidth=0.2,curvature=1.0 0.1 0.0](139.38,21.88)(139.38,21.88)(138.75,18.75)(136.88,16.25)(135.62,13.12)(135.62,10.62)(135.62,5.62)(135.62,4.38)
\newrgbcolor{userFillColour}{1 1 0.4}
\pspolygon[linewidth=0.22,fillcolor=userFillColour,fillstyle=solid](137.5,14.38)
(149.38,14.38)
(149.38,10.38)
(141.38,10.38)(137.5,14.38)
\rput(144.38,12.5){$\coalgebra$}
\rput[l](14.01,40.12){$\objA$}
\rput[l](14.01,64.5){$\objA$}
\rput[l](148.75,29.38){$\objA$}
\rput[l](148.75,5){$\objA$}
\rput[r](3.38,40.12){$\objX$}
\rput[r](134.38,5.62){$\objX$}
\rput[r](4,64.5){$\objX$}
\rput[r](134.38,29.38){$\objX$}
\newrgbcolor{userFillColour}{1 1 0.4}
\pspolygon[linewidth=0.22,fillcolor=userFillColour,fillstyle=solid](149.38,21.25)
(137.38,21.25)
(142.38,25.25)
(149.38,25.25)(149.38,21.25)
\rput(145,23.12){$\algebra$}
\rput(24.38,53){$\eqls$}
\psline[linewidth=0.22](55,35.62)(55,66.25)
\rput[l](55.62,37.38){$\objA$}
\psecurve[linewidth=0.2,curvature=1.0 0.1 0.0](46.88,43.12)(46.88,43.75)(45.88,45.38)(43.38,48.5)(42.62,51)(42.62,53)(42.62,53)(42.62,60)
\psline[linewidth=0.22](38,35.75)(38,49.5)
\rput[r](36.75,37){$\objX$}
\newrgbcolor{userFillColour}{1 1 0.4}
\psline[linewidth=0.22,fillcolor=userFillColour,fillstyle=solid](34.88,49.5)
(41.12,55.12)
(47.38,49.5)(34.88,49.5)
\rput(41.12,52){$\monoid$}
\newrgbcolor{userFillColour}{1 1 0.4}
\pspolygon[linewidth=0.22,fillcolor=userFillColour,fillstyle=solid](44.12,59.88)
(56,59.88)
(56,55.88)
(48,55.88)(44.12,59.88)
\psline[linewidth=0.22](46.12,59.88)(46.12,65.88)
\rput(51.12,58.25){$\coalgebra$}
\rput[l](55.5,64.5){$\objA$}
\rput[r](45.5,64.5){$\objX$}
\newrgbcolor{userFillColour}{1 1 0.4}
\pspolygon[linewidth=0.22,fillcolor=userFillColour,fillstyle=solid](14.12,45.26)
(2.12,45.26)
(7.12,49.26)
(14.12,49.26)(14.12,45.26)
\rput(9,47.62){$\algebra$}
\newrgbcolor{userFillColour}{1 1 0.4}
\pspolygon[linewidth=0.22,fillcolor=userFillColour,fillstyle=solid](44.24,44)
(56.12,44)
(56.12,40)
(48.12,40)(44.24,44)
\rput(51.25,42.38){$\coalgebra$}
\rput(71.88,16.88){$\eqls$}
\psline[linewidth=0.22](100.62,0)(100.62,30.62)
\rput[l](101.25,0.62){$\objA$}
\psecurve[linewidth=0.2,curvature=1.0 0.1 0.0](92.88,7.12)(92.88,7.75)(91.88,9.38)(86.25,12.5)(80.62,17.5)(80,23.12)(80,30.62)(80,41.88)
\newrgbcolor{userFillColour}{1 1 0.4}
\pspolygon[linewidth=0.22,fillcolor=userFillColour,fillstyle=solid](89.75,18.63)
(101.63,18.63)
(101.63,14.63)
(93.63,14.63)(89.75,18.63)
\psline[linewidth=0.22](91.88,18.75)(91.88,24.38)
\rput(96.75,17){$\coalgebra$}
\rput[l](101.88,28.75){$\objA$}
\rput(120.62,17.5){$\eqls$}
\newrgbcolor{userFillColour}{1 1 0.4}
\pspolygon[linewidth=0.22,fillcolor=userFillColour,fillstyle=solid](89.87,9.62)
(101.74,9.62)
(101.74,5.62)
(93.74,5.62)(89.87,9.62)
\rput(96.88,8.01){$\coalgebra$}
\psline[linewidth=0.22](85,0)(85,22.5)
\rput[r](83.75,0.62){$\objX$}
\newrgbcolor{userFillColour}{1 1 0.4}
\psline[linewidth=0.22,fillcolor=userFillColour,fillstyle=solid](82.38,22.62)
(88.62,28.25)
(94.88,22.62)(82.38,22.62)
\rput(88.62,25.12){$\monoid$}
\psline[linewidth=0.22](104.38,35.62)(104.38,66.88)
\rput[l](105,37.38){$\objA$}
\psecurve[linewidth=0.2,curvature=1.0 0.1 0.0](50.62,12.5)(50.62,14.38)(50,16.25)(43.75,20)(43.12,23.12)(43.12,23.12)(43.12,23.12)(43.12,32.5)
\rput[r](84.38,37.5){$\objX$}
\psline[linewidth=0.22](95.62,42.5)(95.62,48.5)
\rput[l](105,65.62){$\objA$}
\newrgbcolor{userFillColour}{1 1 0.4}
\pspolygon[linewidth=0.22,fillcolor=userFillColour,fillstyle=solid](93.62,44)
(105.5,44)
(105.5,40)
(97.5,40)(93.62,44)
\rput(100.62,42.38){$\coalgebra$}
\psecurve[linewidth=0.2,curvature=1.0 0.1 0.0](43.12,10.62)(43.25,14.75)(43.75,16.25)(49.38,20)(50,21.88)(50,24.38)(50,30.62)(50,32.5)
\psline[linewidth=0.22](92.5,51.25)(92.5,58.75)
\psline[linewidth=0.22](98.75,51.25)(98.75,66.88)
\newrgbcolor{userFillColour}{1 1 0.4}
\pspolygon[linewidth=0.22,fillcolor=userFillColour,fillstyle=solid](89.76,51.62)
(101.76,51.62)
(97.76,47.62)
(93.76,47.62)(89.76,51.62)
\rput(96.26,49.62){$\comonoid$}
\psline[linewidth=0.22](85,35)(85,57.5)
\newrgbcolor{userFillColour}{1 1 0.4}
\psline[linewidth=0.22,fillcolor=userFillColour,fillstyle=solid](82.38,57)
(88.62,62.62)
(94.88,57)(82.38,57)
\rput(88.62,59.5){$\monoid$}
\rput[r](97.5,66.25){$\objX$}
\rput(70.62,53.12){$\eqls$}
\rput(117.5,53.12){$\eqls$}
\psline[linewidth=0.22](55.62,-0.62)(55.62,30.62)
\rput[l](56.25,1.13){$\objA$}
\rput[r](35.62,1.25){$\objX$}
\psline[linewidth=0.22](46.88,6.25)(46.88,12.25)
\rput[l](56.25,29.38){$\objA$}
\newrgbcolor{userFillColour}{1 1 0.4}
\pspolygon[linewidth=0.22,fillcolor=userFillColour,fillstyle=solid](44.87,7.75)
(56.75,7.75)
(56.75,3.75)
(48.75,3.75)(44.87,7.75)
\rput(51.88,6.13){$\coalgebra$}
\newrgbcolor{userFillColour}{1 1 0.4}
\pspolygon[linewidth=0.22,fillcolor=userFillColour,fillstyle=solid](41,15.38)
(53,15.38)
(49,11.38)
(45,11.38)(41,15.38)
\rput(47.5,13.38){$\comonoid$}
\psline[linewidth=0.22](36.25,-1.25)(36.25,21.25)
\newrgbcolor{userFillColour}{1 1 0.4}
\psline[linewidth=0.22,fillcolor=userFillColour,fillstyle=solid](33.62,20.75)
(39.88,26.37)
(46.12,20.75)(33.62,20.75)
\rput(39.88,23.25){$\monoid$}
\rput[r](49.38,30){$\objX$}
\rput(21.88,16.88){$\eqls$}
\end{pspicture}

\end{center}
using Lemma~\ref{meas-lemma}, and the commutativity of $\cmn$.

{\bf (ii$\Longrightarrow$i)} is a variation on the same theme:
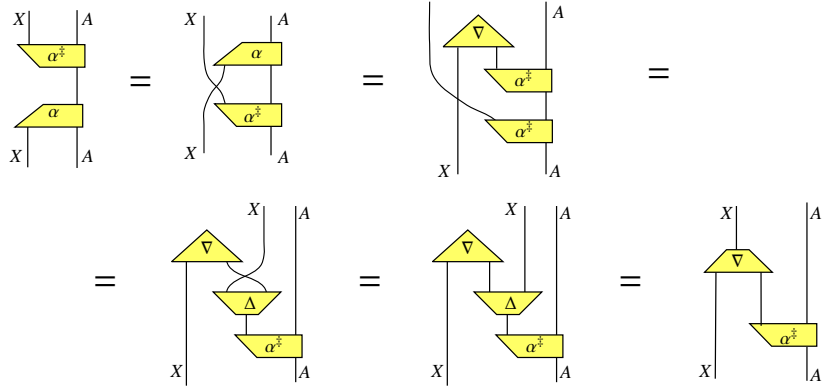
\begin{center}
\def\JPicScale{0.75}
\ifx\JPicScale\undefined\def\JPicScale{1}\fi
\psset{unit=\JPicScale mm}
\psset{linewidth=0.3,dotsep=1,hatchwidth=0.3,hatchsep=1.5,shadowsize=1,dimen=middle}
\psset{dotsize=0.7 2.5,dotscale=1 1,fillcolor=black}
\psset{arrowsize=1 2,arrowlength=1,arrowinset=0.25,tbarsize=0.7 5,bracketlength=0.15,rbracketlength=0.15}
\begin{pspicture}(0,0)(144.24,67.5)
\psline[linewidth=0.22](4.38,38.12)(4.38,45.62)
\psline[linewidth=0.22](13.12,38)(13.12,66)
\newrgbcolor{userFillColour}{1 1 0.4}
\pspolygon[linewidth=0.22,fillcolor=userFillColour,fillstyle=solid](2.62,59.88)
(14.51,59.88)
(14.51,55.88)
(6.5,55.88)(2.62,59.88)
\psline[linewidth=0.22](4.62,59.88)(4.62,65.88)
\rput(9.75,47.88){}
\rput(9.62,58.25){$\coalgebra$}
\psline[linewidth=0.22](47.5,65)(47.5,60.38)
\psline[linewidth=0.22](47.5,49.38)(47.5,56.38)
\psline[linewidth=0.22](47.5,40.25)(47.5,46.25)
\rput(44.5,58.38){}
\psecurve[linewidth=0.2,curvature=1.0 0.1 0.0](39.38,49.38)(39.38,49.38)(38.75,51.25)(36.25,54.38)(35.62,57.5)(35.62,60)(35.62,65)(35.62,66.26)
\psecurve[linewidth=0.2,curvature=1.0 0.1 0.0](39.38,56.88)(39.38,56.88)(38.75,53.75)(36.88,51.25)(35.62,48.12)(35.62,45.62)(35.62,40.62)(35.62,39.38)
\newrgbcolor{userFillColour}{1 1 0.4}
\pspolygon[linewidth=0.22,fillcolor=userFillColour,fillstyle=solid](37.5,49.38)
(49.38,49.38)
(49.38,45.38)
(41.38,45.38)(37.5,49.38)
\rput(44.38,47.5){$\coalgebra$}
\rput[l](14.01,40.12){$\objA$}
\rput[l](14.01,64.5){$\objA$}
\rput[l](48.75,64.38){$\objA$}
\rput[l](48.75,40){$\objA$}
\rput[r](3.38,40.12){$\objX$}
\rput[r](34.38,40.62){$\objX$}
\rput[r](4,64.5){$\objX$}
\rput[r](34.38,64.38){$\objX$}
\newrgbcolor{userFillColour}{1 1 0.4}
\pspolygon[linewidth=0.22,fillcolor=userFillColour,fillstyle=solid](49.38,56.25)
(37.38,56.25)
(42.38,60.25)
(49.38,60.25)(49.38,56.25)
\rput(45,58.12){$\algebra$}
\rput(24.38,53){$\eqls$}
\newrgbcolor{userFillColour}{1 1 0.4}
\pspolygon[linewidth=0.22,fillcolor=userFillColour,fillstyle=solid](14.12,45.26)
(2.12,45.26)
(7.12,49.26)
(14.12,49.26)(14.12,45.26)
\rput(9,47.62){$\algebra$}
\rput(65.62,53.12){$\eqls$}
\psline[linewidth=0.22](96.25,36.88)(96.25,67.5)
\rput[l](96.88,37.5){$\objA$}
\psecurve[linewidth=0.2,curvature=1.0 0.1 0.0](88.5,44)(88.5,44.62)(87.5,46.25)(81.88,49.38)(76.25,54.38)(75.62,60)(75.62,67.5)(75.62,78.75)
\newrgbcolor{userFillColour}{1 1 0.4}
\pspolygon[linewidth=0.22,fillcolor=userFillColour,fillstyle=solid](85.38,55.5)
(97.26,55.5)
(97.26,51.5)
(89.26,51.5)(85.38,55.5)
\psline[linewidth=0.22](87.5,55.62)(87.5,61.25)
\rput(92.38,53.88){$\coalgebra$}
\rput[l](97.5,65.62){$\objA$}
\rput(116.25,54.38){$\eqls$}
\newrgbcolor{userFillColour}{1 1 0.4}
\pspolygon[linewidth=0.22,fillcolor=userFillColour,fillstyle=solid](85.49,46.5)
(97.37,46.5)
(97.37,42.5)
(89.37,42.5)(85.49,46.5)
\rput(92.5,44.88){$\coalgebra$}
\psline[linewidth=0.22](80.62,36.88)(80.62,59.38)
\rput[r](79.38,37.5){$\objX$}
\newrgbcolor{userFillColour}{1 1 0.4}
\psline[linewidth=0.22,fillcolor=userFillColour,fillstyle=solid](78,59.5)
(84.25,65.12)
(90.5,59.5)(78,59.5)
\rput(84.25,62){$\monoid$}
\psline[linewidth=0.22](98.12,0)(98.12,31.25)
\rput[l](98.75,1.75){$\objA$}
\psecurve[linewidth=0.2,curvature=1.0 0.1 0.0](46.88,13.12)(46.88,15)(46.25,16.88)(40,20.62)(39.38,23.75)(39.38,23.75)(39.38,23.75)(39.38,33.12)
\rput[r](78.12,1.88){$\objX$}
\psline[linewidth=0.22](89.38,6.88)(89.38,12.88)
\rput[l](98.75,30){$\objA$}
\newrgbcolor{userFillColour}{1 1 0.4}
\pspolygon[linewidth=0.22,fillcolor=userFillColour,fillstyle=solid](87.37,8.38)
(99.24,8.38)
(99.24,4.38)
(91.24,4.38)(87.37,8.38)
\rput(94.38,6.76){$\coalgebra$}
\psecurve[linewidth=0.2,curvature=1.0 0.1 0.0](39.38,11.25)(39.5,15.38)(40,16.88)(45.62,20.62)(46.25,22.5)(46.25,25)(46.25,31.25)(46.25,33.12)
\psline[linewidth=0.22](86.25,15.62)(86.25,23.12)
\psline[linewidth=0.22](92.5,15.62)(92.5,31.25)
\newrgbcolor{userFillColour}{1 1 0.4}
\pspolygon[linewidth=0.22,fillcolor=userFillColour,fillstyle=solid](83.5,16)
(95.5,16)
(91.5,12)
(87.5,12)(83.5,16)
\rput(90,14){$\comonoid$}
\psline[linewidth=0.22](78.75,-0.62)(78.75,21.88)
\newrgbcolor{userFillColour}{1 1 0.4}
\psline[linewidth=0.22,fillcolor=userFillColour,fillstyle=solid](76.12,21.38)
(82.38,27)
(88.62,21.38)(76.12,21.38)
\rput(82.38,23.88){$\monoid$}
\rput[r](91.25,30.62){$\objX$}
\rput(65,17.5){$\eqls$}
\rput(111.25,17.5){$\eqls$}
\psline[linewidth=0.22](51.88,0)(51.88,31.25)
\rput[l](52.5,1.75){$\objA$}
\rput[r](31.88,1.88){$\objX$}
\psline[linewidth=0.22](43.12,6.88)(43.12,12.88)
\rput[l](52.5,30){$\objA$}
\newrgbcolor{userFillColour}{1 1 0.4}
\pspolygon[linewidth=0.22,fillcolor=userFillColour,fillstyle=solid](41.12,8.38)
(53,8.38)
(53,4.38)
(45,4.38)(41.12,8.38)
\rput(48.12,6.75){$\coalgebra$}
\newrgbcolor{userFillColour}{1 1 0.4}
\pspolygon[linewidth=0.22,fillcolor=userFillColour,fillstyle=solid](37.25,16)
(49.25,16)
(45.25,12)
(41.25,12)(37.25,16)
\rput(43.75,14){$\comonoid$}
\psline[linewidth=0.22](32.5,-0.62)(32.5,21.88)
\newrgbcolor{userFillColour}{1 1 0.4}
\psline[linewidth=0.22,fillcolor=userFillColour,fillstyle=solid](29.88,21.38)
(36.12,27)
(42.38,21.38)(29.88,21.38)
\rput(36.12,23.88){$\monoid$}
\rput[r](45.62,30.62){$\objX$}
\rput(18.12,17.5){$\eqls$}
\psline[linewidth=0.22](142.5,10.26)(142.5,31.88)
\psline[linewidth=0.22](130,31.25)(130,23.12)
\psline[linewidth=0.22](126.25,0.62)(126.37,19.5)
\newrgbcolor{userFillColour}{1 1 0.4}
\pspolygon[linewidth=0.22,fillcolor=userFillColour,fillstyle=solid](132.36,10.38)
(144.24,10.38)
(144.24,6.38)
(136.24,6.38)(132.36,10.38)
\psline[linewidth=0.22](142.5,6.25)(142.5,0.25)
\newrgbcolor{userFillColour}{1 1 0.4}
\pspolygon[linewidth=0.22,fillcolor=userFillColour,fillstyle=solid](136.36,19.5)
(124.36,19.5)
(128.24,23.5)
(132.24,23.5)(136.36,19.5)
\psline[linewidth=0.22](134.38,10)(134.36,19.5)
\rput(130.24,21.5){$\monoid$}
\rput(139.36,8.12){$\coalgebra$}
\rput[l](143.12,30){$\objA$}
\rput[l](143.12,1.26){$\objA$}
\rput[r](125.62,1.88){$\objX$}
\rput[r](129.38,30){$\objX$}
\end{pspicture}

\end{center}

Finally, if the $X$-action $\alpha$ is normal, then postcomposing (i) with $\cun\otimes A$ gives condition \ref{meas-lemma}(b), and hence (a).

$\alpha\circ\alpha^\ddag = \id_A$ is left as an exercise.
\epr

\be{prop}
The category $\Mes{\CCC}{\tp{x}{X}}$ of measurements over $X$ is equivalent with the category $\Ac\CCC X$ of $X$-actions.
%
%The category of normal $X$-actions is equivalent with the subcategory of the category  of Eilenberg-Moore coalgebras for the comonad induced by $(X,\cmn,\cun)$, spanned by the coalgebras that satisfy the equivalent conditions \ref{Frobcoalg}(a-f). Such coalgebras are just the retracts of the cofree coalgebras $\cmn A: XA\to XXA$. 
\ee{prop}

\subsection{Measuring the outcome}
In general, the measurement outcome corresponding to a basis vector is the pure projector that represents it. In order to perform the measurement in the first component of $S|x,y>$ from sec.~\ref{Unbiased}, we use a partial representation of this vector.

\be{lemma} $\sigma_y(x) = (\mnd_m \otimes \id_n) \circ S|x,y>$ is a measurement on  ${\rect}^{\otimes (m+n)}$ in $\Mes{\pw\Rel\left[\tp{|y>}{{\rect}^{\otimes n}}\right]}{\tp{|x>}{{\rect}^{\otimes m}}}$.
\ee{lemma}
Substituting the basis vectors for $x$ in $\sigma_y(x)$ gives the projectors on ${\rect}^{\otimes (m+n)}$, from which the information about the period $c$ is extracted like before.

\section{Conclusions and future work}
Simon's algorithm required three operations:
\begin{description}
\item[abstraction: ] to represent classical functions and classical data in a quantum universe;
\item[transform to a complementary basis: ] to entangle classical data and make use of quantum parallelism;
\item[measurement: ] to extract the classical outcomes of quantum computation.
\end{description}
The abstraction operations shape the classical interfaces of quantum computers. Our analysis of the general abstraction operations uncovered a rich structure, that may be of interest beyond quantum computation. Are there other computational resources, besides entanglement, that provide exponential speedup when suitably combined with the general abstraction operations?

The other two operations that we formalized are typically quantum. Complementary bases provide access to entanglement, as the main resource of quantum computation, and thus enable quantum parallelism. The varied interactions among the different classical structures and with measurements give rise to the wealth of quantum algorithms that remain to be explored. 

Our abstract model uncovered some abstract entanglement structures, and made them available for quantum computation in non-standard mathematical models. The algorithmic consequences of this semantical result need to be carefully explored. 

%The robust notion of computability, postulated by Church's Thesis served as a solid and useful foundation of Computer Science in its first phase, understanding the

%The {\bf essence of functional programming} is that {\em all data can be copied or deleted}. Abstraction is an unrestricted operation.

%The {\bf essence of quantum programming} is that {\em data that cannot be copied or deleted can be entangled}. This is a powerful computational resource.

%\paragraph{Acknowledgements.} Without Samson Abramsky and Bob Coecke, I would never attempted to understand anything quantum. Bob asked me to present a part of this work at his TANCL workshop in 2007, and Samson to present another part at his Clifford Lectures in 2008. Any shortcomings are a consequence of the fact that no further invitations were received. 

\bibliographystyle{plain}
\bibliography{ref-Qabs,../PavlovicD}

\end{document}